\documentclass[nobib, reprint, amsmath,amssymb, aps]{revtex4-2}
\usepackage{graphicx}
\usepackage{bm}
\usepackage[colorlinks]{hyperref}
\usepackage{stackrel} 
\usepackage{xcolor} 
\usepackage{natbib}

\usepackage{subcaption}
\newcommand{\be}{\begin{equation}}
\newcommand{\ee}{\end{equation}}
\newcommand{\bea}{\begin{eqnarray}}
\newcommand{\eea}{\end{eqnarray}}
\newcommand{\lb}{\label}

\newcommand{\bv}{{\bf v}}
\newcommand{\bu}{{\bf u}}
\newcommand{\bw}{{\bf w}}
\newcommand{\bk}{{\bf k}}

\newcommand{\bx}{{\bf x}}
\newcommand{\br}{{\bf r}}
\newcommand{\by}{{\bf y}}
\newcommand{\bz}{{\bf z}}

\newcommand{\BF}{{\bf F}}

\newcommand{\bG}{{\bf G}}

\newcommand{\bD}{{\bf D}}

\newcommand{\bR}{{\bf R}}

\newcommand{\bgamma}{{\mbox{\boldmath $\gamma$}}}

\newcommand{\boeta}{{\mbox{\boldmath $\eta$}}}

\newcommand{\bsigma}{{\mbox{\boldmath $\sigma$}}}

\newcommand{\bcA}{{\mbox{\boldmath ${\mathcal A}$}}}
\newcommand{\bcP}{{\mbox{\boldmath ${\mathcal P}$}}}
\newcommand{\bcU}{{\mbox{\boldmath ${\mathcal U}$}}}
\newcommand{\bcV}{{\mbox{\boldmath ${\mathcal V}$}}}

\newcommand{\grad}{{\mbox{\boldmath $\nabla$}}}
\newcommand{\bdot}{{\mbox{\boldmath $\cdot$}}}

\newcommand{\bzed}{{\mbox{\boldmath $0$}}}

\newcommand{\baro}{\bar{{\rm o}}}

\newcommand{\red}[1]{{{\color{black}#1}}}

\begin{document}



\title{High Schmidt-Number Turbulent Advection \\ and Giant Concentration Fluctuations}

 \author{Gregory Eyink${\,\!}^{1,2}$}
 \email{eyink@jhu.edu}
 
\author{Amir Jafari${\,\!}^1$}
 \email{ajafari4@jhu.edu}

\affiliation {${\,\!}^1$Department of Applied Mathematics and Statistics\\ 
and ${\,\!}^2$Department of Physics \& Astronomy, \\
The Johns Hopkins University, Baltimore, MD, USA}

\begin{abstract}
We consider the effects of thermal noise on the Batchelor-Kraichnan theory of high Schmidt-number mixing in the viscous dissipation range of turbulent flows at sub-Kolmogorov scales. Starting with the nonlinear Landau-Lifschitz fluctuating 
hydrodynamic equations for a binary fluid mixture at low Mach numbers, we justify linearization around the deterministic Navier-Stokes solution in the dissipation range. For the latter solution we adopt the standard Kraichnan model, a Gaussian random velocity with spatially-constant strain but white-noise in time. Then, following prior work of Donev, Fai \& vanden-Eijnden, we derive asymptotic high-Schmidt limiting equations for the concentration field, in which the thermal velocity fluctuations are exactly represented by a Gaussian random velocity that is likewise white in time. We obtain the exact solution for concentration spectrum in this high-Schmidt limiting model, showing that the Batchelor prediction in the viscous-convective range is unaltered. Thermal noise dramatically renormalizes the bare diffusivity in this range, but the effect is the same as in laminar flow and thus hidden phenomenologically. However, in the viscous-diffusive range at scales below the Batchelor length (typically micron scales) the predictions based on deterministic Navier-Stokes equations are drastically altered by thermal noise. Whereas the classical theories predict rapidly decaying spectra in the viscous-diffusive range, either Gaussian or exponential, we obtain a $k^{-2}$  power-law spectrum over a couple of decades starting just below the Batchelor length. This spectrum corresponds to non-equilibrium giant concentration fluctuations (GCF's), which are due to the imposed concentration variations being advected by thermal velocity fluctuations and which are experimentally well-observed in quiescent fluids. At higher wavenumbers, the concentration spectrum instead goes to a $k^2$ equipartition spectrum due to equilibrium molecular fluctuations. We work out detailed predictions for water-glycerol and water-fluorescein mixtures. Finally, we discuss broad implications for turbulent flows and novel applications of our methods to experimentally accessible laminar flows.  
\end{abstract}

\pacs{Valid PACS appear here}
\maketitle


\section{\label{sec:level1}Introduction}

Recent work \cite{bandak2021thermal,eyink2021dissipation,bell2021thermal,gallis2021turbulence} 
has sparked renewed interest in the effects of thermal noise on turbulent flow, a problem 
much neglected since the pioneering work of Betchov more than 60 years ago 
\cite{betchov1957fine,betchov1961thermal,betchov1964measure}. These new studies 
have confirmed Betchov's insight that the dissipation range of turbulent flows
must be strongly affected by thermal noise. In particular, the energy spectrum 
below the Kolmogorov dissipation scale \cite{kolmogorov1941local}, which has 
long been expected to exhibit an exponential decay 
\cite{kraichnan1959structure,frisch1981intermittency,foias1990empirical,frisch1991prediction,sirovich1994energy,khurshid2018energy,gorbunova2020analysis,buaria2020turbulence}, 
instead demonstrates an equilibrium equipartition energy spectrum in numerical 
simulations which incorporate molecular noise \cite{bandak2021thermal,eyink2021dissipation,bell2021thermal}.
The question remains which turbulent processes at sub-Kolmogorov length scales 
can be essentially altered by such noise. Prominent among the candidates 
for essential modifications is the turbulent mixing of a high Schmidt-number passive 
concentration field. On the one hand, the viscous-convective and viscous-diffusive 
ranges in the Batchelor theory \cite{batchelor1959small} of turbulent mixing 
occur entirely at scales below the Kolmogorov scale. On the other hand, thermal 
fluctuations have long been known to produce striking effects in the diffusion of scalar 
concentration in laminar flows, including a renormalization of the diffusion 
constant \cite{bedeaux1974renormalization,mazur1974renormalization,hynes1979molecular} 
and large, long-range nonequilibrium correlations of concentration fluctuations 
\cite{nieuwoudt1990theory,law1989noncritical,segre1993nonequilibrium}. We shall therefore focus in 
this work on the effects of thermal noise in the turbulent mixing of a binary fluid 
mixture at high Schmidt numbers. 

Past research on turbulent scalar mixing has rested upon the assumption that the 
relevant advecting velocity field must solve the deterministic Navier-Stokes equation 
below the Kolmogorov scale $\ell_K=\nu^{3/4}\varepsilon^{-1/4}$ (with $\nu$ the kinematic viscosity 
and $\varepsilon$ the energy dissipation rate per mass) and down to nearly the 
mean-free-path of the fluid. Assuming that the smoothing 
effects of viscosity would produce a velocity field with nearly constant gradient
at lengths well below the Kolmogorov scale, Batchelor in \cite{batchelor1959small}
constructed a model of distortion of small blobs of scalar field by straining motions which were 
assumed to be statistically sharp and time-independent in a coordinate system fixed in the blobs. 
The chief prediction of Batchelor's model with velocity-gradient frozen in time was a cascade 
of scalar fluctuations \red{$c'$} at wavenumbers $k\ell_K\gtrsim 1$ with Fourier spectrum  
\be E_c(k)\sim C_B(\chi/\gamma)k^{-1}\exp(-C_B(k\ell_B)^2/2) \lb{Ba59} \ee
where \red{$\int_0^\infty E_c(k)dk=(1/2)\langle |c'|^2\rangle$} and $\chi$ is the rate of injection of scalar fluctuations 
(or, in a statistical steady-state, the rate of dissipation $\chi=D\langle |\grad c|^2\rangle$
for diffusivity $D$), $\gamma=(\varepsilon/\nu)^{1/2}$ is the strain-rate at the Kolmogorov scale, 
and $\ell_B=(D/\gamma)^{1/2}$ is the scalar dissipation length-scale, now termed 
the ``Batchelor length.''  Note that $\ell_K/\ell_B=Sc^{1/2}$ where $Sc=\nu/D$ is the Schmidt number. 
The constant prefactor $C_B$ in the power-law spectrum $E_c(k)\sim C_B (\chi/\gamma)k^{-1}$ 
in the viscous-convection range for $1/\ell_K\lesssim k\lesssim 1/\ell_B$ is often presumed 
universal and called the ``Batchelor constant.'' In his subsequent works, Kraichnan \cite{kraichnan1968small,kraichnan1974convection} argued that the turbulent velocity-gradient 
in the sub-Kolmogorov scales, while constant in space, was rapidly varying in time and 
he proposed another soluble model in which the advecting random velocity field was taken 
to be Gaussian, white-noise in time. Kraichnan's model predicts a spectrum at
wavenumbers  $k\ell_K\gtrsim 1$ of the form 
\be  E_c(k)\sim C_B (\chi/\gamma k)(1+\sqrt{6C_B}k\ell_B) \exp\left(-\sqrt{6C_B} k\ell_B\right), 
\lb{Kr68} \ee 
reproducing Batchelor's $1/k$ spectrum in the viscous-convective range but exhibiting  
exponential decay in the viscous-diffusive range for $k\ell_B\gtrsim 1.$ Such spectra 
have been widely expected in the turbulence community to hold down to nearly molecular scales. 

Later studies have largely verified these predictions of Batchelor and Kraichnan. 
A recent article of Sreenivasan succinctly reviews observations 
both by experiment and by simulation \cite{sreenivasan2019turbulent}. 
Experiments have been performed in laboratory flows or by field measurements in 
the ocean, both for concentration fields  and for temperature fields. Note 
that it has largely been assumed in the turbulence literature that, when buoyancy 
effects are negligible, advection of concentration at high Schmidt numbers
and of temperature at high Prandtl number will be equivalent. Experiments
supporting the Batchelor $k^{-1}$ spectrum have been performed for concentration 
\cite{nye1967scalar,jullien2000experimental,iwano2021power}, for temperature 
\cite{grant1968spectrum,oakey1982determination}, and for both simultaneously 
\cite{gibson1963universal}. These studies either did not measure spectra in the 
viscous-diffusive range or did not resolve with enough accuracy to discriminate 
between the distinct predictions of Batchelor and Kraichnan. The experimental
picture is a bit unclear, furthermore, as some other laboratory experiments 
with high-Schmidt dye as solute have reported spectra different than 
the predicted $k^{-1}$ \cite{miller1996measurements,williams1997mixing}. 
Most recent studies have resorted instead to numerical simulations of incompressible 
Navier-Stokes turbulence \cite{yeung2004simulations,donzis2010batchelor,gotoh2015spectrum,
clay2017strained}, obtaining thereby increasingly long intervals of $k^{-1}$ spectra in 
the viscous-convective range and furthermore strong evidence in favor of 
Kraichnan's exponential decay spectrum \eqref{Kr68} over Batchelor's prediction 
\eqref{Ba59} in the viscous-diffusion range. The Batchelor regime of high-Schmidt 
scalar advection has achieved an iconic status as ``a rare thing in turbulence theory'' 
\cite{kraichnan1974convection} where exact results are possible. The theory has 
since been extended in various ways, e.g. to allow for finite correlation time 
of the advecting velocity (see \cite{balkovsky1999universal} and further 
references in \cite{falkovich2001particles}), and recently a rigorous mathematical 
proof has even been given of the Batchelor $k^{-1}$ spectrum in a forced 2D 
Navier-Stokes flow \cite{bedrossian2019batchelor}. 

There is reason to believe, however, that thermal fluctuations at sub-Kolmogorov 
scales will fundamentally change the picture of high Schmidt-number turbulent 
advection. Indeed, thermal fluctuations have long been known to have profound 
effects on high-Schmidt mixing, prominently diffusion in liquids. One of the oldest 
pieces of evidence is the Stokes-Einstein relation 
\be D=k_BT/6\pi\eta \sigma, \lb{SE} \ee
which connects the diffusivity $D$ to fluid temperature $T,$ 
shear viscosity $\eta$ and the radius $\sigma$ of a spherical particle in solution. 
Significantly, this relation has long been known to be empirically 
valid quite generally for solutes in liquids, with $\sigma$ close to the 
particle radius. To quote from a seminal 1945 paper of Onsager on liquid diffusion:
\begin{quote}
``the ratio 
$ l= kT/D\eta$ 
is a length of the order of magnitude of molecular dimensions, normally smaller 
than the value $6\pi a$ ... 

From the point of view of molecular theory, viscous flow and diffusion present parallel problems. It would seem that for an exact theory of either, we should have to analyze the cooperative character of the molecular motion involved; but this difficult analysis has not yet been developed further than the hydrodynamic approximation.'' \cite{onsager1945theories}
\end{quote}
The appearance of shear viscosity in the empirical Stokes-Einstein relation 
thus hints that mass diffusion and momentum diffusion are strongly coupled processes.  
Indeed, the observed diffusivity $D$ in liquids generally differs considerably from the ``bare'' collisional diffusivity $D_0$ predicted by Enskog kinetic theory \cite{hynes1979molecular}. 

A second striking piece of evidence for the importance of thermal noise is the 
``giant concentration fluctuations'' observed both in free diffusive mixing \cite{vailati1997giant,brogioli2000giant} and in non-equilibrium steady-states 
with an imposed concentration gradient \cite{li1998concentration,vailati2011fractal}. 
These effects were predicted using linearized fluctuating hydrodynamics
\cite{nieuwoudt1990theory,law1989noncritical,segre1993nonequilibrium}
and are a particular instance of the spatial long-range correlations of fluctuations  
which are generic for systems away from global thermodynamic equilibrium  \cite{kirkpatrick1982fluctuations,ronis1982nonlinear,dorfman1994generic,
dezarate2006hydrodynamic}. The basic prediction involves the static structure function 
$S_{cc}(k)$ defined in terms of the Fourier transform
of concentration fluctuations $\widehat{c}'(\bk)$  by 
\be \langle \widehat{c}'(\bk)\widehat{c}'(\bk')\rangle =(2\pi)^3\delta^3(\bk\red{+}\bk')
S_{cc}(k).  \lb{Scc-def} \ee 
It should be noted here that this structure function is related to the 
scalar spectrum commonly considered in turbulence theory by the relation 
$E_c(k)= \frac{1}{(2\pi)^2}k^2 S_{cc}(k).$ 
The result which has been confirmed by experiment 
is a power-law scaling 
\be  S_{cc}(k)\sim  \frac{k_B T}{D\eta} |\grad c|^2 k^{-4} \lb{GCF} \ee 
down to very low wavenumbers, limited only by the fluid domain size or by buoyancy 
effects \cite{zarate2001fluctuations,zarate2002boundary}. In low-gravity environments
these fluctuations are truly ``giant'', growing to macroscopic scales and with 
amplitudes orders of magnitude larger than equilibrium concentration fluctuations 
\cite{vailati1997giant,vailati2011fractal}. The scale-invariance corresponds to 
fractality of the concentration isosurfaces, which are being  advected by thermal 
velocity fluctuations with long-range correlations induced by pressure forces. 
These striking non-equilibrium fluctuation effects have been the subject 
of many experimental investigations, including the up-coming NEUF-DIX microgravity 
experiment of the European Space Agency \cite{baaske2016neuf,vailati2020giant}.
It is worth remarking that these long-range fluctuation correlations and the 
Stokes-Einstein relation for diffusivity are not necessarily independent 
manifestations of thermal noise but may be connected by heuristic arguments 
\cite{brogioli2000diffusive}. 

A remarkable link of such thermal effects with turbulence theory has been discovered 
in the work of Donev, Fai \& vanden-Eijnden \cite{donev2014reversible} (hereafter, DFV). 
In most liquids, a large separation of time scales exists between the fast viscous 
dynamics of the thermal velocity fluctuations and the slow diffusive dynamics of solute 
molecules, i.e. momentum diffusion proceeds much faster than mass diffusion. DFV 
exploited this fact to develop an exact high-$Sc$ asymptotic reduction of the 
equations of fluctuating hydrodynamics for a binary fluid mixture under the condition 
of incompressible, isothermal flow. Importantly, the DFV theory does not linearize 
the equation for the concentration field and treats nonlinear advection exactly. 
The conclusion of the DFV analysis is a reduced stochastic equation for individual 
realizations of the concentration field on long, diffusive time-scales in which the 
scalar is advected by a modified thermal velocity field which is Gaussian, white-noise 
in time. Thus, the long-time, high-$Sc$ limiting equation for the concentration field 
is a version of the exactly soluble Kraichnan model \cite{kraichnan1968small,kraichnan1974convection,
falkovich2001particles} which has been widely used to study turbulent scalar advection. 

As a result, the DFV theory yields exact closed equations for the correlation functions 
of all orders in 
the scalar concentration field. In particular, DFV showed that the equation for the  
ensemble-average concentration field exhibits a renormalization of the bare molecular 
diffusivity $D_0$ and yields naturally the Stokes-Einstein diffusivity $D$ as 
a renormalized ``eddy-diffusivity'' due to advection by thermal velocity fluctuations. 
As discussed in \cite{donev2014reversible}, the effective stochastic equations for 
individual realizations of the concentration field are furthermore more efficient to solve numerically 
than the original fluctuating hydrodynamics equations, by a factor of $Sc,$ since the fast viscous 
dynamics of the thermal velocity fluctuations has been eliminated. DFV demonstrated 
in numerical simulations of free diffusive mixing that these model equations produce 
the fractal scalar interfaces which are observed experimentally and also 
power-law GCF's of the concentration. They did not, however, observe clearly the 
$k^{-4}$ scaling \eqref{GCF} of the concentration structure function, as predicted by linearized theory, but instead observed a scaling closer to $k^{-3}$ in the 
quasi-steady regime of decay. It has therefore been unclear how to reconcile the 
DFV asymptotic theory with the experimental observations verifying the prediction \eqref{GCF}. 

In this paper we shall illuminate the latter issue and, furthermore, we  
generalize the DFV theory to include turbulent advection by combining 
it with the original approach of Kraichnan \cite{kraichnan1968small,kraichnan1974convection}.
In this manner, we can study analytically the effects of thermal noise in the sub-Kolmogorov scales on high-Schmidt turbulent advection. We choose to consider here 
a statistically stationary turbulent cascade with injection 
of concentration fluctuations at a constant rate $\chi$ at a length-scale 
$L\gtrsim \ell_K$ by a stochastic source field. 
We find that the  Batchelor $k^{-1}$ scalar spectrum in the viscous-convective 
interval is unaffected by thermal noise, despite the rapid decay of 
kinetic energy spectrum in sub-Kolmogorov scales being replaced by a $k^2$
equipartition spectrum. Working in physical space, we find more precisely that 
the steady-state concentration correlation function $C(r)=\langle c'(\br)c'(\bzed)\rangle$, \red{with $r=|{\bf r}|$,}
exhibits the logarithmic scaling 
\be
C(r)\sim C(\ell_K) + C_B\frac{\chi}{\gamma} \ln(\ell_K/r),\qquad 
\ell_B\lesssim r\lesssim \ell_K \lb{VC} \ee
whose Fourier transform yields exactly the Batchelor-Kraichnan $k^{-1}$ spectrum. 
Our key finding, however, is that giant concentration fluctuations with a $k^{-2}$
power-law scalar spectrum occur in the viscous-diffusive range, replacing the 
rapidly decaying spectra predicted by Batchelor and Kraichnan. In physical space we get 
\be C(r)\sim C(0)-\frac{\chi}{2D}\left( r^2+3\sigma r+\cdots\right),
\qquad \sigma\lesssim r\lesssim \ell_B \lb{VD} \ee
where $\sigma$ is a length of order of the radius of the solute particle. 
The first term in \eqref{VD} is the one $\propto \langle |\grad c|^2\rangle r^2$ 
which is expected for a smooth concentration field and which would arise from 
the rapidly decaying spectra of Batchelor and Kraichnan. The second term appears 
to be subleading and negligible until $r\simeq \sigma.$ However, this term 
is {\it non-analytic} in $r$ and on Fourier transforming produces a $k^{-2}$ 
power-law which dominates the spectrum for $k\ell_B\gtrsim 1.$ We obtain 
an exact solution for the scalar spectrum $E_c(k)$ of our model in terms 
of known special functions, which exemplifies this behavior. 
Note using the Stokes-Einstein relation \eqref{SE} and 
$\chi=D\langle|\grad c|^2\rangle$ that the Fourier transform of the 
term $\propto r$ in \eqref{VD} yields the concentration spectrum 
\bea \lb{key}
&& E_{c}(k) \sim  \frac{\chi\sigma}{\pi D} k^{-2} 
\sim \frac{1}{6\pi^2} \frac{k_BT}{D\eta} \langle |\grad c|^2\rangle k^{-2},  \\\nonumber
&& \hspace{120pt} \quad k\ell_B\gtrsim 1  \nonumber
\eea 
which, except for being smaller by a factor of 2/3, corresponds exactly to the structure 
function scaling in \eqref{GCF} associated to the giant concentration 
fluctuations observed experimentally in laminar flows. 
Eventually, at higher wavenumbers, thermal equilibrium fluctuations of the 
concentration field must begin to dominate and an equipartition 
spectrum $E_c(k)\propto k^2$ should appear; \red{for a detailed discussion of the equilibrium spectrum see \S\ref{physics} and Appendix \ref{FDT}}. Because of this effect 
of molecular fluctuations, the concentration gradients $\grad c$ become dependent 
upon a high-wavenumber cut-off $\Lambda$ in the model and the estimates 
by the relation $\langle|\grad c|^2\rangle=\chi/D$ must be interpreted 
as ``effective gradients'' holding over a certain range of scales. This somewhat 
subtle issue will be discussed at length in the following.

\begin{figure}
\includegraphics[width=9cm]{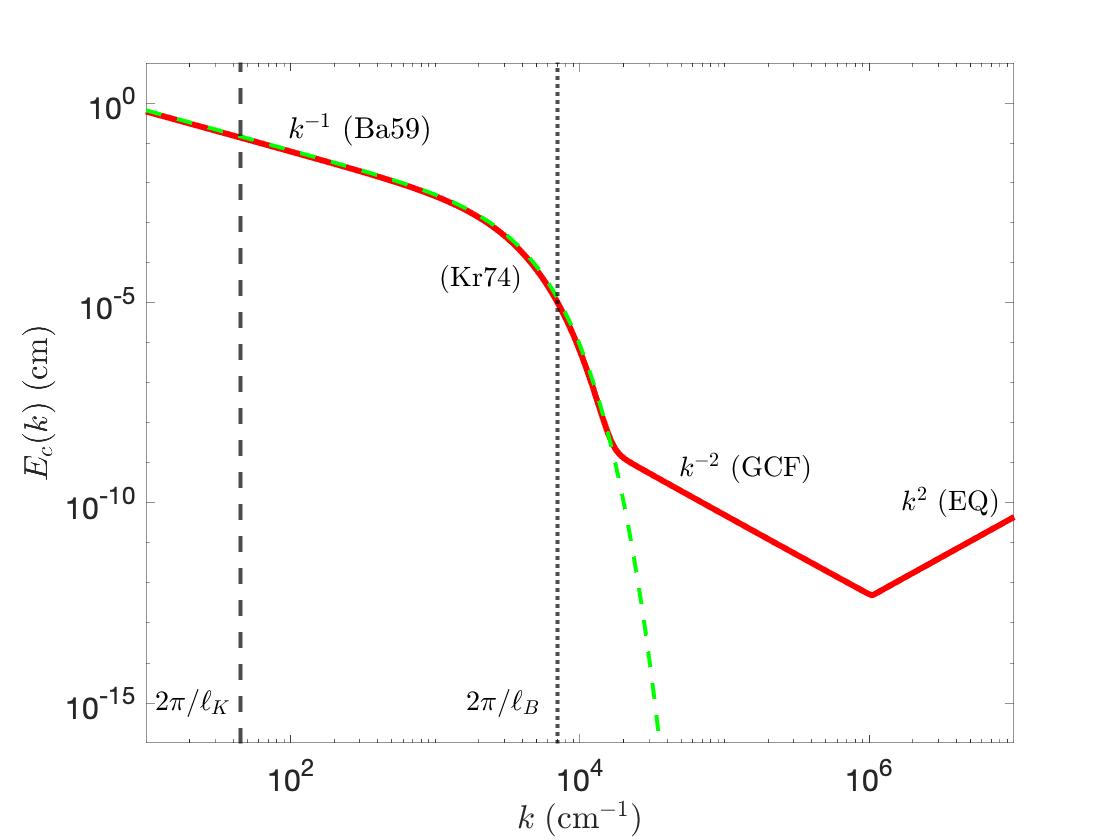}
\caption {\footnotesize {Our predicted scalar concentration spectrum (red solid line, 
\textcolor{red}{ $\boldsymbol -$}) and the prediction of Kraichnan \cite{kraichnan1974convection} 
(Kr74; green dashed line, \textcolor{green}{{\bf -$\,$-$\,$-}}), 
for a water-glycerol solution at temperature $T=25^\circ$C, pressure $p=1$ bar 
and mean concentration of glycerol $\bar{c}=0.5$, with $\gamma=10^2\;{\rm s}^{-1}$
and $\chi=10^2\;{\rm s}^{-1}.$ Distinct ranges of the concentration spectrum are labelled: 
Batchelor's $k^{-1}$ spectrum (Ba59); $k^{-2}$ power-law associated to giant concentration 
fluctuations (GCF); $k^2$ equipartition spectrum (EQ). \red{The vertical dashed ({\bf -$\,$-$\,$-})
and dotted lines ({\bf $\cdot\!\cdot\!\cdot\!\cdot\!\cdot$}) indicate, respectively, the Kolmogorov wavenumber 
$2\pi/\ell_K=44.7\;\text{cm}^{-1}$ and Batchelor wavenumber $2\pi/\ell_B=7034.5\;\text{cm}^{-1}$.} }}
\label{Intro}
\end{figure}

The theoretical predictions of 
our analysis for the concentration spectrum are illustrated in 
Fig.\ref{Intro} for the specific case of a water-glycerol 
mixture. The Kolmogorov turnover rate $\gamma=10^2\;{\rm s}^{-1}$ 
is chosen very close to that in recent fluid turbulence experiments with 
water-glycerol solutions in a von K\'arm\'an flow \cite{debue2018experimental}.
No experiments have been performed on turbulent high-Schmidt mixing with 
water-glycerol mixtures, as far as we are aware, so that we have chosen 
$\chi=10^2\;{\rm s}^{-1}$ from one of the \red{more} recent laboratory experiments 
with a water solution of disodium fluorescein \cite{jullien2000experimental}. 
In this hypothetical experiment, we predict more than two decades of power-law spectrum 
$E_c(k)\propto k^{-2}$ associated to giant concentration fluctuations appearing  
at scales just below the Batchelor length, which is here $\ell_B=1.98\,\mu$m. 

It is important to emphasize that, even if there is no turbulent velocity field 
and the scalar is advected by thermal velocity fluctuations only, then the power-law
\eqref{key} still holds, extending down now to wavenumbers $k\simeq 1/L$ in our steady-state model 
with a random, spatially-distributed source of concentration fluctuations.
Including turbulent shear, these GCF's of thermal origin are supplanted 
by the Batchelor-Kraichnan $k^{-1}$ spectrum of concentration fluctuations at the 
wavenumbers $k\ell_B\lesssim 1$ in the viscous-convective range. This effect 
is similar to the ``shear-quenching'' of GCF's predicted for small departures 
from global equilibrium (weak shear) using linearized fluctuating hydrodynamics 
\cite{wada2004shear}, although the turbulent $k^{-1}$ spectrum differs substantially 
from the $k^{2/3}$ spectrum predicted for the weakly sheared case. There has been 
some question whether such shear-quenching will hold in experimentally realizable
flows, with gravity and finite-size effects argued instead to limit the GCF's 
at low waveumbers \cite{zarate2006comment}. We have not included buoyancy in 
our analysis, but this was done in the work of DFV and gravity effects can thus
be considered, in principle, within our framework. Unfortunately, including gravity
in the asymptotic mode reduction of DFV for $Sc\gg 1$ produces an effective 
equation for concentration with a term {\it quadratic} in $c,$ due 
to advection of concentration by self-induced velocity fluctuations arising from buoyancy.
(See eq.\eqref{bouyc} later in the text.) Because of this quadratic nonlinearity, closed
equations are no longer obtained for the correlation functions of the concentration, fundamentally complicating mathematical analysis. We shall comment more 
on this issue below. 

The main message of our work for turbulence theory is that thermal noise completely 
alters the character of the viscous-diffusive range of high Schmidt-number turbulent advection, leading to fundamentally different predictions than those based on deterministic
Navier-Stokes dynamics. This is likely to be true also for other physical processes
in turbulent flows that involve essentially the sub-Kolmogorov scale motions, 
such as combustion \cite{sreenivasan2004possible,driscoll2008turbulent,echekki2010turbulent}, condensation \cite{saito2018turbulence,elghobashi2019direct,milan2020sub} 
and locomotion of micro-organisms \cite{durham2013turbulence,wheeler2019not,michalec2020efficient}, 
not to speak of the intrinsic nonlinear turbulent dynamics itself. 
The presence of giant concentration fluctuations 
in turbulent flows should not have been unexpected, because they are a generic 
feature of diffusive mixing far from global equilibrium. To quote from the paper 
of Vailati and Giglio: 
\begin{quote}
``So the orders-of-magnitude increase of the fluctuations above the equilibrium value (the most 
prominent feature that can be captured experimentally) is to be expected for any non-equilibrium 
fluid that has macroscopic concentration variations comparable to those in this experiment.''
\cite{vailati1997giant}
\end{quote} 
It will, unfortunately, be probably very difficult in the near future to observe these effects in laboratory or field experiments on high-Schmidt turbulent advection, because 
the Batchelor length $\ell_B$ is generally near micron scales and no current 
experimental techniques can probe such small scales in a turbulent flow
\red{with the required accuracy}. \red{The most recent experiment on turbulent 
high-Schmidt mixing of which we are aware \cite{iwano2021power} measures
concentration fluctuations via laser-induced fluorescence with an optical fiber probe 
having a spatial resolution of 2.8 $\mu$m, which is close to the Batchelor scale. 
However, instrumental noise dominates the measurements before even getting to this scale; 
see \cite{iwano2021power}, Fig.3}. 
It should be quite possible, on the other hand, to test our predictions by means of numerical simulations 
of high Schmidt-number turbulent mixing with existing codes for low Mach-number 
fluctuating Navier-Stokes equations of multi-component \cite{donev2014low} 
and binary \cite{nonaka2015low} mixtures. 

The main message of our work for statistical physics is that methods from turbulence theory 
provide effective tools to study nonequilibrium thermal fluctuations more generally, 
as originally suggested by DFV \cite{donev2014reversible}. The DFV theory applies 
not only to turbulent flows but also to laminar flows, such as the 
free diffusive mixing of an initial blob of concentration in a quiescent (zero Reynolds-number) fluid. 
The DFV theory treats nonlinear advection of scalar concentration exactly without need 
for linearization, and, as seen from our result \eqref{key}, it is able to recover GCF's with the 
scaling $S_{cc}(k)\sim k^{-4}$ which is observed experimentally. This approach is thus 
able to deal with large-amplitude fluctuations driven by strong gradients, high concentrations, 
and non-steady transient diffusion processes, which are difficult 
theoretical problems driving current empirical investigations such as NEUF-DIX. We therefore expect that the first experimental
tests of validity of these methods will come from novel applications to diffusive mixing in laminar flows.  

Because our paper is somewhat lengthy, it is useful to briefly outline 
its contents. The next section \ref{S2} discusses the DFV asymptotic theory 
of the high Schmidt-number limit, first reviewing the original work for zero 
mean flow (\ref{rest}) and then extending that analysis to turbulent 
flows, assuming Kraichnan's standard model for the dissipation-range velocity field 
(\ref{S2A}). In the main mathematical section \ref{S3} we exactly solve the limiting 
model for the concentration spectrum in a statistical steady-state with constant 
injection of scalar fluctuations. We first review necessary background on the 
Kraichnan white-noise advection model and mathematical methods employed in its 
solution (\ref{SKraichnan}). We apply these methods to solve for the 
static 2-point correlation function of concentration fluctuations (\ref{ourmodel}) 
and then compute analytically its Fourier transform to obtain the 
concentration spectrum (\ref{SEnergySpectrum}). In section \ref{physics} 
we develop concrete predictions of our theory for turbulent mixing of 
water-glycerol and water-fluorescein, and in the final section \ref{final} 
we discuss implications and possible extensions of our work. 
Several appendices (\ref{FDT}-\ref{2DGCF}) provide technical details 
of the derivations, background material for easy reference, and 
numerical methods for plotting our analytical results. 

\section{DFV Theory and High $Sc$ Limit}\label{S2}

In this section, we shall first concisely review the work of DFV  \cite{donev2014reversible} 
on diffusion of scalar concentration in the asymptotic limit of large Schmidt numbers. 
DFV considered the problem where the fluid is at rest, in global equilibrium, and performed a formal 
adiabatic mode-elimination procedure for the fast thermal velocity degrees of freedom.  In the limit
they obtained reduced model equations for the scalar concentration field in which the effective 
advecting velocity is Gaussian and white-noise in time (Kraichnan velocity), so that closed equations 
follow for all scalar correlation functions. We shall here extend the asymptotic analysis of DFV to 
a turbulent fluid in the Kolmogorov dissipation range, adopting further Kraichnan's white-noise 
velocity approximation for the turbulent velocity field. The result of the adiabatic elimination 
is another Kraichnan model for the scalar concentration field, in which the Gaussian, white-in-time 
velocity field has two independent contributions representing advection by turbulent fluctuations 
and by thermal fluctuations. The resulting closed equations for the scalar 2-point correlations in this 
reduced model will be solved exactly in the following section \ref{S3}. 

\subsection{Fluid at Rest}\lb{rest} 

In a fluid at rest, i.e., with no large-scale motion, thermal fluctuations produce the entire 
velocity field $\bv=\bv_\theta.$ For low Mach-number, isothermal fluids, DFV adopted 
a standard model of linearized incompressible fluctuating Navier-Stokes equation for the 
velocity field $\bv$
\bea\label{Momentum100}
\rho\partial_t {\bf v}&=&-\grad p+ \eta \triangle {\bf v}+\grad\bdot \Big(\sqrt{ 2\eta k_B T}\; {\boldsymbol\eta}({\bf x}, t) \Big)  \cr 
&=&\bcP\Big[ \eta \triangle {\bf v}+\grad\bdot \Big(\sqrt{ 2\eta k_B T}\; {\boldsymbol\eta}({\bf x}, t) \Big)  \Big]
\eea 
where $\rho$, $\eta$, and $T$, represent, respectively, the mass density,  shear viscosity, and temperature, 
all assumed to be constant, and $k_B$ is Boltzmann's constant. Also, $p$ is the kinematic pressure, which may be 
replaced by the tensor operator ${\cal P}_{ij}=\delta_{ij}-\partial_i\partial_j\triangle^{-1},$ the Leray-Hodge projection 
onto the space of divergence-free velocity fields, so that  the incompressibility constraint 
\be \grad\bdot \bv=0 \lb{incomp} \ee 
is maintained.  The white-noise symmetric, traceless tensor field ${\boldsymbol \eta}({\bf x}, t)$ represents 
a thermal fluctuating stress, with mean zero and covariance
\begin{eqnarray}\lb{etav} 
\langle  \eta_{ij}({\bf x}, t) \eta_{kl}({\bf x'}, t')\rangle&=&(\delta_{ik}\delta_{jl}+\delta_{il}\delta_{jk}-{2\over 3} \delta_{ij}\delta_{kl})\cr 
&&\times \delta^3({\bf x-x'})\delta(t-t').
\end{eqnarray}
The prefactor $\sqrt{ 2\eta k_B T}$ is chosen according to the standard fluctuation-dissipation relation 
so that the correct Gibbs equilibrium distribution is obtained for the equal-time velocity statistics, with energy equipartition
among wave-number modes. For example, see \cite{eyink2021dissipation}, Appendix A, for a careful discussion. 

For the scalar concentration field $c({\bf x}, t)$ in a binary mixture of two identical fluids of molecular mass $m$,
DFV adopted the fluctuating advection-diffusion equation 
\be\label{Passive110}
\partial_t c=-\bu\bdot\grad c +\grad\bdot \left(D_0\grad c+\sqrt{2mD_0 \rho^{-1} c(1-c)} \; {\boldsymbol\eta}_c({\bf x}, t) \right),
\ee
where $\bu$ is a smoothed advection velocity (see below), $D_0$ is the bare molecular diffusivity, and ${\boldsymbol\eta}_c({\bf x}, t)$ 
is a white-noise vector field representing a thermal fluctuating mass flux, with zero mean and covariance
\begin{eqnarray}\lb{etac} 
&&\left\langle  {\eta_c}_i({\bf x}, t) {\eta_c}_j({\bf x'}, t') \right\rangle=\delta_{ij}\delta(t-t')\delta^3({\bf x-x'}).
\end{eqnarray}
See also \cite{donev2011enhancement}.  
Here again the factor $\sqrt{2mD_0 \rho^{-1} c(1-c)} $ in the noise term is dictated by the fluctuation-dissipation 
relation, so that the concentration fluctuations have their equilibrium equal-time statistics given by the 
Boltzmann-Einstein formula determined from the entropy of mixing. See  Appendix \ref{FDT}
for this standard argument in statistical thermodynamics. Note that DFV considered 
only the limit of low concentrations $c\ll 1$ in their work, so that they took $c(1-c)\doteq c$, 
but in our analysis we allow arbitrarily high concentrations. Finally, a key step in the theory of DFV 
was to assume that the concentration field of tracer particles (tagged particles of the fluid, solute 
molecules, colloidal particles, etc.)  is advected by a coarse-grained velocity $\bf u$ obtained by convolving 
$\bf v$ with a smoothing kernel $\boldsymbol\sigma$, 
\begin{eqnarray}\label{regularized-velocity100}
\bu(\bx, t)&\equiv& {\boldsymbol\sigma}\star{\bf v}=\int {\boldsymbol\sigma}(\bx, \bx'){\bf v}(\bx', t) \, d^3x'.
\end{eqnarray}
This convolution filters out features at scales below a cutoff scale $\sigma$, taken to be of order of the typical linear 
size of a tracer particle. 

The theoretical justification for the starting equations \eqref{Momentum100},\eqref{Passive110}
of the DFV theory must be discussed briefly. The fluctuating hydrodynamic equations of a general 
binary mixture with non-constant density and temperature fields have been derived in \cite{morozov1984langevin} 
by the phenomenological arguments of statistical thermodynamics, based upon the corresponding hydrodynamic equations 
(see \cite{zubarev1974nonequilibrium}, \S 22.7). In principle, these stochastic equations should be derivable by the 
Zwanzig-Mori projection methods which have been applied to obtain fluctuating hydrodynamics for simple,
single-component fluids \cite{zubarev1983statistical,espanol2009microscopic}. An important point 
which becomes clear from these derivations is that the equations of fluctuating hydrodynamics such as 
\eqref{Momentum100},\eqref{Passive110} should not be considered as continuum stochastic partial 
differential equations. Instead, they are low-wavenumber effective theories which describe the physics
only of modes at wavenumbers less than some cutoff $\Lambda,$ generally taken to be of order 
the inverse of the mean-free-path length. Thus, the spatial delta functions which appear in 
the covariances \eqref{etav},\eqref{etac} should in fact be interpreted as ``cutoff delta-functions''
$\delta_\Lambda(\bx-\bx');$ see \eqref{cutoff-del} in Appendix \ref{FDT} and \cite{zubarev1983statistical}. 

To obtain the final form of the fluctuating hydrodynamic equations \eqref{Momentum100},\eqref{Passive110}, 
the low Mach number, isothermal limit must be taken. This has been carefully considered for a binary or 
a general multi-component fluid in \cite{donev2014low,nonaka2015low}.
This analysis leads to equations close to \eqref{Momentum100},\eqref{Passive110}, except that 
the incompressibility constraint differs from \eqref{incomp} and the momentum equation \eqref{Momentum100}
contains the nonlinear advection term $(\bu\bdot\grad)\bu.$ As to the first, the constraint on the velocity derived 
in \cite{nonaka2015low} for the binary fluid mixture in the low-Mach limit is 
$$ \grad\bdot \bv=-\grad\bdot (\beta \BF) $$
where $\BF=D_0\grad c+\sqrt{2mD_0 \rho^{-1} c(1-c)} \; {\boldsymbol\eta}_c$ and where $\beta(c)=(1/\rho)
(\partial\rho/\partial c)_{P_0,T_0}$ is the solutal expansion coefficient at background 
pressure $P_0$ and temperature $T_0.$ The above constraint thus reduces to \eqref{incomp}
if either volume changes little with concentration ($\beta$ near zero)
or if the bare diffusivity $D_0$ is negligible, as DFV explicitly assume. The second 
difference, the neglect of the nonlinear advection term, is justified by the renormalization group 
analysis of \cite{forster1976long,forster1977large}, which implies that in thermal equilibrium 
the nonlinearity becomes negligible at sufficiently low wavenumbers and frequencies. A quantitative 
estimate provided in \cite{eyink2021dissipation} implies that the nonlinear coupling should be weak 
except for length scales of order the radius of the fluid molecules, where no hydrodynamic 
description is valid in any case. Finally, the key assumption of DFV that tracer particles are 
advected by the smoothed velocity field \eqref{regularized-velocity100} is intuitively plausible,  
since such particles can feel only a resultant velocity averaged over fluctuations at a scale smaller 
than their size $\sigma.$ This hypothesis is  further motivated in \cite{donev2014reversible}, with 
reference to earlier works such as \cite{hynes1979molecular} on the modeling of fluid-tracer interactions 
in diffusive mixing. 
 
%


The essential result of DFV is an exact analysis of the high Schmidt-number limit, $Sc_0\equiv {\eta\over D_0 \rho}\gg 1$,
for the model equations \eqref{Momentum100},\eqref{Passive110}. Motivated by the empirical success of the Stokes-Einstein 
relation $D\sim k_B T/ \eta\sigma$, DFV introduced a small parameter $\epsilon\ll 1$ to order quantities 
for formal asymptotics and adopted a scaling 
\be \eta\mapsto \epsilon^{-1}\eta, \quad D_0\mapsto \epsilon D_0 \lb{trans-to} \ee
so that $D_0 \eta\simeq (const.)$ and $Sc_0\sim \epsilon^{-2}.$  In the limit $\epsilon\ll 1$ there is a 
separation of time scales between the fast viscous dynamics, governing the thermal velocity fluctuations ${\bf v}$, 
and the slow diffusive evolution of the concentration field $c.$ DFV formalized this separation by introducing 
a ``macroscopic'' diffusive time $\tau$ which is related to the ``microscopic'' 
viscous time $t$ of Eqs. \eqref{Momentum100},\eqref{Passive110} 
by \red{$t=\epsilon^{-1} \tau,$} or, equivalently, by the scaling
\be t\mapsto \epsilon^{-1}t \lb{time-to} \ee 
with $\tau$ renamed $t$. These scalings can be used in a formal 
adiabatic mode-elimination of the fast velocity degrees of freedom, which will be discussed in detail in 
section \ref{S2c} for the more general case of a turbulent flow. The result is a limiting stochastic advection-diffusion 
equation for the concentration field in the ``macroscopic'' time:
\be\label{PassiveStratonovich100}
\partial_t c=-{\bf w}\odot\grad c+D_0 \triangle c+\grad\bdot  \Big(\sqrt{2mD_0\rho^{-1} c(1-c)} \; {\boldsymbol\eta}_c \Big).
\ee
Here $\odot$ represents a Stratonovich dot product and ${\bf w}({\bf x}, t)$ is an incompressible, advecting 
random velocity field which is white noise in time, with zero mean and covariance 
\bea \label{SmoothVelocityCovariance101}
&& \langle {\bf w}({\bf x}, t) \otimes {\bf w}({\bf x'}, t') \rangle={ \cal {\bf R}} ({\bf x, x'}) \delta(t-t'), \cr 
&& { \cal {\bf R}} ({\bf x, x'}):=2 \int_0^\infty \langle {\bf u(x}, t) \otimes {\bf u(x'}, t+t')   \rangle dt'
\eea
and which can thus be shown to be given by 
\be 
{ \cal {\bf R}} ({\bf x, x'})=\frac{2 k_B T}{\eta}(\bsigma\star\bG\star\bsigma^\top)(\bx, \bx') \label{SmoothVelocityCovariance102}
\ee
where $\bf G$ is the Green's function of the linear Stokes operator $\bcA=-\bcP\triangle,$ or the so-called {\it Oseen tensor}. For example, 
in unbounded 3D space 
\be G^{ij}(\bx, \bx')=G^{ij}(\bx- \bx')={1\over 8\pi r}(\delta^{ij}+{r^ir^j\over r^2}) \lb{oseen} \ee 
with $\br=\bx-\bx'.$  Note that $\bG$ is singular for $\bx=\bx'$, but the smoothed tensor $\bR$ is regular at coinciding points. 
The spatial realizations of $\bw$ are obtained from the stationary Stokes equation with smoothed thermal forcing 
\bea \lb{wStokes}
&&\bcP\left[ \nu\triangle\bw+\grad\bdot\left( \sqrt{\frac{2\nu k_BT}{\rho}} \boeta_\sigma\right)\right] \cr 
&& \hspace{10pt} =-\grad q+ \nu\triangle\bw+\grad\bdot\left( \sqrt{\frac{2\nu k_BT}{\rho}} \boeta_\sigma \right)=\bzed
\eea 
with $\boeta_\sigma=\bsigma\star\boeta$ and $q$ determined by $\grad\bdot\bw=0.$ This equation expresses the physics that viscous diffusion and 
smoothed thermal fluctuations are in instantaneous balance for the effective velocity $\bw,$ with long-range spatial correlations induced 
by the incompressibility constraint.

%

Although the change from equation \eqref{Passive110} to \eqref{PassiveStratonovich100} in the limit $\epsilon\to 0$ 
for the concentration field may seem minor, a crucial physical contribution is obscured in (\ref{PassiveStratonovich100})
by the stochastic calculus. Converting instead to the equivalent It$\baro$ form, which is most appropriate to calculate 
ensemble averages, produces 
\begin{eqnarray}\nonumber
\partial_t c&=& -{\bf w}\bdot\grad c+D_0 \triangle c+\grad\bdot ({\bf D(x)}\grad c )\\\label{PassiveIto100}
&&   +\grad\bdot \Big(\sqrt{2mD_0\rho^{-1} c(1-c)} \; {\boldsymbol\eta}_c \Big),
\end{eqnarray}
with an additional drift term $\grad\bdot ({\bf D(x)}\grad c )$ which contains a {\it renormalized diffusivity} 
\be {\bf D(x)}=\frac{1}{2} { \cal {\bf R}} ({\bf x, x}). \lb{D-def} \ee 
The physical origin of this addition to diffusivity is advection by the eliminated thermal velocity 
fluctuations, similar to an ``eddy-diffusivity'' due to eliminated turbulent eddies. Under further assumptions 
of homogeneity and isotropy, $\bD$ becomes independent of $\bx$ and $D_{ij}=D\delta_{ij},$ with the 
enhanced scalar diffusivity $D$ calculated in \cite{donev2014reversible} for a particular choice of filter 
kernel $\bsigma$ as 
\be
D={k_B T\over 6\pi\eta\sigma}\Big(1-{\sqrt{2}\over 2}{\sigma\over L}  \Big), \lb{D-sig} 
\ee
where $L$ is the linear dimension of the flow domain  and $\sigma\ll L$ is the size of the tracer particle 
and where space dimension is 3. The size-dependent correction of order 
$O(\sigma/L)$ is tiny for macroscopic systems but is a well-known effect 
in molecular dynamics (MD) studies of diffusion coefficients \cite{yeh2004system,celebi2021finite}. 

Importantly,  \eqref{D-sig} corresponds exactly to the Stokes-Einstein formula for the renormalized diffusivity $D$
which, since $D_0\ll D,$ dominates in the total effective diffusivity $D_{e\! f\! f}=D_0+D\doteq D.$ Although the 
filter kernel $\bsigma$ was specially selected in \cite{donev2014reversible} to produce the precise 
numerical prefactor in the original Stokes-Einstein relation for hard spheres, an arbitrary kernel 
yields \eqref{D-sig} with $6\pi$ replaced by some other numerical constant of order unity. Thus the 
DFV theory explains the empirical success of the Stokes-Einstein formula as the effect of strong 
renormalization of a small bare diffusivity due to advection of tracer particles by thermal 
velocity fluctuations. This is one of the significant results of the DFV theory.  Note that similar augmented 
diffusivities have been obtained in renormalization group studies of a passive scalar advected by thermal velocity 
fluctuations, without the assumption of high Schmidt numbers \cite{forster1976long,forster1977large}.
However, those studies did not incorporate the fluctuation-dissipation relation for the scalar and thus have 
uncertain relevance to physical diffusion processes.

A further important consequence of the DFV theory, which, however, was not fully utilized 
in \cite{donev2014reversible}, is the existence of closed equations for the scalar correlation functions 
of any order. This result shall be exploited in section \ref{S3} where it will be shown that the DFV theory
predicts the well-known giant concentration fluctuations, but without the usual approximation of linearizing 
the advection term in the concentration equation. It is this capability to deal with nonlinear advection 
which makes the DFV approach particularly useful to study the scalar concentration field in a turbulent flow.

\subsection{Turbulent Flow}\label{S2A}

For turbulent flow, the full nonlinear form of the fluctuating hydrodynamics equation 
\cite{forster1976long,forster1977large,donev2014low,nonaka2015low,bell2021thermal}
must be used  
\begin{eqnarray}\lb{FNS} 
\partial_t{\bf v}  &=&\bcP\Big[ -({\bf v}\bdot\grad){\bf v}+\nu\triangle\bv+\grad\bdot \Big( \sqrt{2\nu k_B T/\rho } \; {\boldsymbol \eta}({\bf x}, t)\Big)\Big], \cr 
\, & \, 
\end{eqnarray}
where $\nu=\eta/\rho$ is kinematic viscosity and where the white-noise term $\boeta$ has covariance 
\eqref{etav}, just as before. See \cite{eyink2021dissipation}, Appendix A. 
We can, however, decompose the velocity into a ``turbulent part" and a ``thermal part" as 
$${\bf v}={\bf v}_{T}+{\bf v}_{\theta}$$
where the turbulent velocity ${\bf v}_{T}$ satisfies the deterministic Navier-Stokes equation and ${\bf v}_{\theta}$ 
represents the small thermal fluctuation around that solution. An equation for ${\bf v}_{\theta}$ follows
by the standard approach of linearization: 
\begin{eqnarray}\nonumber
\partial_t{\bf v}_{\theta}&=&\bcP\Big[ -{\bf v}_{T}\bdot\grad {\bf v}_{\theta} -{\bf v}_{\theta}\bdot\grad {\bf v}_{T}+\nu \triangle{\bf v}_{\theta}\\\label{Momentum102a}
&&+\grad\bdot  \Big( \sqrt{2\nu k_B T/\rho  } \; {\boldsymbol \eta}({\bf x}, t)\Big)\Big],
\end{eqnarray} 
The neglect of the non-linear term ${\bf v}_{\theta}\bdot\grad{\bf v}_{\theta}$ assumed in this approximation is valid because 
it is small compared with the viscous term $\nu \triangle{\bf v}_{\theta}$ in the turbulent dissipation range
below the Kolmogorov scale $\ell_K,$ which is our focus of interest here. The matter was discussed 
in \cite{eyink2021dissipation}, section II.B where it was noted that the ratio of this nonlinear term to the 
viscous term (which is a kind of scale-dependent ``thermal Reynolds number" $Re_\ell^\theta$)  is of order $\theta_K^{1/2}$ at the \red{Kolmogorov scale
$\ell=\ell_K$, where $\theta_K=k_BT/\rho v_K^2\ell_K^3$} is the ratio of fluid thermal energy to the kinetic energy of a Kolmgorov-scale
eddy. Since $\theta_K\sim 10^{-6}-10^{-9}$ in realistic flows, the neglect of the nonlinear term is justified throughout 
the turbulent dissipation range and down to nearly molecular scales. 
 
We shall take for the scalar concentration field the same equation as did DFV 
\begin{eqnarray}
&& \partial_t c=-\bu\bdot\grad c +\cr 
&& \grad\bdot \left(D_0\grad c+\sqrt{2mD_0 \rho^{-1} c(1-c)} \; {\boldsymbol\eta}_c({\bf x}, t) \right), \label{Passive120a}
\end{eqnarray}
but with the crucial difference that now 
\be\label{ScaleSeparation100}
{\bf u}={\bf u}_{T}+{\bf u}_{\theta}
\ee
where $ {\bf u}_{T}={\boldsymbol\sigma}\star{\bf v}_{T}$ and $ {\bf u}_{\theta}={\boldsymbol\sigma}\star{\bf v}_{\theta}$. We emphasize that 
eq.(\ref{Passive120a}) is fundamentally different from eq.(\ref{Passive110}), since it includes the effects of turbulent advection 
as well as advection by thermal velocity fluctuations. 
Because we consider here the turbulent dissipation range at scales 
below the Kolmogorov length $\ell_K,$ where the velocity $\bv_T$ is smooth, and because $\ell_K\gg \sigma,$ we should expect that 
$\bsigma\star{\bf v}_{T}\simeq {\bf v}_{T}.$   However, the coarse-graining at scale $\sigma$ remains crucial for the much rougher thermal component.  

The equations \eqref{Momentum102a}-\eqref{ScaleSeparation100} are the basis of all of our subsequent
analysis. To make the problem mathematically tractable, however, we shall follow Kraichnan \cite{kraichnan1968small,kraichnan1974convection} 
in further modeling the Navier-Stokes solution $\bv_T$  in the turbulent dissipation range as a Gaussian random velocity field, white noise 
in time, with zero mean and covariance 
\be\label{w-correlation1}
 \langle {\bf v}_{T}({\bf x}, t)\otimes {\bf v}_{T}({\bf x'}, t')\rangle=\bcV_{T}({\bf x-x'})\delta(t-t'),
 \ee
 where 
\begin{eqnarray}
{\cal V}_{T,ij}(\br)&=&2{\cal V}_{T0} \delta_{ij}-2 {\Gamma}\left(2 r^2\delta_{ij}-r_ir_j\right)\label{w-correlation2}
\end{eqnarray}
and $2{\cal V}_{T0}\delta_{ij}={\cal V}_{T,ij}(0)$. We consider here statistically homogeneous flows (periodic 
domains or infinite space) so that the covariance depends only upon the difference $\br=\bx-\bx'.$
The constant ${\cal V}_{T0}$ with units of $({\rm length})^2/({\rm time})$ represents the sweeping effects 
of large integral-scale eddies, while the constant $\Gamma$ has units of $1/({\rm time})$ and its magnitude should be taken to be of order of the inverse Kolmogorov time 
(the eddy turn-over rate at the Kolmogorov scale)\footnote{Our constant $\Gamma$ is chosen to coincide with $D_1$ in Eq.(48) 
of \cite{falkovich2001particles} for the case $\xi=2,\, d=3$ and thus equals $A/30$ in terms of the constant $A$ introduced 
in the original work of Kraichnan \cite{kraichnan1968small,kraichnan1974convection}.}. 
The equations \eqref{Momentum102a}, \eqref{Passive120a} now become
\begin{eqnarray}\nonumber
\partial_t{\bf v}_{\theta}&=&\bcP\Big[ -{\bf v}_{T}\odot\grad{\bf v}_{\theta} -{\bf v}_{\theta}\odot\grad{\bf v}_{T}+\nu \triangle{\bf v}_{\theta}\\\label{Momentum102}
&&+\grad\bdot \Big( \sqrt{2\nu k_B T/\rho  } \; {\boldsymbol \eta}({\bf x}, t)\Big)\Big],
\end{eqnarray} 
\begin{eqnarray}\nonumber
\partial_t c&=&-\bu\odot\grad c \cr
&& +\grad\bdot \left(D_0\grad c+\sqrt{2mD_0 \rho^{-1} c(1-c)} \; {\boldsymbol\eta}_c({\bf x}, t) \right).\\\label{Passive120}
\end{eqnarray}
Because  of the white-noise character of ${\bf u}_{T},$ we must specify the stochastic calculus and $\odot$ indicates a Stratonovich 
dot product. This is the standard choice for the Kraichnan model, because it is considered as the zero-correlation limit of a model with 
a stochastic advecting velocity field that has finite time correlation. By taking space-derivatives of 
\eqref{w-correlation2}, one obtains 
\begin{eqnarray}\label{gradw-correlation}
&& \langle \partial_k v_{Ti}({\bf x}, t) \partial_\ell v_{Tj}({\bf x'}, t')\rangle=\\\nonumber
&& \hspace{30pt} 2 {\Gamma}\left(4\delta_{ij}\delta_{k\ell}-\delta_{ik}\delta_{j\ell} -\delta_{i\ell}\delta_{jk}\right)\delta(t-t'),
 \end{eqnarray}
which makes clear that Kraichnan's model of the velocity field in the turbulent dissipation range corresponds 
to a spatially uniform random straining field, rapidly varying in time. As a consequence, the smoothed turbulent velocity field 
${\bf u}_T=\bsigma\star\bv_T$ has a spatial covariance $\bcU_T$ which differs from $\bcV_T$ in \eqref{w-correlation1} only by the replacement 
of ${\cal V}_{T0}$ with a constant ${\cal U}_{T0}$ larger by an amount $\sim \Gamma\sigma^2.$ \red{To avoid possible 
confusion, we note that this constant-strain model introduced by Kraichnan \cite{kraichnan1968small,kraichnan1974convection}
was later generalized by him \cite{Kraichnan94} to involve arbitrary spatial covariance and this generalization is now more commonly known as the ``Kraichnan model''; see \S \ref{SKraichnan}.}


%

The high Schmidt-number limit of our turbulent advection problem can be studied by the 
same formal asymptotics with the small parameter $\epsilon\ll 1$ used by DFV, 
with identical rescalings \eqref{trans-to},\eqref{time-to}. The key physical issue is the 
ordering to be adopted for the white-noise turbulent field $\bv_T.$ The correct ordering
can be motivated by the observation that for Navier-Stokes turbulence
$$ \frac{d}{dt} \frac{1}{2}\langle v^2\rangle =\nu \langle|\grad {\bf v}|^2\rangle\sim \nu \Gamma^2,$$ 
where we used the fact that $\Gamma$ is of the order of the inverse Kolmogorov time. Now 
invoking the rescalings $t\mapsto\epsilon^{-1}t,$ $\nu\mapsto\epsilon^{-1}\nu$ we see that $\Gamma$
must be rescaled as $\Gamma\mapsto \epsilon \Gamma.$
Stated equivalently, the turbulent velocity gradient must be ordered so that, with large viscosity,
a finite total amount of energy is dissipated in a unit ``macroscopic'' time. Together with the 
covariance of the white-noise field \eqref{w-correlation1},\eqref{w-correlation2} and the scaling 
$t\mapsto\epsilon^{-1}t,$ one then obtains 
$$ {\bf v}_{T}(\bx,t)\mapsto \epsilon {\bf v}_{T}(\bx,t). $$
We shall see below that this ordering leads to turbulent advection making an $O(1)$ contribution 
in the high Schmidt-number limit. 

Before discussing the formal $\epsilon\to 0$ limit, we first consider the inclusion of additional
terms into the concentration equation (\ref{Passive120}) to enable a steady-state scalar cascade. 
As written, equation \eqref{Passive120} corresponds to a freely decaying scalar. While this set-up 
permits a Batchelor regime (e.g. see \cite{kraichnan1974convection,jullien2000experimental}), it 
entails the complication of non-trivial time-dependence. It is simpler to analyze instead 
a steady-state cascade with some external source included to inject concentration  fluctuations, 
so that a time-independent balance can be achieved with diffusive dissipation. That is what
we shall do in this work. The simplest source 
to include for this purpose is a constant mean concentration gradient $\bgamma=\langle\grad c\rangle,$
which contributes an additional term $-\bgamma\bdot\bu$ to  \eqref{Passive120} that injects 
scalar fluctuations. Such a constant concentration-gradient has been often considered, both 
in turbulence \cite{shraiman1994lagrangian,donzis2010batchelor} 
and in statistical physics \cite{segre1993nonequilibrium,wada2004shear,vailati2011fractal}, 
having the motivation that it is realizable in laboratory experiments. An alternative source of scalar fluctuations more amenable to mathematical analysis is a spatially 
distributed source $s(\bx,t),$ which may be either deterministic or random. In the latter case, 
a Gaussian random source field which is white-noise in time is especially convenient, because it gives 
exact control of the rate of injection $\chi$ of scalar fluctuations. In particular, with covariance 
\be\label{SourceCorrelation}
\langle s(\bx, t) s(\bx', t') \rangle= \delta(t-t') S\left({\bx-\bx'\over L}\right),
\ee
it follows that in the statistically stationary state 
\be \frac{1}{2}S(\bzed)=D_0\langle|\grad c |^2\rangle=\chi \label{Ec-bal} \ee
and rate of input of scalar fluctuations matches the dissipation rate by diffusion \cite{novikov1965functionals}.
Here $S$ is a smooth, positive-definite function and thus $L$ gives the length-scale 
of injection of the scalar fluctuations. It is such stochastic white-noise forcing which we shall analyze
in this work, but for future applications we shall derive the reduced equations for all of the various 
scalar sources.  

What is important to consider in including scalar source terms into equation \eqref{Passive120} is their 
ordering in the small parameter $\epsilon\ll 1.$ We shall rescale these quantities so that they appear 
at O(1) in the final equation for the concentration and, thus, a strong scalar cascade is obtained in macroscopic time.  
The final equations we consider, with all quantities properly ordered but expressed in original microscopic time units, is 
\begin{eqnarray}\nonumber
\partial_t{\bf v}_{\theta}&=&\bcP\Big[ -\sqrt{\epsilon} {\bf v}_{T}\odot\grad{\bf v}_{\theta} 
-{\bf v}_{\theta}\bdot\sqrt{\epsilon} \grad{\bf v}_{T}+\nu \epsilon^{-1} \triangle{\bf v}_{\theta}\\\label{Momentum102b}
&&+\grad\bdot \Big( \sqrt{2\nu \epsilon^{-1} k_B T/\rho  } \; {\boldsymbol \eta}({\bf x}, t)\Big)\Big],
\end{eqnarray} 
\begin{eqnarray}\nonumber
\partial_t c'&=&-\sqrt{\epsilon}\bu_T\odot (\grad c' +\bgamma) - \bu_\theta\bdot(\grad c'+\bgamma)  
+\epsilon D_0\triangle c' \\\label{Passive10b5}
&& + \epsilon s_0(\bx,\epsilon t) + \sqrt{\epsilon} s(\bx,t) \\\nonumber
&&+\grad\bdot \Big(\sqrt{2\epsilon \, m D_0 \rho^{-1} c(1-c) } {\boldsymbol \eta}_c({\bf x}, t) \Big).
\end{eqnarray}
Here $c'$ denotes the {\it scalar fluctuation}, so that $\langle c'\rangle=0,$ and thus $c=c'+\bgamma\bdot\bx.$
Note that $\sqrt{\epsilon} \bv_T(\bx,\epsilon^{-1}t)=\epsilon \bv_T(\bx,t)$ because of the scaling properties 
of white noise, in agreement with our previous argument. Likewise, the white-noise random scalar source 
satisfies $\sqrt{\epsilon} s(\bx,\epsilon^{-1}t)=\epsilon s(\bx,t).$ A deterministic distributed source, denoted 
here by $s_0,$ is assumed to have zero space average and is scaled as $\epsilon s_0(\bx,\epsilon t),$ which 
amounts to the assumption that it is weak and slowly varying in microscopic time units. 

Finally, changing to macroscopic/diffusive time by the substitution $t\mapsto \epsilon^{-1}t$ we get the equations
\begin{eqnarray}\nonumber
\partial_t{\bf v}_{\theta}&=&\bcP\Big[ -{\bf v}_{T}\odot\grad{\bf v}_{\theta} 
-{\bf v}_{\theta}\bdot \grad{\bf v}_{T}+\nu \epsilon^{-2} \triangle{\bf v}_{\theta}\\\label{Momentum200}
&&+\grad\bdot \Big( \sqrt{2\nu \epsilon^{-2} k_B T/\rho  } \; {\boldsymbol \eta}({\bf x}, t)\Big)\Big],
\end{eqnarray} 
\begin{eqnarray}\nonumber
\partial_t c'&=&-\bu_T(\bx,t)\odot (\grad c' +\bgamma) - \epsilon^{-1} \bu_\theta(\bx,\epsilon^{-1}t) \bdot(\grad c'+\bgamma)  
\\\label{Passive200}
&& +D_0\triangle c' + s_0(\bx,t) + s(\bx,t) \\\nonumber
&&+\grad\bdot \Big(\sqrt{2 m D_0 \rho^{-1} c(1-c) } {\boldsymbol \eta}_c({\bf x}, t) \Big).
\end{eqnarray}
In Appendix \ref{S2c} we start with these equations and perform a standard adiabatic mode elimination 
of the fast velocity degrees of freedom, following closely the argument in \cite{donev2014reversible}. 
Here, we present the main result of that analysis for the readers who wish to skip 
the mathematical details. First, the thermal velocity field $\bv_\theta,$ to leading order in the 
asymptotics, is unaffected by the background turbulent velocity $\bv_T$ and evidences thermal equilibrium 
statistics  at the temperature $T.$ This result is in agreement with numerical simulations of the 
full nonlinear fluctuating hydrodynamic equation \eqref{FNS}, where the velocity in the far dissipation 
range of turbulence exhibits Gaussian equipartition statistics \cite{bell2021thermal}.  The equation for 
the concentration fluctuation field on the macroscopic time-scale reduces simply to  
 \begin{eqnarray}\nonumber
\partial_t c' &=&-({\bf u}_{T}+{\bf w}_\theta)\odot(\grad c'+{\boldsymbol\gamma})   \\\label{LangevinPassive1a}
&&+D_0\triangle  c' + s_0(\bx,t) + s(\bx,t) \\\nonumber
   &&+\grad\bdot  \left(\sqrt{2mD_0 \rho^{-1}c(1-c)} {\boldsymbol \eta}_c({\bf x}, t) \right),
\end{eqnarray}
where $\bw_\theta$ is the Gaussian, white-noise random field $\bw$ which appeared in the DFV theory 
for a fluid at rest, defined by \eqref{SmoothVelocityCovariance101} with the space covariance 
\eqref{SmoothVelocityCovariance102}. An equivalent It$\bar{{\rm o}}$ form of the equation is 
\begin{eqnarray}\nonumber
\partial_t c' &=&-({\bf u}_{T}+{\bf w}_\theta)\bdot(\grad c'+{\boldsymbol\gamma})   \\\label{LangevinPassive1}
&& +D_{e\! f\! f}\triangle  c'+ s_0(\bx,t) + s(\bx,t) \\\nonumber
   &&+\grad\bdot  \left(\sqrt{2mD_0 \rho^{-1}c(1-c)} {\boldsymbol \eta}_c({\bf x}, t) \right),
\end{eqnarray}
where $D_{e\! f\! f}=D_0+D_T+D$ with $D_T={\mathcal U}_{T0}$ the turbulent eddy-diffusivity 
and with $D$ the diffusivity contribution from thermal fluctuations calculated by DFV. In general,  
$D_{e\! f\! f}\simeq D_T$ because the turbulent diffusivity dominates. The important result, 
which shall form the basis of all of our conclusions in this paper, is that the equation for the concentration 
field in the dissipation range reduces to a Kraichnan model with white-noise advecting velocity 
$\bv=\bu_T+\bw_\theta$ which is an additive sum of contributions from turbulence and from thermal fluctuations.

\section{Solution of the Model}\lb{S3} 

Having developed our reduced model of high-Schmidt turbulent mixing by the asymptotic method
of DFV, we now solve it exactly for the concentration correlation function $C(r)$ and the 
corresponding spectrum $E_{c}(k).$ We consider the simplest situation of a statistically 
homogeneous and isotropic steady state
with concentration fluctuations injected at a constant rate $\chi$ at a length-scale $L\gtrsim\ell_K$
via a random spatially distributed source $s(\bx,t),$ as in \eqref{SourceCorrelation}. In the 
first subsection below we briefly review the Kraichnan model in general and some of the important  mathematical  results concerning it. Then, in the following subsections, we employ those standard results 
as part of the exact solution of our specific problem.

\subsection{Pr\'ecis of the Kraichnan Model}\label{SKraichnan}

The standard review of the Kraichnan model is the 2001 article of Falkovich et al. 
\cite{falkovich2001particles}. Very good discussions for beginners (and even 
for experts) are contained in published conference lectures by Gaw\c{e}dzki  \cite{Gawedzki1997,gawedzki2002easy,gawedzki2002soluble}. In particular, these 
reviews discuss the theoretical breakthrough of the exact calculation of anomalous 
scaling exponents of the passive scalar in the inertial-convective range, which,  
however, plays no role in our analysis here. For a recent comparison of the Kraichnan model
predictions with experimental data and with simulations of incompressible 
Navier-Stokes turbulence, see \cite{sreenivasan2019turbulent}. We draw upon 
all of these sources for the presentation below. 
 
What is now termed the ``Kraichnan model'' is described mathematically by an equation for a passive scalar field $c({\bx}, t)$ which is advected by a Gaussian random velocity 
field $\bv(\bx,t)$ which is white-noise in time:  
\be\label{ScalarDiff}
\partial_t c=-\bv\odot\grad c+D_0 \triangle c+s,\ee
where the velocity field has zero mean and covariance
\be\label{Kraichnan1}
\langle v_i({\bf x}, t)v_j({\bf x}', t')\rangle={\cal V}_{ij}(\bx,\bx')\delta(t-t'). 
\ee
As in Sec.\ref{S2}, the symbol $\odot$ denotes a Stratonovich dot product and 
$D_0$ is the molecular (bare) diffusivity. We have included a random Gaussian source
 $s({\bf x}, t)$ (independent of velocity) with zero mean and covariance given by eq.(\ref{SourceCorrelation}). We consider here only a solenoidal 
 (incompressible) velocity field, so that $\nabla_{x^i}{\mathcal V}_{ij}(\bx,\bx')=0,$ 
 although the compressible case has been studied in the literature 
 \cite{gawedzki2002soluble,falkovich2001particles}. 
 In our specific problem, we consider a statistically homogeneous flow, so that ${\cal V}_{ij}$ depends only upon the difference $\br=\bx-\bx'$. However, we present all of the results for general inhomogeneous flows
 in this summary, since many important future applications will involve situations with boundary conditions or initial conditions for velocity and concentration that break homogeneity. Converting to It$\bar{{\rm o}}$ calculus, the Langevin equation (\ref{ScalarDiff}) gains a noise-induced drift term, hence in It$\bar{{\rm o}}$ interpretation the equivalent equation is 
\be\label{ItoScalarDiff}
\partial_t c=-\bv\bdot\grad c+D_0\triangle c+\grad\bdot\left(\frac{1}{2}\bcV(\bx,\bx)\bdot\grad c\right) +s.
\ee
The additional term has the physical meaning of a tubulent ``eddy-diffusivity''
$\bD_{eddy}(\bx)=\frac{1}{2}\bcV(\bx,\bx)$ which is induced by the random advection. 

One of the important features of the Kraichnan model is that there is no closure problem for correlation functions of the advected scalar $c({\bf x}, t)$ and in fact the $N$-point,
equal-time correlation function $C_N({\bf x}_1, \dots, {\bf x}_N;t):=\langle c({\bf x}_1, t)c({\bf x}_2, t)...c({\bf x}_N, t)\rangle$ satisfies an exact differential equation \cite{falkovich2001particles}: 
\begin{eqnarray}\nonumber
&& \partial_t C_N({\bf x}_1,\dots, {\bf x}_N)={\cal M}_N C_N({\bf x}_1,\dots, {\bf x}_N)\\\nonumber
&& \hspace{20pt} +\sum_{n<m} C_{N-2}({\bf x}_1,\dots,\stackrel[n]{}{\wedge},\dots,
\stackrel[m]{}{\wedge},\dots, {\bf x}_N) S(\bx_n,\bx_m),\\\label{GeneralCKraichnan}
\end{eqnarray}
where the notation ``$\stackrel[n]{}{\wedge}$'' indicates that the variable $\bx_n$
is omitted, where we have introduced the second-order many-particle diffusion 
operator 
\be\label{LN}
{\cal M}_N={1\over 2} \sum_{n, m=1}^N 
\grad_{x_n^i}\left[{\cal V}_{ij}(\bx_n,\bx_m)\grad_{x_m^j}(\cdot)\right] +D_0 \sum_{n=1}^N \triangle_{{\bf x}_n},
\ee
and where $S$ is the spatial covariance function of the random source $s({\bf x}, t)$ introduced in eq.(\ref{SourceCorrelation}) in Sec.\ref{S2}. Note that eq.(\ref{GeneralCKraichnan}) is in fact a triangular system of equations with no closure problem since, with lower-point functions at hand, the $N$-point function is governed by a closed differential equation. While the closed equations (\ref{GeneralCKraichnan}) can be obtained in many ways (see e.g. \cite{Gawedzki1997, falkovich2001particles}), one particularly straightforward approach, which follows directly from our considerations in Sec.\ref{S2}, is simply 
to use the backward Kolmogorov operator corresponding to the  It$\bar{{\rm o}}$ equation (\ref{ItoScalarDiff}): 
\begin{eqnarray}\nonumber
{\bf L}_c&:=&  
\int d^3x\ 
\left[ D_0\triangle c+\grad\bdot\left(\frac{1}{2}\bcV(\bx,\bx)\bdot\grad c\right)\right] 
{\delta \over \delta c({\bf x})}\\\nonumber 
             &+&{1\over 2}\iint d^3x \,d^3x'\  {\cal V}_{ij}({\bf x, x'})\,  \partial_i c({\bf x})  \partial'_j c({\bf x'}){\delta^2 \over \delta c({\bf x})\delta c({\bf x'})}\\\nonumber
        &+&{1\over 2}\iint d^3x \, d^3x'\ S(\bx,\bx'){\delta^2  \over \delta c({\bf x})\delta c({\bf x'})},
\end{eqnarray}
which appears also in the Markov operator ${\bf L}_0$ given by eq.(\ref{L0}) of Appendix \ref{S2c}, in a slightly different notation. Applying this operator ${\bf L}_c$ to the 
product functional $F[c]=c({\bf x}_1) c({\bf x}_2)\dots c({\bf x}_N)$ immediately yields (\ref{GeneralCKraichnan}). 

For $N=1$, the sum on the RHS of eq.(\ref{GeneralCKraichnan}), which contains lower order correlations, vanishes and we get $
\partial_t C_1({\bf x})={\mathcal M}_1 C_1({\bf x})$ with $C_1({\bf x}):=\langle c({\bx}, t)\rangle$.
In fact, the closed equation for $C_1$ can be simply obtained by taking the average of 
the  It$\bar{{\rm o}}$ equation (\ref{ItoScalarDiff}), which yields 
\be\label{MeanScalarDiff}
\partial_t \bar{c} = D_0\triangle \bar{c}+\grad\bdot\left(\frac{1}{2}\bcV(\bx,\bx)\bdot\grad \bar{c}\right)
\ee
where we use the more standard notation $\bar{c}$ for $C_1.$ This equation is trivially 
solved for the specific problem in this paper, where we consider a statistically homogeneous
flow (periodic domain or infinite space) in the long-time steady state, so that 
$\bar{c}$ is a space-time constant. In more realistic problems concerning GCF's 
with inhomogeneous statistics 
and involving transient decay, the equation \eqref{MeanScalarDiff} must be solved 
and $\bar{c}$ used as an input to the equation for the concentration cumulant 
(or connected correlation) function $G_2(\bx,\bx')=C_2(\bx,\bx')-C_1(\bx)C_1(\bx').$
We shall leave such studies, which are directly relevant to experimentally realizable 
flows in microgravity, to future work. 

In this paper, we are concerned with $C_N$ for $N=2$, although it would be of interest 
to study also higher-order correlations $C_N$ for $N>2$ and we shall discuss this matter 
later in the  conclusions. The equation (\ref{GeneralCKraichnan}) for $N=2$ simplifies to 
$\partial_tC_2={\mathcal M}_2C_2+S$. We further specialize to the homogeneous case, 
so that all 2-point correlators ($C_2,$ $\bcV,$ etc.) become functions of the difference 
variable $\br=\bx-\bx'$ only.  The correlation $C_2,$ which shall be hereafter denoted 
simply as $C,$ then satisfies the equation 
\begin{eqnarray}\nonumber
     \partial_tC ({\bf r}, t)&=&\left[{\cal V}_{ij}({\bf 0})-{\cal V}_{ij}({\bf r}) \right]\partial_i\partial_j C    +2 D_0 \triangle C+S\left( {\br\over L}\right),\\\label{correlation1}
     \end{eqnarray}
where all differential operators are now with respect to the variable $\br.$ 
In the case of interest to us, the velocity statistics (both for turbulent 
and thermal fluctuations) are in addition isotropic. For such isotropic cases,
the spatial velocity covariance can be written in terms of the Leray projection 
$\bcP$ as 
\be
{\mathcal V}_{ij}(\br)={\mathcal P}_{ij} K(r) \lb{iso-V} \ee
where $K(r)$ is a positive-definite function of the radial variable $r=|\bx-\bx'|.$
See \cite{EyinkXin2000}. The general equation for the 2-point correlation in 
the Kraichnan model with isotropic velocity statistics was implicit in the 
paper \cite{EyinkXin2000}, but not written explicitly there. We thus derive 
this equation in our Appendix \ref{AppKraichnanCorrelation}, where we show 
that for any space dimension $d$ the equation for $C(r,t)$ can be expressed 
in terms of the ball-averaged function 

\begin{eqnarray}\label{H'/r}
    J(r)&=&-{1\over r^d}\int_0^r  K(\rho) \,  \rho^{d-1} d\rho
\end{eqnarray}
as 
\begin{eqnarray}\nonumber
\partial_t{C}&=&{1\over r^{d-1}}{\partial\over\partial r}\left(\Big[2 D_0  - (d-1)\triangle J(r)\Big] r^{d-1} {\partial{C}\over \partial r}\right)\\\label{finalcorr}
&&+S\left( {r\over L}\right), 
\end{eqnarray}
where $\triangle J(r)=J(0)-J(r)$ and we have also assumed that the source covariance 
function $S(r/L)$ is isotropic. This equation is already indicative of a renormalized 
diffusivity $-\triangle J(r)$; cf. eq.(\ref{Jtheta}) in Sec. \ref{ourmodel}. 

In the statistical steady state, which is our focus here, $\partial_t C=0$ and the 
solution of (\ref{finalcorr}) is easy to obtain by straightforward integration:
\be\label{2CorrelationKraichnan}
C(r)=\int_r^\infty { \int_0^\rho  S\Big( {\rho^\prime\over L}  \Big)  \rho^{\prime{d-1}} d\rho^\prime       \over \rho^{d-1}  [2D_0-(d-1)\triangle J (\rho)]        } d\rho,
\ee
where we have applied boundary conditions $\partial_r C(0)=0$ and $C(\infty)=0$ \cite{Gawedzki1997}. The above expression represents the final form of the steady-state $2$-point correlation function for the passive scalar in the Kraichnan model with homogeneous, isotropic statistics. 

Before we proceed with evaluating this expression, we first observe that 
the steady-state balance equation \eqref{Ec-bal} follows directly from \eqref{finalcorr} 
by setting  $\partial_t C=0$ and $r=0,$ which yields
\bea  2D_0\langle |\grad c|^2\rangle &=& -2D_0
\left.{1\over r^{d-1}}{\partial\over\partial r}\left(r^{d-1} {\partial{C}\over \partial r}\right)\right|_{r=0} \cr 
&=& S(\bzed)\, = \, 2\chi,
\eea
and thus $\chi=D_0\langle |\grad c|^2\rangle.$ Although this result is an exact consequence
of our mathematical model when ignoring thermal fluctuations of the concentration field, the 
latter invalidates this result physically. We shall see in the following that a physically 
valid balance equation for the concentration fluctuations involves instead a ``renormalized 
diffusivity'' $D$ and ``effective gradients'' $(\grad c)_{e\! f\! f}$.

\subsection{Concentration Correlation Function}\label{ourmodel}

We now apply these general results on the Kraichnan model to the high-Schmidt limit equations  (\ref{LangevinPassive1}). Throughout this entire subsection we shall neglect the final term 
in that equation representing molecular noise in the dynamics of the concentration field, 
because of the presumed smallness of the molecular mass $m$ and the bare diffusivity $D_0.$ This term, 
however, can become important at sufficiently small scales and we shall evaluate its 
contribution in Section \ref{physics}. With omission of the molecular noise term,
\eqref{LangevinPassive1} becomes a particular case of the Kraichnan model, 
with white-noise advecting velocity $\bu_T+\bw_\theta$, representing statistically 
independent contributions from turbulent fluctuations and from thermal fluctuations. 
Therefore, the equation for the steady-state $2$-point correlation function $C(r)$ 
in our model is exactly eq.\eqref{finalcorr} in $d=3$ spatial dimensions, 
with $\triangle J=\triangle J_T+\triangle J_\theta.$  
Solution of this differential equation requires expressions for the 
scale-dependent diffusivities.  

The turbulent velocity field with space covariance \eqref{w-correlation2} 
for $r\ll \ell_K$ is easily checked to be given by \eqref{iso-V} with $K_T(r)=-5\Gamma r^2$ and thus  
\begin{eqnarray}\nonumber
  \triangle J_T(r)&:=&J_T(0)-J_T(r) \\\label{DeltaJT}
  &=& -J_T(r) \ = \ -{\Gamma} r^2.
\end{eqnarray}
To obtain the scale-dependent diffusivity from thermal fluctuations, we need first to determine 
$K_\theta(r)$ corresponding to the spatial covariance $R_{ij} (\br, \br')$ given by \eqref{SmoothVelocityCovariance102} and \eqref{oseen} and then calculate the associated 
$\triangle J_\theta(r).$ A suitable choice of the filter kernel must be made in 
\eqref{SmoothVelocityCovariance102}. DFV employed the isotropic kernel 
$\bsigma=\sigma {\bf I}$ with Fourier transform
specified as 
\be\label{DonevKernel} 
\widehat\sigma(k):={k^2L^2\over \sqrt{(1+k^4L^4) (1+k^2 \sigma^2)}   }
\ee
which leads to the thermally renormalized diffusivity $D=k_BT/6\pi\eta\sigma$ given by (\ref{D-sig}). As noted by DFV, the
kernel \eqref{DonevKernel} was chosen for convenience to give the prefactor $(6\pi)^{-1}$  
of the conventional Stokes-Einstein relation for hard-spheres, but any isotropic kernel 
will lead to a similar result with a different prefactor of order unity. Employing (\ref{DonevKernel}), with $\sigma\ll L$, we find (for details see Appendix \ref{AppJtheta})
\be\label{Jtheta}
               \triangle J_\theta(r) = {k_BT\over 6\pi \eta\sigma}  \left[-1+3\left( { {1\over 2}}{1\over {r\over \sigma}} + {e^{-r/\sigma}\over \left({r\over \sigma}\right)^2}-{1-e^{-r/\sigma} \over \left({r\over \sigma}\right)^3}\right)\right].
\ee
Most important is the asymptotic limit 
\be \triangle J_\theta(r) \sim -D\left(1-\frac{3\sigma}{2r}\right) \quad r\gg \sigma 
\lb{J-large-r} \ee 
which will be universal for any  filter kernel which is rapidly decaying in physical-space, 
up to a different choice of $\sigma$ rescaled by a factor of order unity. The physical 
meaning of the result \eqref{J-large-r} is that diffusivity becomes {\it scale-dependent}  
due to the renormalization by thermal velocity fluctuations, with effective 
molecular diffusivity 
\be D(r)=D_0+D\left(1-\frac{3\sigma}{2r}\right) \lb{Dr} \ee  
at length-scale $r.$ Thus, $D(r)\simeq D_0$
for $r\simeq \sigma$ but $D(r)\simeq D_0+D:=D_{e\! f\! f}$ for $r\gg \sigma.$ 
This scale-dependence in the diffusivity $D(r)$ for correlations of concentration 
fluctuations is closely related to the dependence on system size $L$ in the 
effective diffusivity \eqref{D-sig} for the mean concentration. 
Note likewise that the turbulent velocity fluctuations according to \eqref{DeltaJT} contribute an effective turbulent eddy-diffusivity $D_T(r)=\Gamma r^2$ at length-scale $r.$

We shall be concerned in this paper only with the regime $L>\ell_K\gg r\gg \sigma$ 
and hence need only the asymptotic expression \eqref{J-large-r}. However, if one 
considers also $r\simeq \sigma,$ then the specific 
choice of kernel influences the results. Here we note that the Fourier-transformed 
kernel \eqref{DonevKernel} adopted by DFV decays very slowly $\sim 1/k$ at high wavenumbers $k$ 
and the physical space kernel is thus not differentiable in space. This kernel leads to a
renormalized diffusivity $\Delta J_\theta(r)$ which is not smooth in the separation $\br.$ 
Whereas a general filter kernel that is smooth in physical space would produce 
$\Delta J_\theta(r)\propto r^2$ for $r\ll \sigma,$ instead the choice \eqref{DonevKernel} 
of DFV leads to $\Delta J_\theta(r)\propto r,$ as may easily be checked 
from (\ref{Jtheta}). To avoid this undesirable feature, we can choose instead, for example, an exponentially decaying kernel 
\be  \widehat\sigma(k)\sim e^{-k \sigma/\pi}, \quad kL\gg 1, \lb{exp-sig} \ee 
with $\pi$ added to reproduce the Stokes-Einstein relation for $D.$ 
In that case, for $r\ll L$ 
 \begin{eqnarray}
\triangle J_\theta(r)&=&{k_B T\over 6 \pi\eta \sigma}\Big( -1+{3 \sigma\over \pi r^3}\Big[ \Big({4\sigma^2\over \pi^2}+r^2\Big)\arctan\Big({\pi r\over 2\sigma }\Big) \cr 
&&\hspace{30pt} -{2\over \pi}\sigma r \Big]      \Big).
\label{expJ}\end{eqnarray}
as also shown in Appendix \ref{AppJtheta}. Because the physical-space kernel corresponding 
to \eqref{exp-sig} is $C^\infty,$ then $\triangle J_\theta(r)\propto r^2$ for $r\ll \sigma$ 
as may be verified from \eqref{expJ}. On the other hand, \eqref{J-large-r} is recovered 
for $r\gg \sigma.$
 
 
We consider only the regime $L>\ell_K \gg r\gg \sigma$ hereafter and 
we restrict attention also to the steady-state correlation $C(r)$ given by the integral 
\eqref{2CorrelationKraichnan}. We find it easiest to study the derivative $\partial_r {C}(r)$
and, because $r\ll L,$ we can take  $S(r/L)\simeq S(0).$ These approximations 
together with the asymptotic expression \eqref{J-large-r} yield 
\begin{eqnarray}\label{Theta>}
   \partial_r {C}(r)  \cong -{ \chi \over 3  D }{ r\over 1-{3\over 2}{\sigma\over r}+{r^2\over \ell_B^2} },\\\nonumber
    \end{eqnarray}
where we used $\chi=S(0)/2$ and neglected the bare diffusivity under the assumption 
that $D_0\ll D.$ We have also introduced the Batchelor dissipation length with 
convenient definition $\ell_B^2:=D/\Gamma,$ which agrees with the standard definition 
up to a constant of order unity since $\Gamma$ is assumed to be of order $\gamma=(\varepsilon/\nu)^{1/2}.$ We may now consider separately the two relevant
subranges, the viscous-convective range $\ell_K\gg r\gg \ell_B$ and the viscous-diffusive 
range $\ell_B\gg r\gg \sigma.$
    
In the viscous-convective range we can keep only the $r^2$-term in the denominator 
of (\ref{Theta>}) obtaining thus by integration over $r$
\be\label{BatchelorCorr}
C(r) \cong const.+ {\chi\over 3\Gamma} \ln\left(\frac{\ell_K}{r}\right), \quad
\ell_K\gg r\gg \ell_B.
\ee
Taking the Fourier transform with the standard isotropic relation
(\cite{hinze1975turbulence}, Eq.(3-229))
\be E_c(k) = \frac{1}{\pi} \int_0^\infty kr\sin(kr) \, C(r) \, dr \lb{3DFT} \ee 
yields 
\be\label{Batchelor1k}
E_c(k)\simeq  {\chi\over 6\Gamma} {1\over k}, \quad 1/\ell_K\ll k\ll 1/\ell_B
\ee 
in exact agreement with standard theory. Note that our factor $6\Gamma$ 
corresponds to the mean least-rate-rate-of-strain $\gamma$ of Batchelor, 
\cite{batchelor1959small}, Eq.(4.9) and to the corresponding factor $\langle a\rangle=A/5$ 
of Kraichnan \cite{kraichnan1968small,kraichnan1974convection}. 

It may at first sight be  
surprising that we recover this standard theoretical result. As we have noted at the 
end of section \ref{S2A} and in Appendix \ref{S2c}, the thermal velocity fluctuations 
in our model to leading order in the high Schmidt-number limit exhibit Gaussian 
thermal equilibrium statistics at temperature $T,$ in agreement with the theoretical 
arguments and numerical results of \cite{bandak2021thermal,eyink2021dissipation,bell2021thermal}
for the turbulent dissipation range. 
The velocity field thus has the equilibrium equipartition spectrum $E_v(k)\sim
(k_BT/2\pi^2\rho)k^2,$ drastically different from the steep exponential-decay-type 
spectrum assumed by Batchelor and Kraichnan in the dissipation range. 
One might wonder why this drastically different spectrum for $k\ell_K\gg 1$ appears 
to have no observable effect on the behavior of the advected concentration field. 
In fact, there is a very large effect which is hidden from view. Since 
the classical theory of the Batchelor range was developed without considering 
thermal fluctuations, the molecular diffusivity in the works 
\cite{batchelor1959small,kraichnan1968small,kraichnan1974convection}
(which was denoted $\kappa$ in those papers) in fact corresponds to the {\it bare
diffusivity} $D_0$ in our work. Because $D\gg D_0,$ the Batchelor length 
$\ell_B$ is thus greatly increased by thermal velocity fluctuations. Note, however,
that this effect is purely theoretical because DFV obtained exactly the same 
diffusivity renormalization in an equilibrium fluid at rest as we do in a turbulent 
flow and thus the diffusivity measured in most macroscopic experiments will coincide 
with the effective diffusivity $D_{e\!f\!f}$ calculated here. Thus, the effect of thermal velocity 
fluctuations in renormalizing the bare diffusivity is very large but not apparent 
from a phenomenological point of view. 
 
In the viscous-diffusive range, on the contrary, the effects of the thermal velocity 
fluctuations are very large and should be directly observable in future experiments 
that can probe such small scales. We can Taylor expand the righthand side of (\ref{Theta>}) 
in the small quantities $\sigma/r$ and $r/\ell_B$ to obtain 
$$\partial_r C\cong -{\chi\over 3 D} \Big(r-{r^3\over\ell_B^2}+{3\over 2}\sigma\Big),
\quad \ell_B\gg r\gg \sigma $$
which upon integration yields 
\be\label{GCF100}
C(r)\cong C(0)- {\chi\over 6 D} \left[r^2- {1\over 2} {r^4\over\ell_B^2}+3 r \sigma \right].
\ee
The $r^2$ term due to asymptotic diffusivity $D$ is the dominant one in physical space, 
with the $r^4/\ell_B^2$ term arising from turbulent diffusivity and the $r\sigma$ term 
from thermally-induced scale-dependence of diffusivity both subleading. 
Note, however, that the first two terms are polynomials in $r^2$ and under 
Fourier transform contribute terms formally $\propto \delta^3(\bk),$ consistent 
with the rapidly decaying spectra \eqref{Ba59} and \eqref{Kr68} in the viscous-diffusive 
range predicted by Batchelor and Kraichnan. In fact, we shall see in the following 
subsection that setting $\sigma=0$ in our model yields exactly the scalar spectrum 
\eqref{Kr68} of Kraichnan. However, the $\sigma r$-term, although sub-dominant in 
physical space, is non-analytic in $\br$ and the scalar spectrum calculated from 
\eqref{3DFT} is thus a power-law in wavenumber   
\be\label{GCF101}
E_c(k)\simeq {1\over \pi}  {\chi\sigma \over D} {1\over k^2}, \quad
1/\ell_B \ll k\ll 1/\sigma. 
\ee
This is exactly the result \eqref{key} announced in the Introduction. 
As pointed out there, this power-law and the dimensional coefficients multiplying 
it correspond exactly to the giant concentration fluctuations ubiquitously 
observed in diffusive mixing \cite{vailati1997giant,brogioli2000giant,li1998concentration,vailati2011fractal}.
We note again that the numerical prefactor $1/\pi$ is smaller by 2/3 than that predicted by 
linearized fluctuating hydrodynamics \eqref{GCF}. We would not expect linearized 
theory to be adequate to treat with perfect fidelity our physical situation, which
involves nonlinear advection of concentration and large, fluctuating gradients 
of the concentration field. 

There is a subtlety, in fact, in obtaining this agreement with linearized
theory, because it depends upon the relation $\chi=D\langle |\grad c|^2\rangle.$
However, we have seen that the ``correct'' result for the model is instead 
$\chi=D_0\langle |\grad c|^2\rangle$ which involves the {\it bare} diffusivity. 
To explain the discrepancy, we use Eq.\eqref{GCF100} to obtain 
\bea 
&&\langle \grad c(\br)\cdot\grad c(\bzed) \rangle \, =\, -\triangle C(r)\cr 
&& \hspace{20pt} =\frac{\chi}{D}\left(1+O\left(\left(\frac{r}{\ell_B}\right)^2\right)+ 
O\left(\frac{\sigma}{r}\right)\right), \lb{gradc-eff} 
\eea
which implies that there is a long range of length-scales $\ell_B\gg r\gg \sigma$
over which the ``effective gradients'' $(\grad c)_{e\! f\! f}$ are independent of $r$ and given by 
$(\grad c)_{e\! f\! f}^2\simeq \chi/D.$ More precisely, these are the gradient magnitudes
that would be observed for a coarse-grained field $\grad \bar{c}_\ell$ which has been low-pass filtered 
to contain contributions only from length-scales $>\ell$ for some $\ell_B\gg \ell \gg \sigma.$  These are the gradients that would be seen experimentally by measurements 
with a space resolution $\ell.$
In fact, these are the only physically meaningful gradients, because the effects 
of molecular noise on the concentration field, which we have so far ignored, invalidate 
the ``exact'' result $\chi=D_0\langle |\grad c|^2\rangle.$ We shall discuss this latter 
point in more detail in the following section \ref{physics}. Using the notion 
of ``effective gradients'', we simply note for now that the result \eqref{GCF101} 
can be rewritten in the form 
\be
E_c(k)\simeq  {k_BT \over 6\pi^2 D\eta} (\grad c)_{e\! f\! f}^2 k^{-2}, \quad
1/\ell_B \ll k\ll 1/\sigma, \lb{GCF101-eff}\ee 
in formal agreement with linearized theory.
 
As we now argue, the power-law \eqref{GCF101} will dominate in the concentration 
spectrum for $k$ just slightly greater than $1/\ell_B$ even when $\sigma\ll\ell_B.$
We may follow a similar argument as that in \cite{bandak2021thermal,eyink2021dissipation}
for the kinetic energy spectrum, by simply equating the Kraichnan exponential decay 
spectrum \eqref{Kr68} and the power-law spectrum \eqref{GCF101} to obtain the 
wavenumber $k_{tr}$ where transition to power-law occurs. Setting to 1 all constants 
of order unity, this estimation yields the condition 
\be   (k\ell_B)^2 \exp(-k\ell_B)=\sigma/\ell_B \lb{ktr} \ee 
which has an exact solution 
\be k_{tr}\ell_B =-2 W_{-1}\left(-\frac{1}{2}\left(\frac{\sigma}{\ell_B}\right)^{1/2}\right) 
\lb{lambw} \ee  
in terms of the branch $W_{-1}(z)$ of the Lambert W-function \cite{corless1996lambertw}. 
This implies only a very slow logarithmic increase of $k_{tr}\ell_B$ with decreasing 
$\sigma/\ell_B,$ corresponding to the asymptotics $W_{-1}(x)\sim \ln(-x)-\ln(-\ln(-x))$
for small negative arguments $x$ \cite{corless1996lambertw}. As a typical example, 
in the experiment of \cite{jullien2000experimental} observing the turbulent Batchelor range 
with a solution of fluorescein in water, the Batchelor wavenumber was estimated 
to be 2800 ${\rm cm}^{-1}$ corresponding to $\ell_B\simeq 22.44\ \mu$m whereas the hydrodynamic 
radius of fluorescein is $\sigma\simeq 0.50$ nm \cite{chenyakin2017diffusion}, giving a ratio $\sigma/\ell_B\simeq 2.23\times 10^{-5}.$
Nevertheless, according to \eqref{lambw} one obtains only the modest value 
$k_{tr}\ell_B\simeq 16.3.$ Even if one assumes a very small value such as $\sigma/\ell_B=10^{-10}$
then $k_{tr}\ell_B$ increases only to $29.8,$ not quite doubled. 

The main prediction of our theory is thus that, whereas Batchelor's $1/k$ spectrum for the viscous-convective range survives, thermal fluctuations in the viscous-diffusive range erase 
entirely the rapidly decaying scalar spectra of Batchelor and Kraichnan and replace those 
with a $k^{-2}$ power-law spectrum due to giant concentration fluctuations. This occurs 
at a wavenumber which is just slightly larger than $1/\ell_B,$ with $\ell_B$ the Batchelor 
dissipation length for the concentration field. 
 
The previous simple arguments are rigorous regarding the asymptotic wavenumber ranges $1/\ell_K\ll k\ll 1/\ell_B$ and $1/\ell_B\ll k\ll 1/\sigma,$ but only crudely treat the critical region near the transition wavenumber $k_{tr}\simeq 1/\ell_B.$ To obtain more precise predictions in this region, 
we shall solve our model exactly for the concentration spectrum over the entire range 
$1/\ell_K\ll k\ll 1/\sigma.$ For this purpose, we note that the integral representation  
\begin{eqnarray}\label{Theta>2}
   {C}(r) =   { \chi \over 3  \Gamma }
    \int_r^\infty  { \rho^2\, d\rho\over \rho^3 +\ell_B^2 \rho-{3\over 2}\sigma \ell_B^2  },\\\nonumber
    \end{eqnarray}
following from \eqref{Theta>} can be evaluated exactly by the method of partial fractions, as: 
\begin{eqnarray}\nonumber
{C}(r)=const. -{\chi \over 3\Gamma}\Big[ { {\cal A}\over 2}\ln(b^2+(r+a)^2))\\\label{Final2PointCorr}
+{ {\cal B} r_1\over b}\arctan\left({r+a\over b}\right)  \Big]+{\cal C}\ln|r-r_1|,
\end{eqnarray}
where ${\cal A}\equiv {2r_1^2+\ell_B^2\over 3r_1^2+\ell_B^2}$; ${\cal B}\equiv  {\ell_B^2\over 2(3r_1^2+\ell_B^2)}$ and ${\cal C}={r_1^2\over 3r_1^2+\ell_B^2} $ are dimensionless constants. Note that the cubic polynomial 
$r^3+\ell_B^2 r-{3\over 2}\sigma\ell_B^2$ has negative discriminant, so that it has one real root, 
$r_1,$ and a complex pair of roots $-(a\pm ib),$ where $a=\frac{1}{2}r_1,$ $b^2=\frac{3}{4}r_1^2
+\ell_B^2,$ and $c^2=a^2+b^2$. The real root is given by Vieta's formula as 
\be r_1=w-\frac{\ell_B^2}{3w}, \quad w=\ell_B\sqrt[3]{\frac{3}{4}\left(\frac{\sigma}{\ell_B}\right)
+\sqrt{\frac{1}{27}+\frac{9}{16}\left(\frac{\sigma}{\ell_B}\right)^2}}. \ee  

From this exact solution we can readily verify our previous limiting results. First we 
observe that for $r\gg\ell_B$ the two logarithmic terms dominate and, using the 
relation  ${\cal A}+{\cal C}=1,$ one recovers \eqref{BatchelorCorr} for the 
viscous-convective range. If instead one sets $\sigma=0,$ then also $r_1=0$ and 
the solution \eqref{Final2PointCorr} reduces to 
\be {C}(r)=const. -{\chi \over 6\Gamma}\ln(r^2+\ell_B^2), \ee
which is the physical-space analogue of the scalar spectrum \eqref{Kr68} 
found by Kraichnan \cite{kraichnan1974convection}. The correction
due to thermal noise in the viscous-diffusive range can be evaluated by a joint 
expansion of the exact solution \eqref{Final2PointCorr} in $r/\ell_B$ and 
$\epsilon=\sigma/\ell_B$ and, using 
$r_1/\ell_B=\frac{3}{2}\epsilon+O(\epsilon^2),$ one recovers the result 
\eqref{GCF100}. As we show in the next subsection, the concentration 
spectrum $E_c(k)$ corresponding to \eqref{Final2PointCorr} by Fourier transform
can be found exactly and this result yields the two limiting power laws, \eqref{Batchelor1k}
and \eqref{GCF101}, thus verifying the giant concentration fluctuations 
in the viscous-diffusive range but further describing in detail the transition 
between the two power-law regimes.

\subsection{Concentration Spectrum}\label{SEnergySpectrum}

In this subsection, we discuss the concentration spectrum $E_c(k)$ of our high Schmidt-number 
model \eqref{LangevinPassive1}, which is obtained from the concentration correlation function $C(r)$ by the isotropic Fourier transform relation \eqref{3DFT}. As in the previous subsection, we shall here neglect 
the molecular noise term in Eq.\eqref{LangevinPassive1}, which allows us to obtain an exact result 
for $E_c(k)$ in the range $1/\ell_K\ll k\ll 1/\sigma$ by Fourier transform of the formula 
\eqref{Final2PointCorr} for $C(r).$
The result is easiest to express in terms of the ``one-dimensional spectrum'' 
given by the Fourier cosine transform
\begin{equation}\label{FFunction100}
F(k)\equiv \frac{1}{\pi} \int_0^\infty \cos(kr){C}(r) dr
\end{equation}
in terms of which 
\begin{equation}\label{exactE}
E_c(k)=-k{\partial\over\partial k}F(k).  
\end{equation}
See \cite{hinze1975turbulence}, Eq.(3-231). The result of a somewhat 
lengthy calculation is that 
\begin{equation}\label{FFunction101}
F(k)=F_{{\cal A}}(k)+F_{{\cal B}}(k)+F_{{\cal C}}(k),
\end{equation}
where the three terms correspond to the three terms in Eq.\eqref{Final2PointCorr} for $C(r)$
and are given explicitly by 
\begin{equation}\label{FFunction102}
F_{{\cal A}}(k)= {\chi \over 3\pi\Gamma} {\cal A} {1\over k} {\rm Re}\Big( {\rm fi}(k(a+i b))\Big),
\end{equation}
\begin{equation}\label{FFunction103}
F_{{\cal B}}(k)=- {\chi \over 3\pi\Gamma} {{\cal B}r_1 \over b} {1\over k} 
{\rm Im}\Big(  {\rm fi}(k(a+i b))\Big),
\end{equation}
and
\begin{equation}\label{FFunction104}
F_{{\cal C}}(k)=- {\chi \over 3\pi\Gamma} {\cal C} {1\over k} 
\Big( {\rm fi}(kr_1)-\pi \cos(k r_1)\Big),
\end{equation}
where ${\rm fi}(z)$ denotes the auxiliary sine integral function 
\be {\rm fi}(z)=\int_0^\infty \frac{\sin t}{t+z} dt = \int_0^\infty \frac{e^{-zt}}{1+t^2} dt, 
\quad {\rm Re}(z)>0. \lb{fi-def} \ee 
See \cite{abramowitz2012handbook}, section 5, formula 5.2.12 and 
\cite{oldham2010atlas}, section 38:13. We consulted many tables of 
integrals and collections of integral transforms (such as \cite{Erdelyi1954}),
but we were unable to find the above results in the published literature. 
We therefore give a complete derivation of \eqref{FFunction102}-\eqref{FFunction104}
in the following subsection \ref{integral}. 

However, the reader who is not 
interested in this derivation can skip to the next subsection \ref{limits} 
where we discuss the behavior of $E_c(k)$ in the three limiting cases $k\ell_B\ll 1,$ 
$\sigma=0,$ and $1/\ell_B\ll k\ll 1/\sigma.$ 
We then put together our various results to obtain a global picture 
of the concentration spectrum and we study systematically the effect of varying 
$\epsilon=\sigma/\ell_B,$ exploiting our exact solution \eqref{FFunction102}-\eqref{FFunction104}
to plot the results for $E_c(k)$ with $\epsilon$ varying over realistic values.

%

\subsubsection{Evaluation of Integrals}\lb{integral} 

\begin{figure}
\includegraphics[scale=.1]{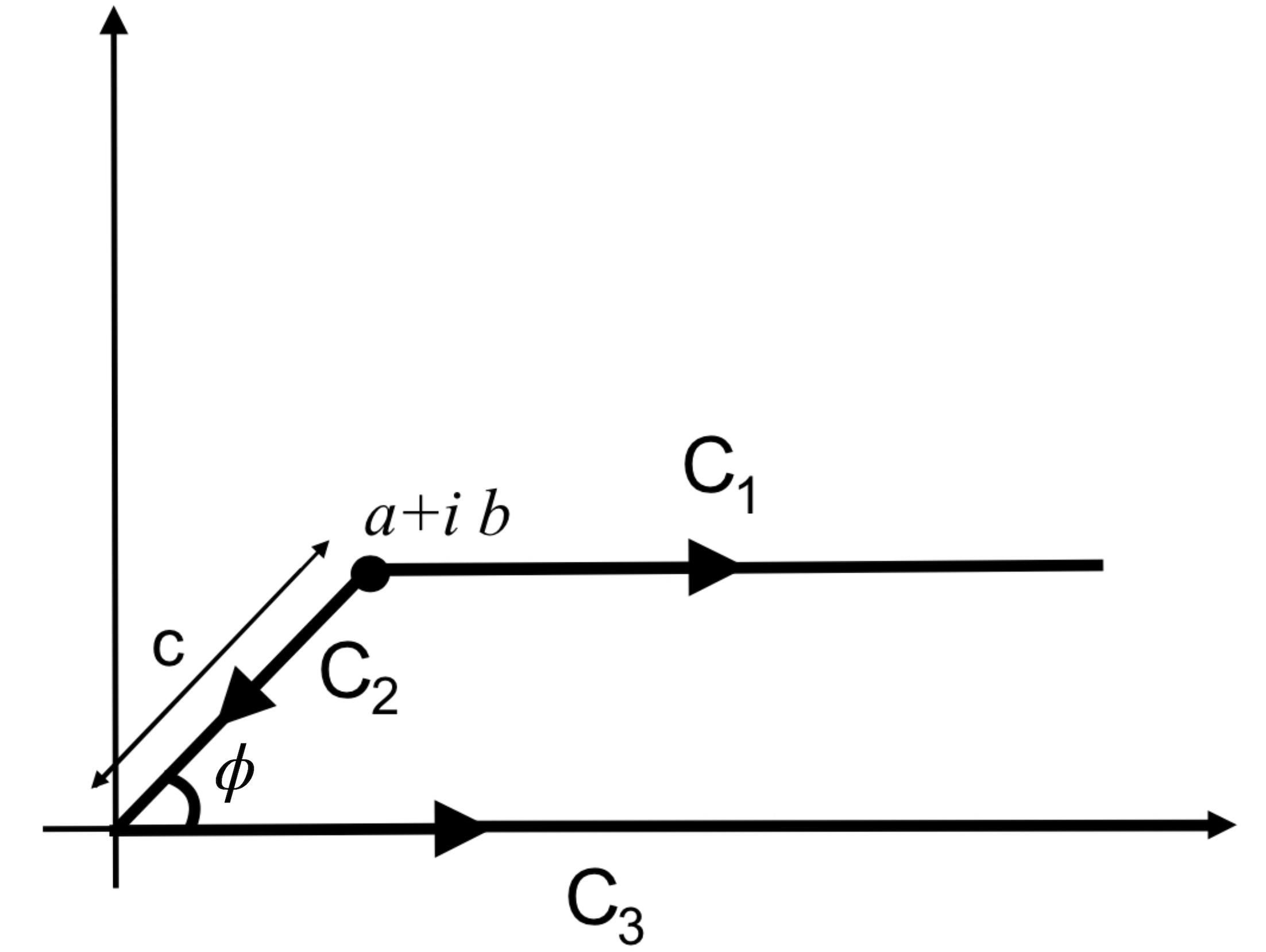}
\centering
\caption {\footnotesize {The contours used in (\ref{contourintegral}),
(\ref{contourintegral2}). Here, $a+ib=c e^{i\phi}$ in terms of the constant 
$c=\sqrt{a^2+b^2}$ and $\phi=\arctan(b/a)$.}}
\label{contours}
\end{figure}

The three formulas \eqref{FFunction102}-\eqref{FFunction104} are direct consequences of 
the following two integrals: 
\begin{eqnarray}\nonumber
&&\lim_{\mu\to 0}\left[{1\over 2}  \int_0^\infty e^{-\mu r} \log \Big( (r+a)^2+b^2 \Big) \cos(kr) dr
\right. \\\nonumber
&&\hspace{40pt} +i \left. \int_0^\infty e^{-\mu r} \arctan\Big({b\over r+a} \Big) \cos (kr) dr\right]\\\label{finalmain}
&& \hspace{80pt} \ =\ -{1\over k} {\rm fi}(k(a+ib)) \lb{first} 
\end{eqnarray}
and
\bea 
&& \lim_{\mu\rightarrow 0} 
\int_0^\infty e^{-\mu r} \ln|r-r_1|\cos(k r)\,dr \cr
&& \hspace{50pt} =\frac{1}{k}\left[{\rm fi}(kr_1)-\pi \cos(kr_1)\right].
\lb{second} \eea 
Here $\mu$ is an infrared regulation scale whose role in the physical problem is played by $1/L$ and 
$1/\ell_K.$ Since the results for $\mu\ll k$ do not depend upon the particular regularization adopted, however, 
we chose the above exponential IR cutoff  for mathematical convenience. The key idea in the 
evaluation of the first integral \eqref{first} was to realize that the two integrands,
$\frac{1}{2}\log \left( (r+a)^2+b^2 \right)$ and $\arctan\left({b\over r+a}\right)=
\frac{\pi}{2}-\arctan\left({r+a\over b}\right),$ are the real and imaginary 
parts respectively of $\ln(r+a+ib)$ and the combined integral may thus be expressed as the 
contour integral of analytic functions $\ln z \,\cos(kz),$ $\ln z\,\sin(kz)$ along the path 
$C_1$ in the complex plane which is illustrated in Fig.~\ref{contours}. A convenient
change of the contour allows us to reduce the integral to standard formulas, which 
we explain in more detail below. The second integral \eqref{second} is simpler and 
will be discussed briefly at the end.

The first integral \eqref{first}, which we call $I_1,$ 
can be directly written as the following complex contour integral: 
\begin{eqnarray}\nonumber
&& I_1\ =\ \lim_{\mu\to 0} \int_{C_1:{\mathbb R}^++a+ib} dz e^{-\mu z}\ln z \\\label{contourintegral}
&&\hspace{10pt}\times \Big[   \cos (kz) \cos(k_\phi c)+\sin(kz)\sin (k_\phi c)\Big],
\end{eqnarray}
where $k_\phi:=k e^{i\phi}$ and $\phi=\arctan (b/a),$ noting that $a+ib=ce^{i\phi}.$
Because the integrand is analytic in the complex plane with a branch cut along the 
negative real axis, the integration contour can be shifted to $C_2+C_3$ as illustrated in Fig.\ref{contours}. This yields directly
\begin{eqnarray}\nonumber
&& I_1= {-1\over c}\int_0^c dr \Big[ (a\ln r -b \phi)+i(b \ln r +a\phi)  \Big] \cr 
&& \hspace{20pt} \times [\cos (k_\phi r)\cos(k_\phi c)+\sin (k_\phi r)\sin(k_\phi c)]\\\nonumber
&& +\lim_{\mu\to 0} \int_0^\infty dr e^{-\mu r} \ln r \cr 
&& \hspace{20pt}\times  [\cos (k_\phi r)\cos(k_\phi c)+\sin (k_\phi r)\sin(k_\phi c)].
\\\label{contourintegral2}
\end{eqnarray}
The first of these integrals resulting from contour $C_2$ can now be evaluated 
using \cite{Erdelyi1954}, formulas 1.5(1) and 2.5(1), while the 
second integral resulting from contour $C_3$ can be evaluated using \cite{Erdelyi1954},
formulas 1.5(6) and 2.5(7), giving 
\begin{eqnarray}\nonumber
&& I_1 \ = \ {1\over k}\Big[ \Big({\rm Si}(k(a+ib))-{\pi\over 2}\Big)\cos(k(a+ib))\\\nonumber
&&+\Big( {\rm Cin}(k(a+ib))-\gamma-\ln (k(a+ib)) \Big)\sin(k(a+ib))  \Big].
\end{eqnarray}
Here we use the trigonometric integral functions ${\rm Si}(z)$ and ${\rm Cin}(z)$
with also ${\rm Cin}(x)=\gamma+\ln (x)-{\rm Ci}(x)$ where $\gamma$ is the Euler-Mascheroni 
constant: see \cite{abramowitz2012handbook}, section 5.2; \cite{oldham2010atlas},
Chapter 38; or \cite{bateman2006higher}, section 9.8. Finally, a standard formula
\be\label{fi}
{\rm fi}(z) = {\rm Ci}(z) \sin (z)+\Big( {\pi\over 2}-{\rm Si} (z)\Big) \cos (z),
\ee
for the auxiliary sine integral function (see \cite{abramowitz2012handbook},
formula 5.2.6 or \cite{oldham2010atlas}, formula 38:13:7) yields \eqref{first}. 

The second integral \eqref{second} can be straightforwardly decomposed 
into two contributions for $r<r_1$ and $r>r_1$: 
\bea 
&& \lim_{\mu\rightarrow 0} 
\int_0^\infty e^{-\mu r} \ln|r-r_1|\cos(k r)\,dr \cr
&& \hspace{20pt} =\int_0^{r_1} \ln r\, \cos(k(r-r_1)) \, dr \cr 
&& \hspace{20pt} + \lim_{\mu\rightarrow 0} 
\int_0^\infty e^{-\mu r} \ln r\, \cos(k (r+r_1))\,dr.
\lb{second2} \eea 
The part for $r<r_1$ can be evaluated as 
\bea 
&& \int_0^{r_1} \ln r\, \cos(k(r-r_1)) \, dr =
\frac{1}{k} \ln r_1 \sin(kr_1) \cr 
&& -\frac{1}{k} \left[{\rm Si}(kr_1)\cos(k r_1)+{\rm Cin}(k r_1)\sin(kr_1)\right] 
\eea 
using 
$\cos(k(r-r_1))=\cos(kr)\cos(kr_1)+\sin(kr)\sin(kr_1)$ and 
\cite{Erdelyi1954}, formulas 1.5(1) and 2.5(1), while the part for $r>r_1$
can be evaluated as 
\bea 
&& \lim_{\mu\rightarrow 0} 
\int_0^\infty e^{-\mu r} \ln r\, \cos(k (r+r_1))\,dr \cr
&& =\frac{1}{k} (\gamma+\ln k)\sin(kr_1)-\frac{\pi}{2k}\cos(kr_1) 
\eea 
using $\cos(k (r+r_1))=\frac{1}{2}\left(e^{ik (r+r_1)}+e^{-ik (r+r_1)}\right)$ 
and \footnote{This result can be obtained, in principle, by combining two 
results in \cite{Erdelyi1954}, formulas 1.5(6) and 2.5(7). Note, however, that 
there is a typographical error in the second of these formulas and, in fact,
a surprising number of misprints for these formulas in standard sources. Thus, 
in \cite{Erdelyi1954}, formula 2.5(7), the quantity $\tan^{-1}(C/\alpha)$ 
should instead be ${\rm tan}^{-1}(y/\alpha).$ Likewise, in \cite{gradshteyn2007table}, formula 
4.441(1), $\ln(p^2-q^2)$ should instead be $\ln(p^2+q^2).$ Finally, in 
\cite{Erdelyi1954}, formula 4.6(1), $\log(\gamma p)$ should instead be $\gamma +\log p.$
The correct result can be easily obtained from the standard integral for Euler's 
$\Gamma$-function 
$$ \int_0^\infty e^{-zt} t^{s-1} dt= \Gamma(s)/z^s, \quad {\rm Re}\,z>0$$
by differentiating both sides with respect to $s$ and setting $s=1.$}
\be \int_0^\infty e^{-zr}\ln r\, dr=-\frac{\gamma+\ln z}{z}, \quad {\rm Re}\,z>0. \ee  
Combining the two parts and again using \eqref{fi} yields the second integral \eqref{second}. 

\subsubsection{Model Spectrum for $1/\ell_K\ll k\ll 1/\sigma$}\lb{limits}

\begin{figure*}
          \begin{subfigure}[t]{0.4\textwidth}
           \includegraphics[width=7.5cm]{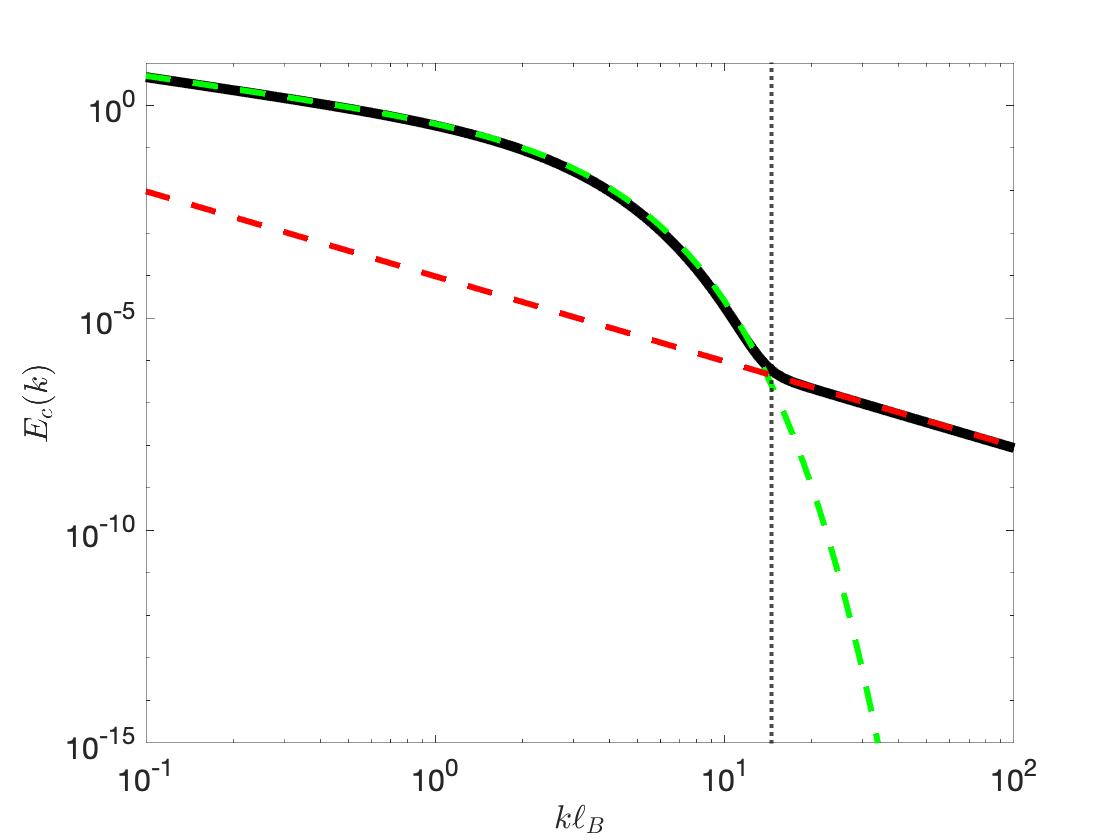}
         \caption{}
         \label{1e4}
     \end{subfigure}
          \begin{subfigure}[t]{0.4\textwidth}
        \includegraphics[width=7.5cm]{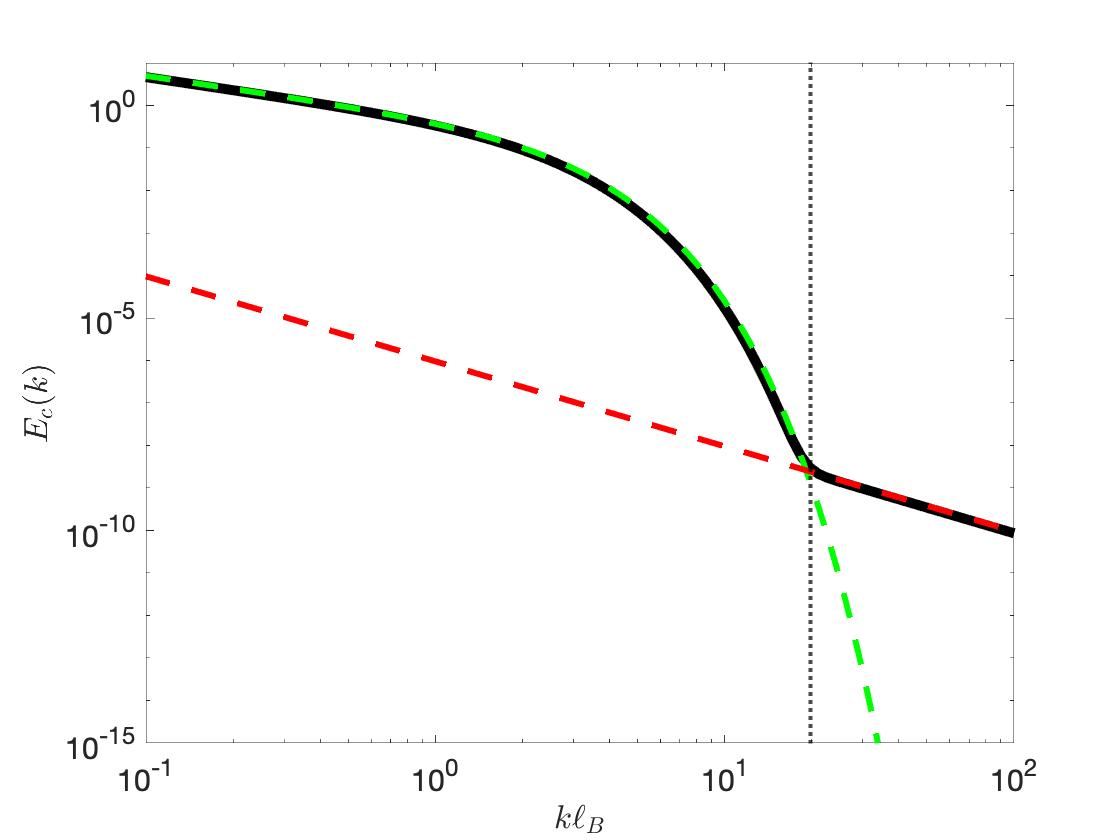}
         \caption{}
         \label{1e6}
     \end{subfigure}
      \\\hfill
   \\
     \begin{subfigure}[b]{0.4\textwidth}
         \includegraphics[width=7.5cm]{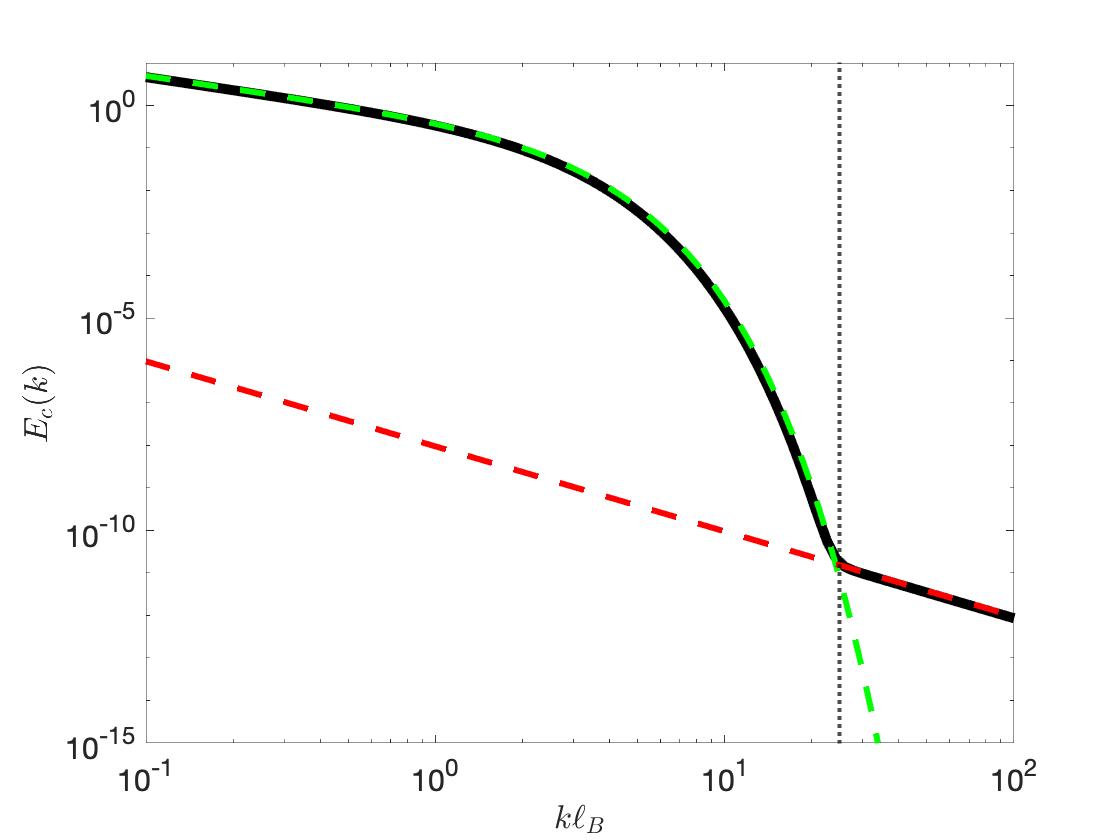}
         \caption{}
         \label{1e8}
     \end{subfigure}
    \begin{subfigure}[b]{0.4\textwidth}
       \includegraphics[width=7.5cm]{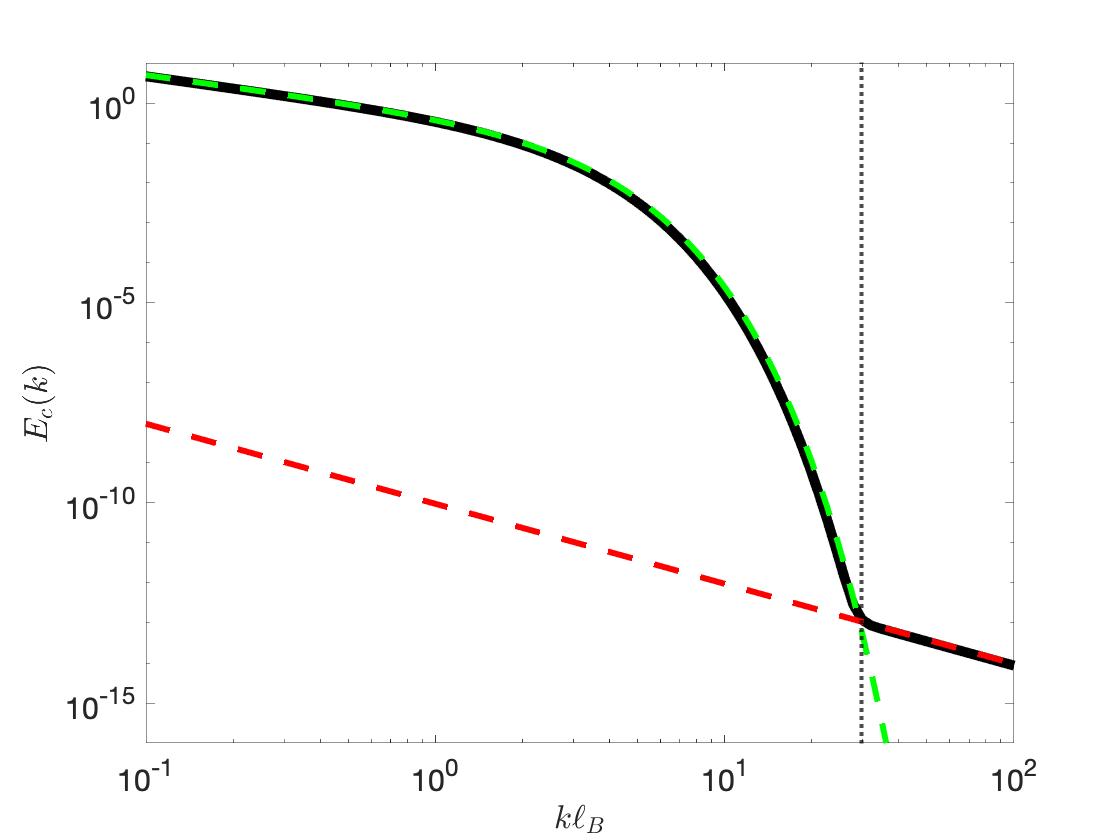}
         \caption{}
         \label{1e8}
     \end{subfigure}
         \caption{Plots of the exact solution for the concentration spectrum 
    $E_c(k)$ with (a) $\epsilon=10^{-4}$, (b) $\epsilon=10^{-6}$, (c) $\epsilon=10^{-8}$, (d) $\epsilon=10^{-10}$. The solid black line ($\boldsymbol -$) is the exact result from \eqref{FFunction102}-\eqref{FFunction104}, the green dashed line (\textcolor{green}{{\bf -$\,$-$\,$-}}) is Kraichnan's spectrum \eqref{Kraichnanexp}, and the red dashed line (\textcolor{red}{{\bf -$\,$-$\,$-}}) is the power-law spectrum \eqref{GCF2} associated to GCF's. The vertical grey dotted line marks the prediction in \eqref{lambw} for the transition wavenumber
    $k_{tr}$.}\label{Spectra}
\end{figure*}

We now consider the limits of our exact solution \eqref{FFunction102}-\eqref{FFunction104}
in the viscous-convective and viscous-diffusive ranges. First, however, we verify 
that our solution recovers the result of Kraichnan \cite{kraichnan1974convection} when 
setting $\sigma=0.$ In this case, $a=\frac{1}{2}r_1=0,$ $b=\ell_B,$ ${\cal A}=1,$ 
${\cal B}=\frac{1}{2},$ ${\cal C}=0,$ and 
\be F(k)=\frac{\chi}{3\pi\Gamma} \frac{{\rm Re}({\rm fi}(ik\ell_B))}{k} 
= \frac{\chi}{6\Gamma} \frac{e^{-k\ell_B}}{k} 
\ee 
using definition \eqref{fi-def} of ${\rm fi}(z)$ and \cite{Erdelyi1954}, 
formula 1.2 (11). The standard concentration spectrum obtained from (\ref{exactE}) 
then reproduces exactly the result of Kraichnan 
\be\label{Kraichnanexp}
E_c(k)={\chi\over 6\Gamma} \Big({1\over k}+\ell_B \Big)e^{-k\ell_B} \qquad (\sigma=0). 
\ee
See equations (2.27),(5.14) in \cite{kraichnan1974convection}. Note that in terms of the notations used in Kraichnan's original work, his constants $A$, $\langle a\rangle $, $\alpha$ and $\kappa$ are related to our constants as $\Gamma= A/30= \langle a \rangle/6$ and $\ell_B^2= 2 \alpha^2=30\kappa /A$, with his bare diffusivity $\kappa$ replaced by our renormalized diffusivity $D.$ We recover also 
\eqref{Kr68} in the Introduction if, following \cite{kraichnan1968small}, we introduce
$C_B=\gamma/6\Gamma$ and if the Batchelor length is redefined in the more conventional 
way as $\ell_B:=(D/\gamma)^{1/2}=(D/6C_B\Gamma)^{1/2}.$

The result of Batchelor for the concentration spectrum in the viscous-convective range 
can be recovered, however, when $k\ell_B\ll 1$ for all finite $\sigma$ without taking 
the limit $\sigma=0.$ This result follows from our exact solution 
\eqref{FFunction102}-\eqref{FFunction104} using 
${\rm fi}(0)=\pi/2,$ which gives 
$$F_{\cal A}(k)\simeq (\chi/6\Gamma){\cal A} k^{-1}, \quad
F_{\cal C}(k) \simeq (\chi/6\Gamma){\cal C} k^{-1} $$
and $F_{\cal B}(k)\simeq o(1/k)$. Using ${\cal A}+{\cal C}=1$ we obtain in general 
$ F(k)=(\chi/6\Gamma)k^{-1}$ so that 
$$ E_c(k)\simeq (\chi/6\Gamma)k^{-1}=C_B (\chi/\gamma)k^{-1} ,\quad k\ell_B\ll 1$$
in agreement with \eqref{Ba59} and \eqref{Batchelor1k}. As already emphasized,
the Batchelor spectrum in the viscous-convective range is unaltered by the 
sub-Kolmogorov-scale thermal velocity fluctuations. 

The behavior of our model in the viscous-diffusive range for small 
finite $\sigma,$ on the other hand, is completely different than that 
obtained by Kraichnan for $\sigma=0.$ The limit $1/\ell_B\ll k\ll 1/\sigma$
of our exact solution \eqref{FFunction102}-\eqref{FFunction104} can be 
obtained from the standard asymptotic expansion of ${\rm fi}(z)$ 
for large arguments $z.$ We are further interested in the limit 
$\epsilon=\sigma/\ell_B\ll 1$, so that 
$$a=\frac{1}{2}r_1=\frac{3}{4}\sigma \left(1+O(\epsilon)\right), \
b=\ell_B\left(1+O(\epsilon^2)\right)$$
and 
$${\cal A}=1+O(\epsilon^2),\ {\cal B}r_1/b 
=\frac{3}{4}\epsilon + O(\epsilon^3),\ {\cal C}=O(\epsilon^2).$$ 
In that case, the contributions to $F(k)$ at leading order in $\epsilon$
are obtained from 
$$ {\rm Re}\Big( {\rm fi}(k(a+i b))\Big) \simeq -ka \,
{\rm Re}\Big( {\rm gi}(i k\ell_B)\Big) +O(\epsilon^2) 
$$
and 
$$ {\rm Im}\Big(  {\rm fi}(k(a+i b))\Big) \simeq 
{\rm Im}\Big(  {\rm fi}(ik\ell_B)\Big) + O(\epsilon), 
$$ 
where we have introduced the auxiliary cosine integral function 
${\rm gi}(z)=-{\rm fi}'(z).$ Invoking the asymptotic expansions 
${\rm fi}(z)\sim 1/z$ and ${\rm gi}(z)\sim 1/z^2$ at large 
arguments $z$ satisfying
$|{\rm arg}\, z|<\pi,$ \cite{abramowitz2012handbook},5.2.34-35 and 
\cite{oldham2010atlas}, 38:13:10-11, we then obtain 
$$F_{\cal A}(k)\simeq F_{\cal B}(k)\simeq {\chi \sigma\over 4\pi\Gamma}k^{-2}\ell_B^{-2}, 
\quad F_{\cal C}(k)\simeq 0  $$ 
and finally 
\be E_c(k) \simeq {\chi \sigma\over \pi D}k^{-2},\quad 
1/\ell_B\ll k\ll 1/\sigma. \lb{GCF2} \ee
Thus, we confirm the spectrum (\ref{GCF101}) with $k^{-2}$ power-law  
associated to giant concentration fluctuations.

The effect of varying $\epsilon=\sigma/\ell_B$ can be 
illustrated by plotting our exact solution over a range of possible values.
See Fig.\ref{Spectra}. For details of the numerical method
used in constructing the plots, see Appendix \ref{AppPlots}. The first 
observation from the figure is that the spectrum $E_c(k)$ given by 
\eqref{FFunction102}-\eqref{FFunction104} is almost perfectly 
represented, on the log-log scale of the plots, by the superposition 
of the Kraichnan spectrum \eqref{Kraichnanexp} and the power-law spectrum 
\eqref{GCF2}, with transition at the wavenumber $k_{tr}$ predicted 
by \eqref{lambw}. The second conclusion is that the Kraichnan exponential 
decay spectrum will generally exist for only a very narrow range 
of wavenumbers with $k\ell_B\gtrsim 1.$ From a review of the past 
experiments on turbulent high-Schmidt mixing, we find that the 
ratio $\epsilon=\sigma/\ell_B$ ranges over values between $10^{-4}$
and $10^{-7}.$ In the latter extreme case, $k_{tr}\ell_B=22.3$ and the 
Kraichnan spectrum exists over just a bit more than a decade of wavenumbers.
However, even if one considers a very unrealistic value $\epsilon=10^{-10},$
then $k_{tr}\ell_B=29.8$ and the extent of the Kraichnan spectrum is barely 
increased. These considerations suggest that an exponential decay 
of the concentration spectrum will generally hold in physical fluid 
mixtures for at most a decade of wavenumbers in the viscous-diffusive range.

\section{Physical Predictions}\lb{physics} 

We now develop concrete predictions of our theory for two specific binary mixtures, water-glycerol and water-fluorescein. We note 
that water-glycerol solutions are very commonly employed in fluid turbulence 
experiments as a means to vary the viscosity by changes in concentration,
e.g. see \cite{cadot1997energy,debue2018experimental}. Giant concentration
fluctuations have also been seen experimentally in water-glycerol 
solutions by a variety of observational techniques \cite{brogioli2000universal,
brogioli2000giant,croccolo2007nondiffusive,ortiz2013nonequilibrium}. 
For prior experiments on the turbulent Batchelor regime, solutions 
of disodium fluorescein (or, in shorthand, fluorescein) in water have been popular 
\cite{sreenivasan1989new,williams1997mixing,miller1996measurements,jullien2000experimental},
because of the ease of visualization by laser fluorescence. It is worth 
emphasizing that the Stokes-Einstein relation for the diffusivity is observed 
to be valid for both mixtures, water-glycerol \cite{chen2006stokes,elamin2015brownian}
and water-fluorescein \cite{mustafa1993dye,chenyakin2017diffusion}, at sufficiently 
high temperatures well above the glass transition. The measured hydrodynamic radii
in water are $\sigma=0.35\,$nm for glycerol \cite{elamin2015brownian} and 
$\sigma=0.50\,$nm for fluorescein \cite{mustafa1993dye}, relatively consistent  
with the molecular volumes. Since the molar masses of water (${\rm H}_2{\rm O}$), 
glycerol (${\rm C}_{3}{\rm H}_{8}{\rm O}_{3}$), and disodium fluorescein 
(${\rm C}_{20}{\rm H}_{12}{\rm Na}_{2}{\rm O}_{5}$) are $18.0\,$ g/mol, $92.1\,$ g/mol, and 
$376.3\,$ g/mol, respectively, the hydrodynamic Stokes-Einstein prediction should 
be expected to be even more accurate for the water-fluorescein mixture. Our 
predictions can be easily extended to other binary mixtures. 

Before we can discuss our detailed predictions, however, we must first discuss the 
equilibrium fluctuations of the concentration which we have so far ignored. 
As is well known, the fluctuations in thermal equilibrium correspond to 
the structure function \eqref{Scc-def} given by 
\be  S_{cc}(k)=\frac{k_B\overline{T}}{\overline{\rho} 
(\partial \mu/\partial \overline{c})}_{\overline{T},\overline{p}}, \lb{Scc-eq} \ee
which is independent of wavenumber $k,$ where $\overline{\rho},$ $\overline{c},$ etc. 
are the mean values in the equilibrium state and $\mu(T,p,c)$ is the 
chemical potential of the mixture for given temperature $T,$ pressure $p,$ 
and mass concentration $c$. See \cite{dezarate2006hydrodynamic},
Eq.(5.34) and references therein. At the hydrodynamic level of description, 
these fluctuations are due to the molecular noise term in the equation 
\eqref{Passive110} for concentration, whose general form is 
\be 
\partial_t c=-\bu\bdot\grad c +\grad\bdot \left(D_0\grad c+\sqrt{\frac{2D_0}
{\rho (\partial \mu/\partial c)_{T,p}}}\; {\boldsymbol\eta}_c({\bf x}, t) \right).\\
\lb{ceq-gen} 
\ee 
See \cite{morozov1984langevin,dezarate2006hydrodynamic,donev2014low,nonaka2015low}.
In the prior discussion, following DFV \cite{donev2014reversible} and\cite{usabiaga2012staggered}, we have considered for simplicity the special case 
of an ideal solution of two equal-mass molecules, e.g. the self-diffusion of 
tagged particles in a single component fluid.  In this setting, we provide
in Appendix \ref{FDT} a self-contained derivation of \eqref{Scc-eq} from 
\eqref{ceq-gen}, as a convenience for readers. It is important to note, however,
that the high-Schmidt asymptotics of DFV in Appendix \ref{S2c} can be carried 
out for the general equation \eqref{ceq-gen} and does not require the special 
assumptions of \eqref{Passive110}. 

The structure function \eqref{Scc-eq} corresponds to a scalar spectrum 
\be  E_{c}(k)=\frac{k_B\overline{T}}{4\pi^2 \overline{\rho} 
(\partial \mu/\partial \overline{c})}_{\overline{T},\overline{p}} k^2. \lb{Ec-eq} \ee
This equilibrium spectrum is growing in wavenumber and it must thus exceed the spectrum (\ref{GCF101}) of the non-equilibrium fluctuations above some sufficiently 
high transition waveumber, which we call $k_{tr}'.$  We can estimate the latter by 
equating the two spectra, which yields 
\be k_{tr}'= \left( \frac{4\pi\chi\sigma}{Dk_BT}
\left( \frac{\partial\mu}{\partial c}\right)_{T,p}\right)^{1/4}. \lb{ktrp} \ee 
To make a quantitative determination of $k_{tr}'$, we must evaluate the 
derivative of the chemical potential.  
For this purpose, we note the general result 
\be \left( \frac{\partial\mu}{\partial c}\right)_{T,p}=
\frac{B\cdot k_BT}{c(1-c)[m_0 c+m_1(1-c)]} \lb{Bfac} \ee
where $m_0$ is the molecular mass of the solute (water), $m_1$ is the 
molecular mass of the solvent (glycerol/fluorescein) and $B=B(T,p,c)$
is a factor which accounts for the non-ideality of the mixture, with 
$B\equiv 1$ in the ideal case. See Appendix \ref{AppMix} for the 
derivation of this equilibrium thermodynamic result. In addition, estimation 
of $k_{tr}'$ requires the diffusion coefficient $D.$

\begin{figure}
\includegraphics[scale=.45]{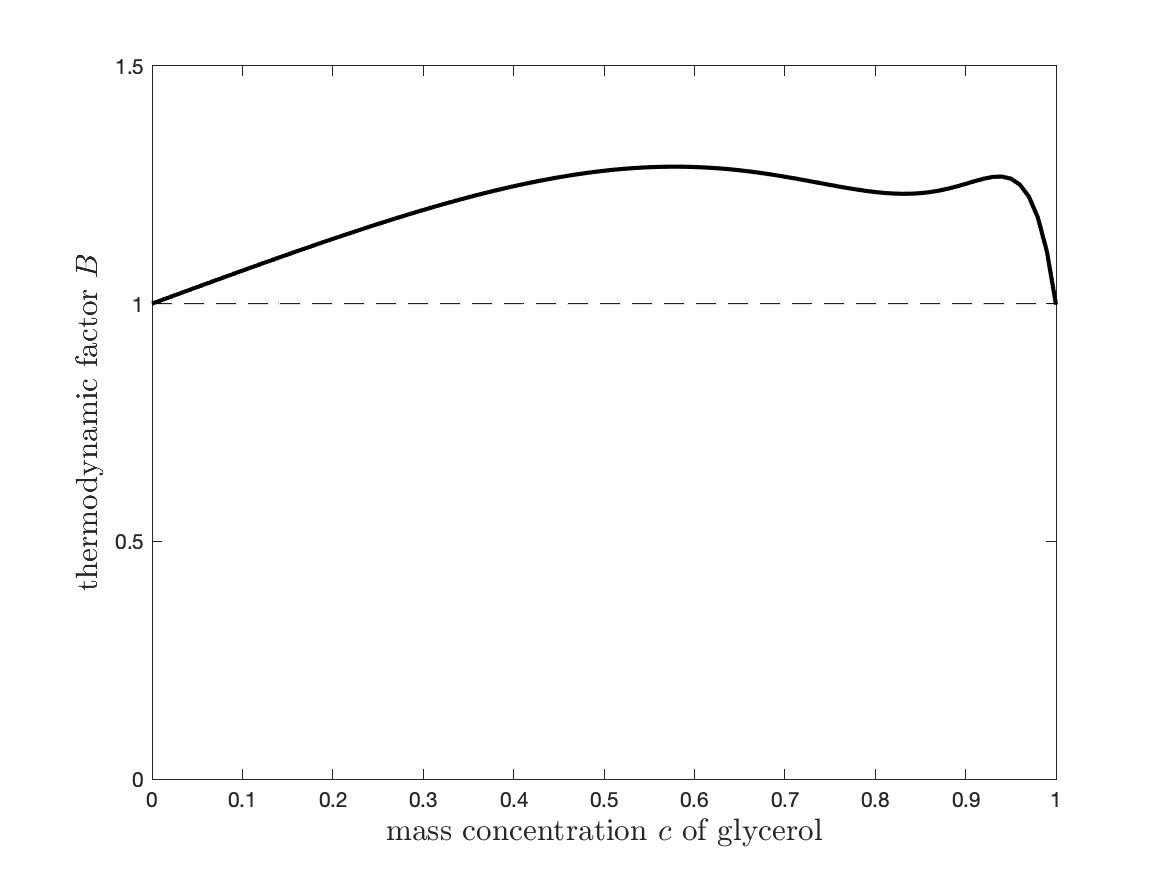} 
\caption {\footnotesize {The non-ideality factor $B$ for a water-glycerol 
solution, from \cite{derrico2004diffusion}, Eq.(11), plotted versus mass 
concentration of glycerol.}}
\label{Bfac:fig}
\end{figure}

For water-glycerol solutions at temperature $25^\circ$C and at atmospheric 
pressure (1 bar) experimental values of both quantities $D$ and $B$ are 
conveniently provided by the paper \cite{derrico2004diffusion}, in 
parameterized form as functions of the molar concentration $n.$
We plot in Fig.\ref{Bfac:fig}  the factor $B$ versus mass concentration $c,$
where it can be observed that the deviation from ideality is at most 
about $28.8\%.$ Because of the detailed information provided in 
\cite{derrico2004diffusion}, we shall present our results for the 
concentration spectrum of water-glycerol solutions at $T=25^\circ$C
and $p=1\,$bar. For water-fluorescein we shall use the value of 
diffusivity $D=5.54\times 10^{-6}\,{\rm cm}^2/{\rm s}$ at $T=30^\circ$C and 
$p=1\,$bar reported in \cite{mustafa1993dye}. There has not been much 
experimental investigation of thermodynamic properties of water-fluorescein 
solutions and we are unaware of any measurements of the non-ideality 
factor $B$ for that mixture. Here we may remark that the thermodynamics 
and diffusive transport of an electrolyte such as disodium fluorescein
in water may be treated as a binary mixture, at least at not too 
low concentrations (see \cite{tyrrell2013diffusion}, Chapter 8). 
In the lack of precise information, one may 
simply take $B\equiv 1.$ In fact, the deviations from ideality must be 
much larger for water-fluorescein than for water-glycerol, because the 
differences between water and fluorescein molecules and their interactions 
are considerable. However, $B$ is unlikely to be orders of magnitude 
different from unity and, since it appears in formula \eqref{ktrp} to the $1/4$ power, 
setting $B\equiv 1$ should result in just slight inaccuracy of $k_{tr}'.$

Before presenting any concrete predictions, we must first note an important 
consequence of the equipartition spectrum \eqref{Ec-eq}. It is easily seen that 
the concentration gradients become dependent upon the UV cut-off $\Lambda$ 
which is necessary for the validity of the fluctuating hydrodynamic 
equation \eqref{ceq-gen}. See Appendix \ref{FDT} for a discussion of this point. 
The concentration gradient develops a large contribution, diverging with $\Lambda,$
of the form 
\be \langle |\grad c|^2\rangle = 2\int_0^\Lambda dk\, k^2 E_c(k) 
\simeq \frac{2}{5} A\Lambda^5, \ee 
where $A=\frac{k_B\overline{T}}{4\pi^2 \overline{\rho} 
(\partial \mu/\partial \overline{c})}_{\overline{T},\overline{p}}$
is the constant prefactor in the equipartition spectrum \eqref{Ec-eq}.  
Note that the estimate of the ``effective gradient'' $\grad \bar{c}_\ell$ 
from equation \eqref{gradc-eff} ignored this contribution and is valid only 
if the filtering scale $\ell$ is chosen so that $\ell k_\nabla \gtrsim 1$
with $\frac{2}{5} A k_\nabla^5 := (\grad c)_{e\! f\! f}^2=\chi/D$. 
It is easy to check from this condition that 
\be k_\nabla/k_{tr}'\simeq 1/(k_{tr}'\sigma)^{1/5} \gtrsim 1, \ee
and thus it will suffice to choose $k_{tr}'\ell\gtrsim 1$ 
(and $\ell\lesssim \ell_B$) to ensure that 
$\grad \bar{c}_\ell\simeq (\grad c)_{e\! f\! f}.$

\begin{figure}
          \begin{subfigure}[t]{0.45\textwidth}
           \includegraphics[width=8cm]{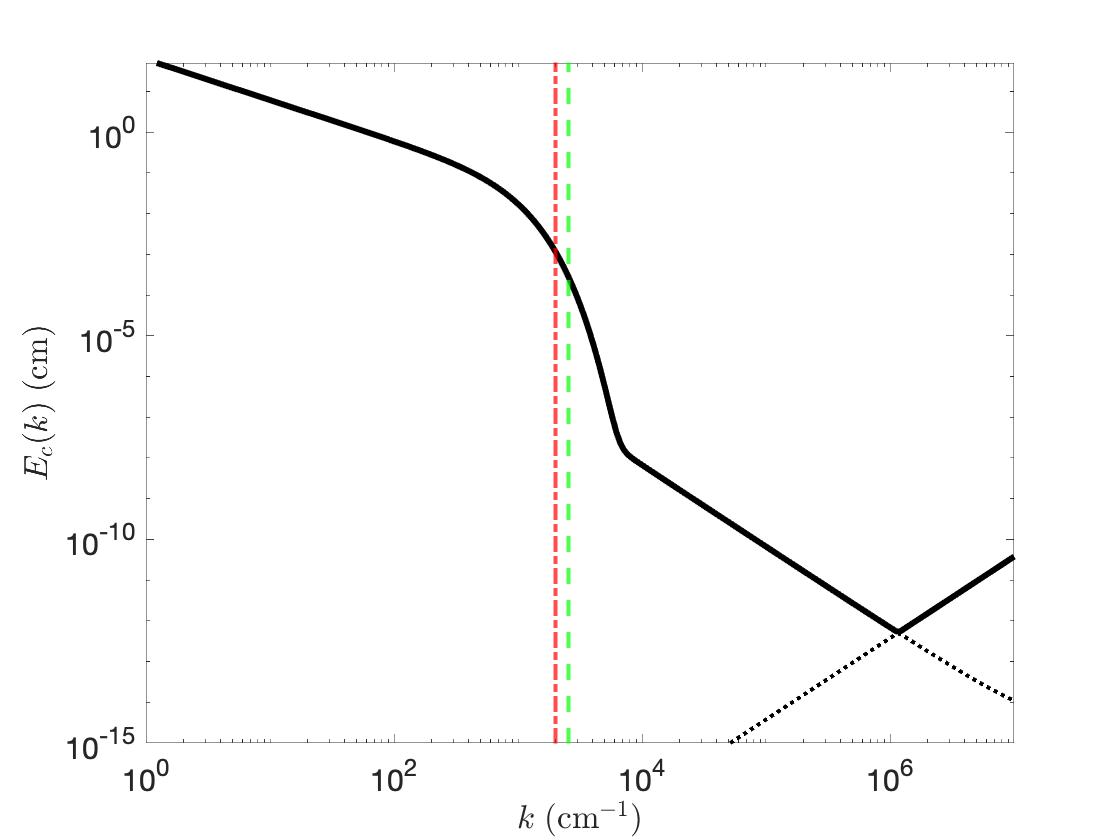}
         \caption{}
         \label{aa}
     \end{subfigure}
          \begin{subfigure}[t]{0.45\textwidth}
        \includegraphics[width=8cm]{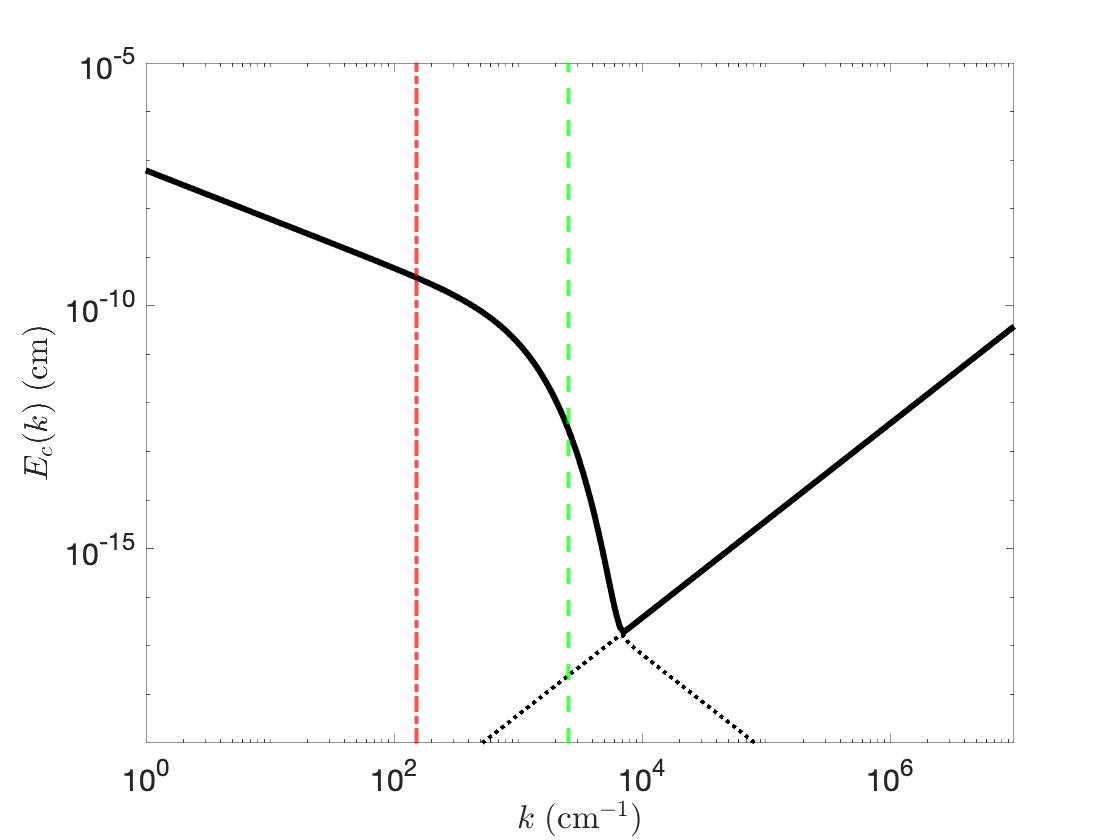}
         \caption{}
         \label{bb}
     \end{subfigure}
      \\\hfill
   \\
     \begin{subfigure}[b]{0.45\textwidth}
         \includegraphics[width=8cm]{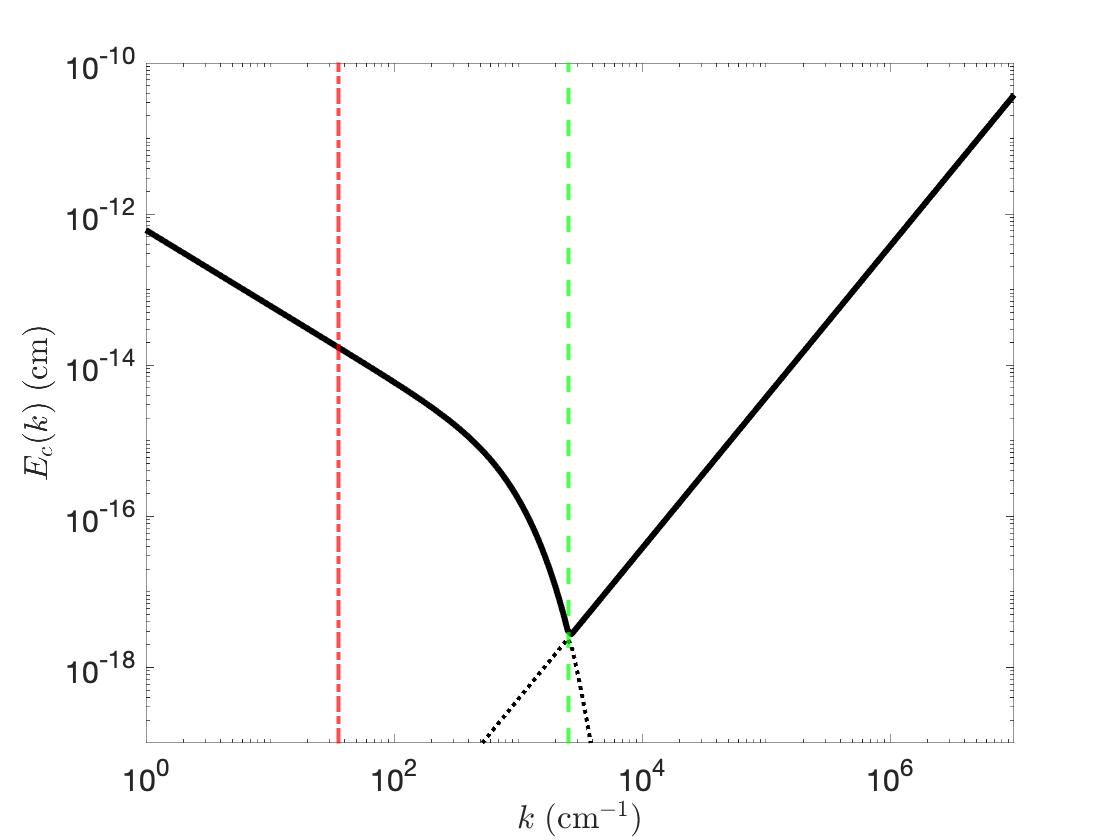}
         \caption{}
         \label{cc}
     \end{subfigure}
             \caption{Plots of the predicted concentration spectrum for water-glycerol mixture ($T=25^\circ\,C,$ $p=1$ bar, $\bar{c}=0.5$) with $\gamma=10\;s^{-1}$ and (a) $\chi=10^2$, (b) $\chi=10^{-7}$, (c) $\chi=10^{-12}\;s^{-1}$. 
             The green dashed line (\textcolor{green}{{\bf -$\,$-$\,$-}}) marks the Batchelor wavenumber $k_B$, and the red dot-dashed line (\textcolor{red}{{\bf -$\,\cdot\,$-}}) represents buoyancy cut-off wavenumber $k_g$.}\label{waterglycerol10}
\end{figure}

\begin{figure}
\begin{centering}
          \begin{subfigure}[t]{0.45\textwidth}
           \includegraphics[width=8cm]{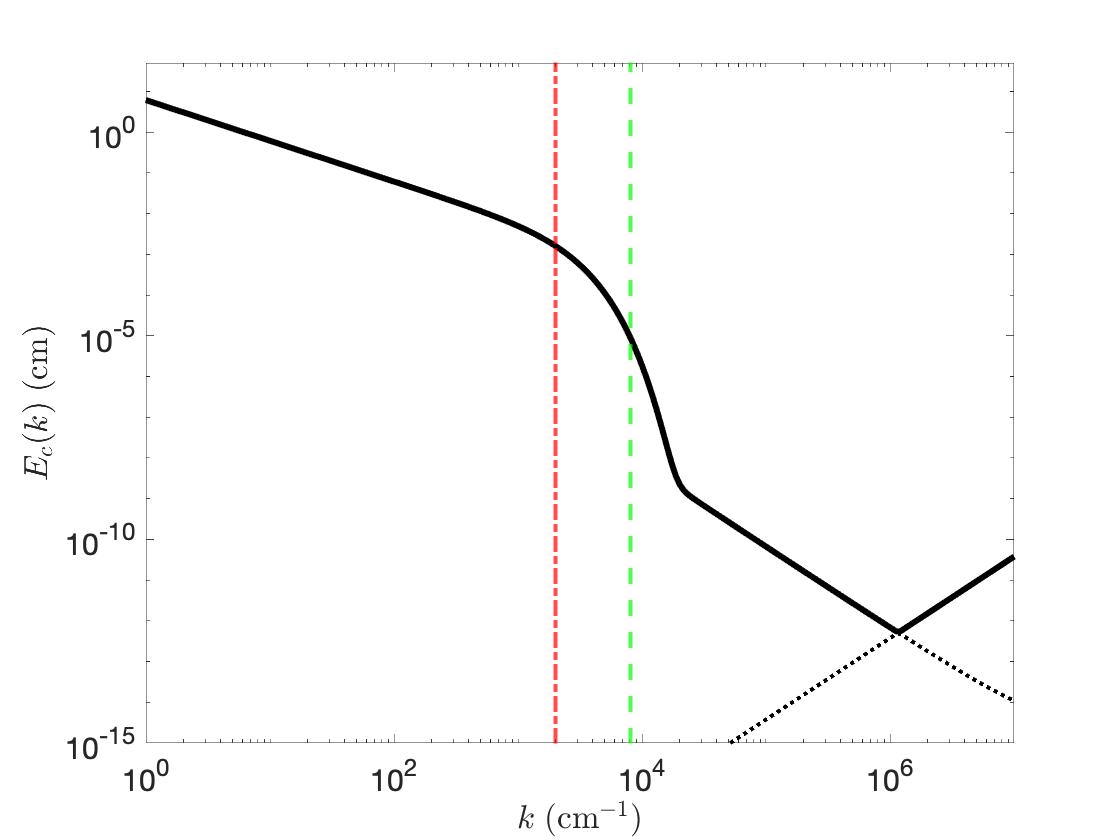}
               \caption{}
     \end{subfigure}
           \\\hfill
   \\
          \begin{subfigure}[t]{0.45\textwidth}
        \includegraphics[width=8cm]{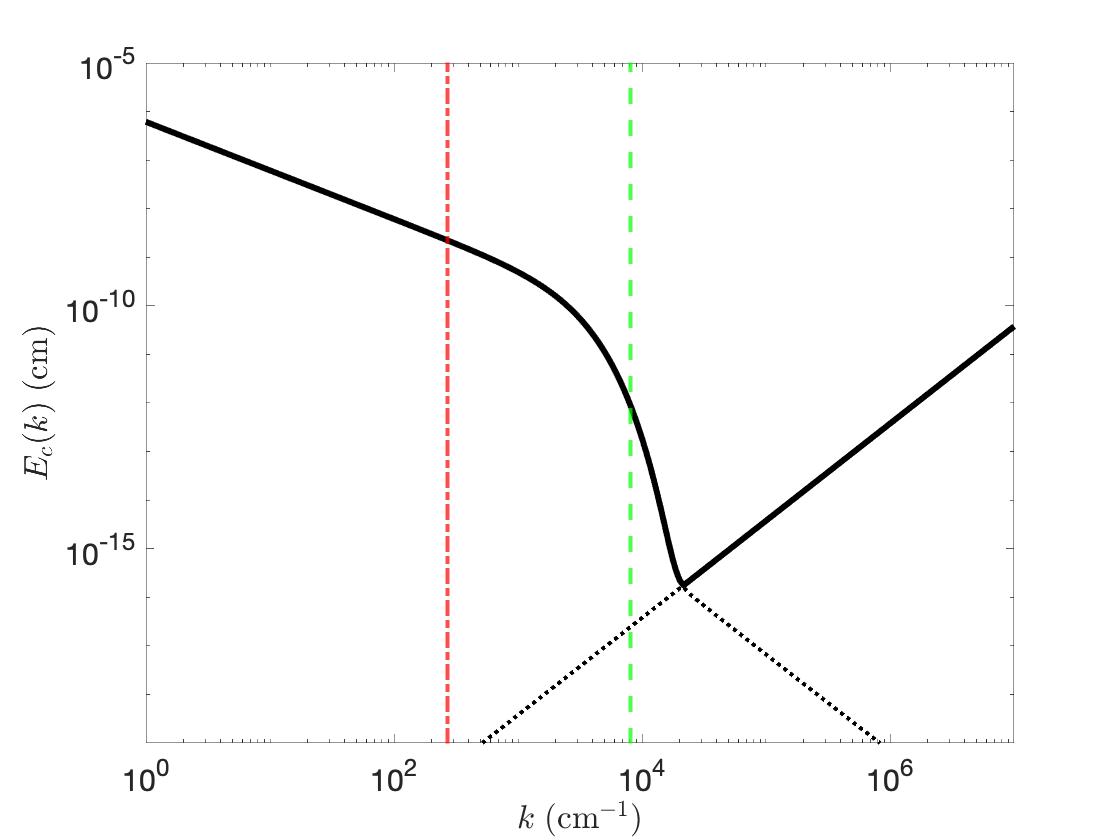}
               \caption{}
     \end{subfigure}
    
     \begin{subfigure}[b]{0.45\textwidth}
         \includegraphics[width=8cm]{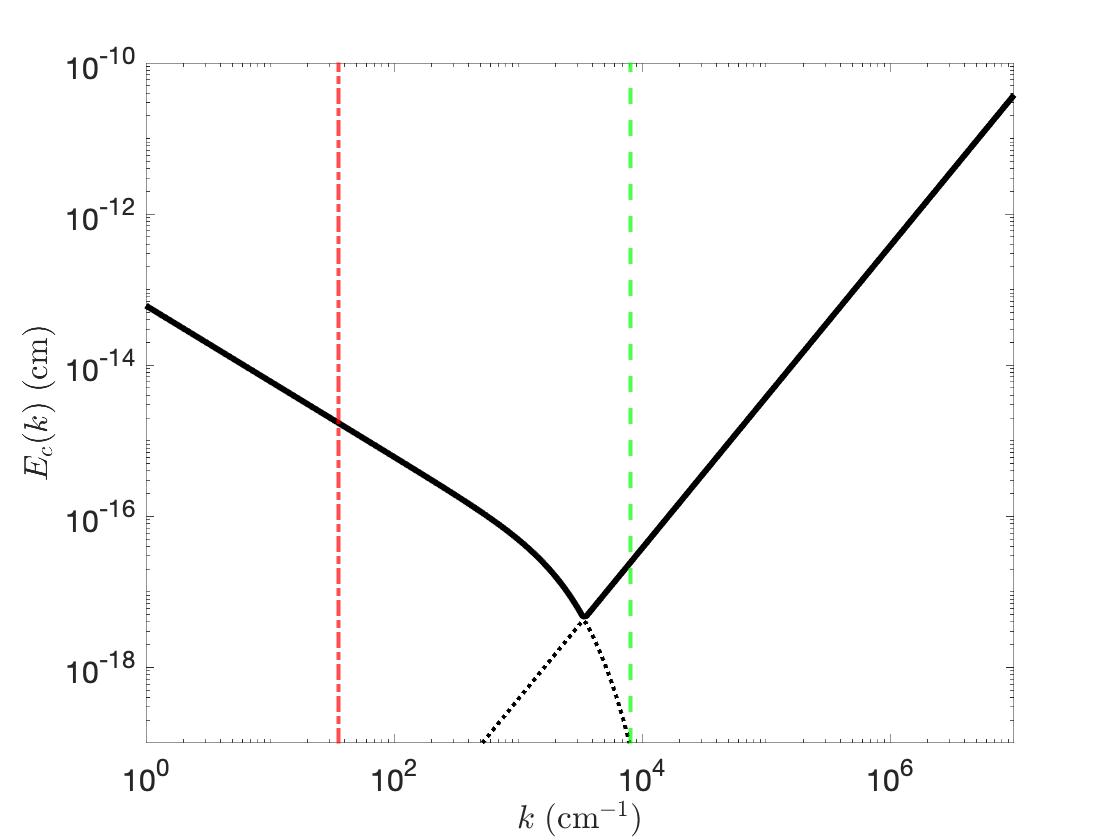}
             \caption{}
     \end{subfigure}
                 \caption{Similar to Fig.(\ref{waterglycerol10}) but with $\gamma=100\;s^{-1}$ and (a) $\chi=10^2\;s^{-1}$, (b) $\chi=10^{-5}\;s^{-1}$, (c) $\chi=10^{-12}\;s^{-1}$.}\label{waterglycerol100}
\end{centering}
\end{figure}

We first present our predictions for possible future experiments 
on turbulent mixing with water-glycerol. In addition to the 
thermodynamic parameters of the mixture, discussed above, the two 
important parameters of the turbulent flow which must be specified 
are the Kolmogorov turnover rate $\gamma$ and the rate of injection 
$\chi$ of concentration fluctuations. 
To identify reasonable ranges for these parameters, 
we reviewed a set of experimental studies of high-Schmidt turbulent mixing
\cite{gibson1963universal, nye1967scalar, grant1968spectrum, miller1996measurements, williams1997mixing, jullien2000experimental}. Extracting data from these references, 
we found a range of values within the intervals $0.1\, {\rm s}^{-1}\leq \gamma 
\leq 10^2\,{\rm s}^{-1}$ and $10^{-12}\, {\rm s}^{-1}\leq \chi\leq 10^2 \, {\rm s}^{-1},$ 
on order of magnitude. \red{See Appendix \ref{survey}}. None of these 
experiments studied water-glycerol mixtures, but recent laboratory experiments on 
fluid turbulence in water-glycerol 
\cite{debue2018experimental} had $\gamma\doteq 129 \, {\rm s}^{-1},$ near the upper
range from the experiments on turbulent mixing. We therefore plot our predictions 
in Fig.\ref{waterglycerol10} for $\gamma=10\, {\rm s}^{-1}$ and 
in Fig.\ref{waterglycerol100} for $\gamma=100\, {\rm s}^{-1},$ which 
are values typical of most of the cited experiments. We also show for both of these 
choices of $\gamma,$ three values of $\chi,$ the smallest value from 
the cited experiments $\chi=10^{-12}\, {\rm s}^{-1}$, the largest value 
$\chi=10^2\,{\rm s}^{-1},$ and one intermediate value. We note that in all 
of our plots the highest wavenumber considered is well below the 
value $k_\sigma=2\pi/\sigma,$ which is $\doteq 1.8\times 10^8\;{\rm cm}^{-1}$
for water-glycerol, and thus within the regime of validity of our theory. 

The solid curves plotted in Figs.\ref{waterglycerol10} \& \ref{waterglycerol100}
are our predicted concentration spectra $E_c(k)$, obtained simply as the maximum of the exact solution from (\ref{exactE})-\eqref{FFunction104} and of the equipartition $k^2$ spectrum (\ref{Ec-eq}). We plot 
the two curves also individually, the exact solution of our model and the equilibrium
spectrum, as black dotted lines, and we indicate the Batchelor wavenumber $k_B=2\pi/\ell_B$
by a vertical green dashed line. The most important conclusion from these plots is that, 
except for very small values of injection rate $\chi$, a few decades of GCF's with spectrum 
$E_c(k)\sim k^{-2}$ should occur. With decreasing $\chi$ the statistics of the concentration 
field become closer to equilibrium and the equipartition $k^2$ spectrum dominates at 
increasingly smaller wavenumbers. In our plots the intermediate value of $\chi$ is chosen 
just small enough so that the range of GCF's entirely disappears, which is $\chi=10^{-7}$ 
for $\gamma=10$ in Fig.\ref{waterglycerol10} and $\chi=10^{-5}$ for $\gamma=100$ in Fig.\ref{waterglycerol100}. Although these values lie within the range of those 
sampled in prior studies, most experiments have larger $\chi$ and thus correspond 
to the upper panels (a) in Figs.\ref{waterglycerol10} \& \ref{waterglycerol100}. 
We can thus expect that future laboratory experiments with choices of parameters $\chi$ 
and $\gamma$ similar to those accessed in prior experiments will exhibit a sizable range 
of power-law spectra $k^{-2}$ associated to GCF's, appearing just above the 
Batchelor wavenumber. With smaller $\chi$ the range of GCF's may be 
very short or disappear entirely, replaced by the equipartition $k^2$ spectrum, 
but, in either case, Kraichnan's exponential decay spectrum \eqref{Kr68} will exist 
over only a very narrow range.

In the typical case where a range of CGF's exists below the Batchelor length, 
these long-range non-equilibrium correlations are quenched in our theory 
by turbulent shear. This shear-quenching is similar to that predicted
for weak perturbations of global thermodynamic equilibrium by 
Wada \cite{wada2004shear}, but, of course, much stronger. Whereas the analysis in \cite{wada2004shear} could rely on linearized fluctuating hydrodynamics, 
our solution for turbulent shear requires the asymptotic method 
of DFV, which treats exactly the nonlinear advection of the concentration field 
by both turbulent and thermal velocity fluctuations. We note that the 
effect of increasing $\gamma,$ with other parameters fixed, is to 
decrease the Batchelor length-scale and to reduce the range of wavenumbers 
where GCF's appear. This effect is seen clearly by comparing the top panels (a)
of Fig.\ref{waterglycerol10} for $\gamma=10$ and of Fig.\ref{waterglycerol100}
for $\gamma=100,$ where the sole effect is to increase $k_B$ and push the 
GCF's to a higher, narrower range of wavenumbers. 

Since we could not include buoyancy in our exact mathematical analysis, there 
remains the possibility that gravity (or finite system-size) could quench 
the GCF's rather than shear, as is typical for laminar experiments 
\cite{zarate2006comment}. Because of the much stronger shear in fluid 
turbulence compared with laminar flow, one may expect that shear  
will in fact be dominant. However, it is useful to make some test of this 
reasonable conjecture, by considering the gravitational wavenumber cutoff 
\be k_g=(\beta g\nabla c/\nu D)^{1/4}, \lb{kgdef} \ee
below which $S_{cc}(k)\sim k^0$ and $E_c(k)\sim k^2$ according to linearized theory. 
Here $\beta=\frac{1}{\rho}\frac{\partial\rho}{\partial c}$ is the solutal 
expansion coefficient and $g$ is the acceleration due to gravity. 
See \cite{segre1993fluctuations,segre1993nonequilibrium,vailati1998nonequilibrium} 
for detailed discussion, but note that the result \eqref{kgdef} follows intuitively 
by equating the damping rates from diffusion and buoyancy as 
$ \gamma_{{\rm diff}}=Dk^2 \sim \gamma_{{\rm grav}}=\beta g\nabla c/\nu k^2.$  
The theory of DFV makes clear how this estimate for $\gamma_{{\rm grav}}$ arises,  
because Eq.(A.27) in \cite{donev2014reversible} shows that buoyancy adds the extra term 
\be (\beta/\nu)(\bG_\sigma\star c){\bf g}\bdot \grad c  \lb{bouyc} \ee
to the righthand side of the asymptotic high-$Sc$ equation \eqref{LangevinPassive1}
for the concentration field. Here ${\bf G}_\sigma={\boldsymbol\sigma}\star {\bf G}$ 
is convolution of smoothing kernel ${\boldsymbol\sigma}$
with the Oseen tensor $\bf G$  (see \eqref{oseen} and Appendix \ref{S2c}). 
To make use of expression \eqref{kgdef} in a forced steady-state with 
continuous injection of concentration fluctuations, we employ the r.m.s. 
gradient from the balance $\chi=D\langle|\grad c|^2\rangle,$ which yields 
\be k_g=(\beta^2g^2\chi/\nu^2D^3)^{1/8}. \lb{kg2} \ee 
We shall adopt this estimate below, but note that balancing the buoyancy term 
\eqref{bouyc} against the diffusive term $D_{e\! f\! f}\triangle c$ in 
\eqref{LangevinPassive1} leads to the much smaller value 
$k_g'=(\beta^2g^2\chi/\nu^2D^3_{e\! f\! f})^{1/8}\ll k_g$ because the effect of 
turbulent diffusivity implies $D_{e\! f\! f}\simeq D_T\gg D.$ Thus, using 
\eqref{kg2} probably greatly overestimates the effect of gravity.

\begin{figure}
\includegraphics[scale=.22]{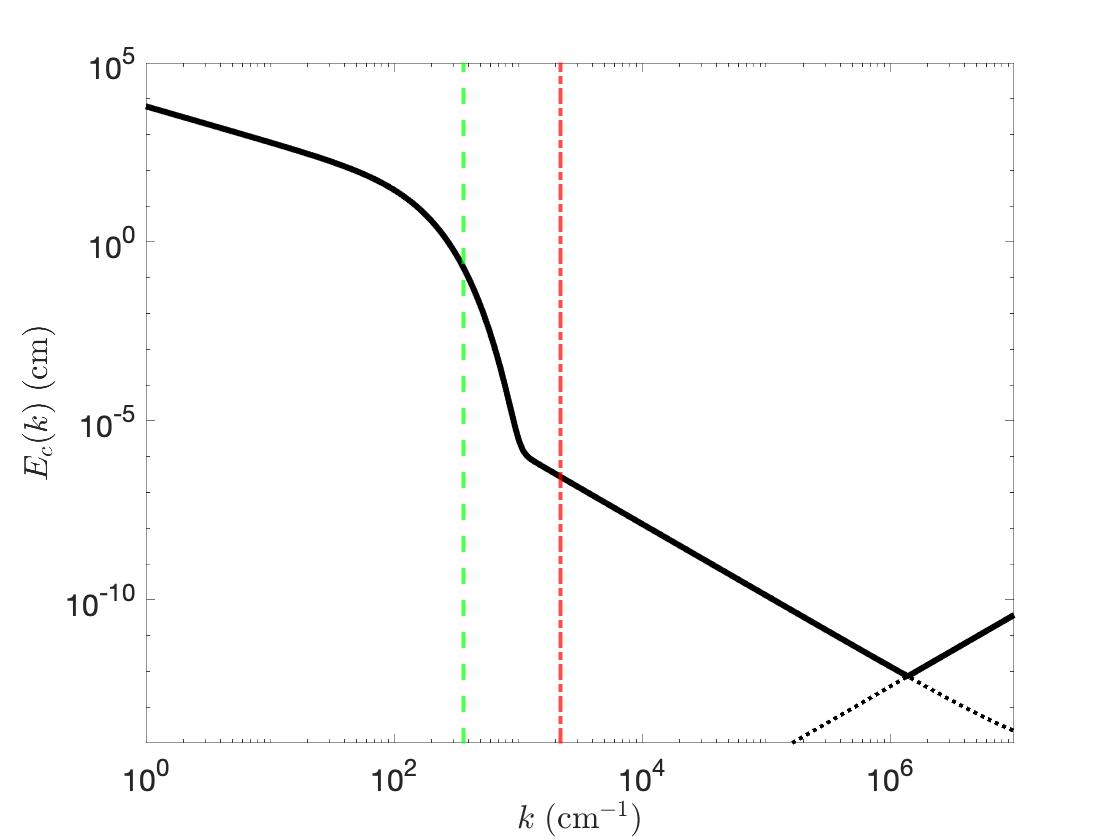} 
\caption {\footnotesize {Predicted concentration spectrum for water-glycerol ($T=25^\circ\,C,$ $p=1$ bar, $\bar{c}=0.5$) for the largest scalar injection rate and smallest strain rate observed in the experiments available to us; $\chi\simeq 2\times 10^2\,s^{-1}$ \cite{jullien2000experimental}, $\gamma \simeq 0.2\,s^{-1}$ \cite{grant1968spectrum}. With these extreme values, buoyancy (red dot-dashed line) might conceivably cut off the $k^{-2}$ power-law spectrum associated to GCF's. Green dashed line marks the Batchelor wavenumber $k_B$.}}
\label{buoyancy}
\end{figure}

The wavenumber $k_g$ is marked in Figs. \ref{waterglycerol10} \& \ref{waterglycerol100}
by the vertical red, dot-dashed line. To calculate \eqref{kg2} we used the 
following convenient parameterizations of the solutal expansion coefficient and 
kinematic viscosity of water-glycerol solutions as functions of concentration:  
\be \beta=0.2246+0.1c-0.125c^2,\quad \nu=0.01\exp(2.06c+2.32c^2), \ee 
see \cite{wolf1988handbook}. It can be seen immediately that $k_g<k_B$ 
in all cases shown, which implies that the GCF's for typical values of 
$\gamma$ and $\chi$ are cut off by turbulent shear rather than by buoyancy. 
To investigate a possible role for gravity one must consider $\gamma$ 
as small as realistic, since $k_B\propto \gamma^{1/2}$. Furthermore, 
one should consider large $\chi,$ because $k_g\propto \chi^{1/8}$ according 
to \eqref{kg2}. Note that also $k_{tr}\propto \chi^{1/4}$ according to 
\eqref{ktrp}, so that increasing $\chi$ in addition increases the wavenumber 
range of the GCF's. In Fig.\ref{buoyancy} we plot our predicted concentration spectrum 
for the smallest value $\gamma\simeq 0.2\,s^{-1}$ \cite{grant1968spectrum} 
and the largest value $\chi\simeq 2\times 10^2\,s^{-1}$ \cite{jullien2000experimental}
that we found in reported experiments on the Batchelor range. With these 
extreme choices we see that $k_g>k_B,$ so that gravitational effects may 
possibly in this case quench the GCF's rather than turbulent shear. Of course,
considering the effects of turbulent diffusivity $D_T$ gives $k_g'\ll k_g$
and thus buoyancy effects even in this extreme parameter range are in fact 
probably small compared with turbulent shear effects.

\begin{figure}
\includegraphics[scale=.22]{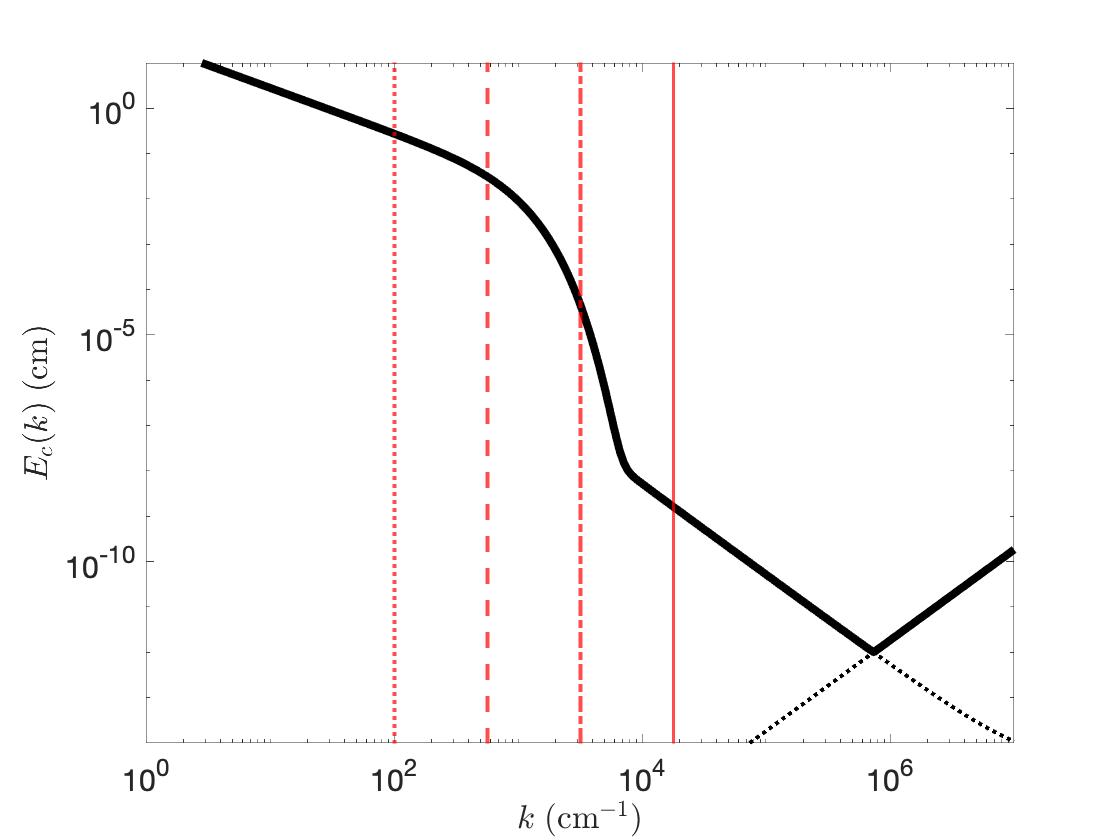} 
\caption {\footnotesize {Predicted concentration spectrum for water-fluorescein 
($T=30^\circ$C, $p=1$ bar). The buoyancy cut-off wavenumber $k_g$ is marked with vertical lines for $\beta g=10^3$G  (\textcolor{red}{{\bf ---}}), $1$G (\textcolor{red}{{\bf -$\,\cdot\,$-}}); $10^{-3}$G (\textcolor{red}{{\bf -$\,$-$\,$-}}) and $10^{-6}$G (\textcolor{red}{{\bf .$\,$.$\,$.}}) in units of Earth gravitational acceleration $1\,G=980\ {\rm cm}/{\rm s}^2$. 
Here, $\gamma\simeq 7.8\,s^{-1}$ and $\chi\simeq 2\times 10^2\, s^{-1}$ corresponding to the experimental parameters in \cite{jullien2000experimental}.}}
\label{Gunits}
\end{figure}

Finally, we shall consider the predictions of our theory for water-fluorescein 
solutions, since this fluid mixture has been the choice of most recent experimental studies 
on high-Schmidt turbulent mixing \cite{sreenivasan1989new,williams1997mixing,
miller1996measurements,jullien2000experimental}. For specificity, we take  
$\gamma\simeq 7.8\,s^{-1}$ and $\chi\simeq 2\times 10^2\, s^{-1}$ from the 
experiment of Jullien et al. \cite{jullien2000experimental}. One 
difficulty in making concrete predictions arises from the poor state of knowledge 
about the thermodynamic properties of water-fluorescein mixtures; in particular,  
the solutal expansion coefficient $\beta$ seems not to be available in the literature. Fortunately, the real parameter of interest is the combination $\beta g,$ in which 
$g$ may be lowered by performing low-gravity space experiments or increased by 
centrifugal effect in a rapidly rotating apparatus. Thus, we present results in 
Fig.\ref{Gunits} on our predicted spectrum for water-fluorescein mixture at 
$T=30^\circ$C, $p=1$ bar, for different values of the quantity $\beta g$ in units of the acceleration due to Earth gravity ($1\,G\simeq 980\, {\rm cm}/{\rm s}^2$). We expect $\beta$ to be of order unity 
(e.g. for water-glycerol $\beta\simeq 0.20-0.25$), so that the values 
$\beta g=10^{-6},$ $10^{-3},$ $1,$ $10^3$G considered in Fig.\ref{Gunits} 
are probably close to the corresponding values of $g.$ Most importantly, 
we see that there are at least two decades of $k^{-2}$ power-law arising 
from GCF's appearing just above the Batchelor wavenumber $k_B.$ Furthermore,
turbulent shear effects cut off the GCF's for wavenumbers lower than $k_B,$
except possibly for $\beta g\geq 40$G, where buoyancy could provide the 
cut-off, but such large $\beta g$ would  be difficult to achieve experimentally. We want 
to emphasize that in the actual experiment reported in \cite{jullien2000experimental},
the concentration fluctuations were strongly damped at wavenumbers $k>7\,{\rm cm}^{-1}$
because of additional shear-enhanced diffusion, and the range above that wavenumber 
in their setup would not be described by our theory. However, in an ideal experiment 
with the same values of $\gamma$ and $\chi$ as  \cite{jullien2000experimental}
but avoiding such enhanced diffusion, a $k^{-2}$ spectrum due to GCF's would 
appear at lengths just below $\ell_B\simeq 22.4\,\mu$m.


\section{Discussion and Conclusions}\lb{final} 

Our theoretical predictions clearly demand empirical verification. As already discussed,
the prospects for direct laboratory experiments appear remote 
because our novel predictions all involve scales below the Batchelor length. 
Since $\ell_B\ll \ell_K$ and it is already difficult to get reliable experimentable measurements 
in the turbulent dissipation range, the difficulties are much more severe for the viscous-diffusive 
range. Even in the viscous-convective range the Batchelor spectrum $\propto k^{-1}$
has not been observed in some experiments \cite{miller1996measurements,williams1997mixing}.
Among experiments which report a Batchelor spectrum \cite{gibson1963universal,nye1967scalar,
grant1968spectrum,oakey1982determination,jullien2000experimental}, only the first 
\cite{gibson1963universal} reported any measurements in the viscous-diffusive range.
That experiment measured fluctuations of both temperature and concentration in 
salt-water and had Batchelor lengths from $14.8-82.5\, \mu$m. 
Nevertheless, the data in \cite{gibson1963universal}, 
Figure 9, for the viscous-diffusive range had large scatter and the authors cautioned 
that ``the high wave-number data may have been affected by noise and/or spatial resolution.''  
It is precisely because of these grave experimental difficulties that most modern 
studies have turned to numerical simulations of deterministic Navier-Stokes equations \cite{yeung2004simulations,donzis2010batchelor,gotoh2015spectrum,clay2017strained}, 
which have verified the Kraichnan-Batchelor predictions in that setting. Likewise, 
the prospects for verification of our novel predictions by numerical simulations 
of Landau-Lifschitz fluctuating hydrodynamics appear excellent, since codes have 
been developed to simulate binary and multi-component mixtures at low Mach numbers \cite{donev2014low,nonaka2015low}, 
especially an overdamped scheme appropriate to high Schmidt numbers \cite{nonaka2015low}. 
Ultimately, of course, laboratory experiments will be absolutely essential to determine 
which of the various theoretical predictions are correct in Nature.  

In addition to empirical studies, our work suggests many interesting further 
theoretical investigations within the DFV approach.  Here we have studied 
only the second-order correlation function and the Fourier spectrum of the
concentration field in a forced steady-state with injection of scalar fluctuations,
but many further generalizations are possible. Techniques exist in the Kraichnan
model to study higher-order correlations \cite{falkovich2001particles,Gawedzki1997,
gawedzki2002easy,gawedzki2002soluble} and even individual realizations of the 
concentration field \cite{gawedzki2002soluble,lototskii2004passive}. Multi-time
correlations such as $C(\bx,t;\bx',t'):=\langle c(\bx, t)c(\bx',t')\rangle$
satisfy also closed equations in the Kraichnan model
\bea\label{Cmultitime} 
&& \partial_t C(\bx,t;\bx',t')\cr 
&& \hspace{10pt} 
= \grad_\bx\bdot\left(\left(D_0+\frac{1}{2}\bcV(\bx,\bx)\right)\bdot\grad_\bx C(\bx,t;\bx',t')\right) 
\hspace{10pt} 
\eea
for $t>t',$ as a direct consequence of \eqref{ItoScalarDiff}. Note that this equation 
expresses the temporal relaxation of fluctuations by the renormalized diffusivity. 
In addition to the statistical steady state, free decay can be studied also 
in the Kraichnan model \cite{EyinkXin2000,chaves2001universal}. This is an important 
problem for further theoretical study because striking experimental observations 
of giant concentration fluctuations have been made in transient decay \cite{vailati1997giant,croccolo2007nondiffusive}. The current analytical theory 
of this problem is based on linearized fluctuating hydrodynamics \cite{vailati1998nonequilibrium}, 
but systematic deviations are observed between linearized theory and experiment 
at early times when concentration gradients are very large: see \cite{croccolo2007nondiffusive},
Figure 8. The DFV approach is not based on linearization and treats nonlinear 
advection of concentration exactly, even if gradients are large. 

Besides analytical theory, the DFV approach yields also an efficient 
numerical scheme to solve fluctuating hydrodynamics of binary mixtures in 
the asymptotic limit of high Schmidt numbers. As emphasized in their original 
work \cite{donev2014reversible}, numerically solving the high-Schmidt 
limit equations \eqref{PassiveStratonovich100},\eqref{wStokes} is more 
efficient by a factor of $Sc$ than solving the standard equations of 
fluctuating hydrodynamics \eqref{Momentum100},\eqref{Passive110}. Unlike 
our analytical approach, the numerical implementation of the DFV limit 
equations has no difficulty incorporating buoyancy effects of gravity 
(see \cite{donev2014reversible}, Appendix A). A certain puzzle does exist 
why DFV failed to observe $S_{cc}(k,t)\sim k^{-4}$ in numerical 
simulation of free diffusive mixing with their high-$Sc$ limit equations 
\eqref{PassiveStratonovich100},\eqref{wStokes}, but instead reported a scaling 
closer to $S_{cc}(k,t)\sim k^{-3}.$ This is curious because experiments 
\cite{vailati1997giant,croccolo2007nondiffusive} and numerical simulations 
with the full fluctuating hydrodynamics equations \cite{donev2014low,gorbunova2020analysis}
both yield $S_{cc}(k,t)\sim k^{-4}$ for free decay, as does our exact solution 
of the DFV correlation equations for the forced steady-state. These 
various results suggest that the DFV theory should yield also such a $k^{-4}$ 
scaling for free decay. \red{It has been suggested to us by A. Donev (private communication) 
that the $k^{-3}$ scaling reported in \cite{donev2014reversible} might be due to the  
fact that those numerical simulations were performed for a 2D fluid. However, we have now solved 
the closed equations of the DFV theory in 2D, for steady-state concentration correlations 
with random injection of fluctuations and with velocity in thermal equilibrium. Because 
the calculation in 2D is more difficult than in 3D, we give details in Appendix \ref{2DGCF}. 
However, we verify the scaling $S_{cc}(k,t)\sim k^{-4}$ also for 2D, in agreement with 
the prediction of linearized fluctuating hydrodynamics.} We cannot advance a definitive 
explanation why the numerical implementation in \cite{donev2014reversible} failed 
to observe this power law, but perhaps the computation ran insufficient time 
or had an insufficient span of wavenumber. In fact, if one fits 
a power-law to the lower range of wavenumbers in \cite{donev2014reversible},
inset of Figure 3, then the result is closer to $k^{-4}.$  


The results that we obtain in this work suggest that, very generally, the effects 
of thermal noise at scales below the Kolmogorov length in turbulent flows will 
be quite similar to those that occur in laminar flows. Although the shear-quenching 
of GCF's in the inertial-convective range is stronger than that found by 
linearized fluctuating hydrodynamics for weak shear \cite{wada2004shear}, they 
are qualitatively similar. Furthermore, at scales below the diffusive length 
analogous to $\ell_B$ (which is the length-scale $\xi_c$ defined in \cite{wada2004shear}, Eq.(45)) 
the predictions for the GCF's in the weakly sheared flow differ from ours only 
by the constant prefactor in front of the power-law. Thus, we imagine that 
effects of thermal noise existing in laminar flows will generally persist, 
in perhaps some modified form, in the sub-Kolmogorov scales of turbulent flows. 
For example, it is known that thermal noise can reduce the efficiency of
combustion in laminar flows, via a noise-induced bifurcation which 
changes the domains of mono- and bi-stability of the 
chemically reacting system \cite{lemarchand2004fluctuation}. Further, 
thermal noise can accelerate the formation and growth of droplets and bubbles
in fluids rapidly cooled or heated in the multiphase regime 
\cite{chaudhri2014modeling,gallo2020nucleation}. Finally, it is 
known that thermal noise is important during collisions of 
self-propelled microorganisms in laminar flows and strongly affects 
the postcollision velocity directions of both swimmers \cite{gotze2010mesoscale}.  
In the sub-Komologorov range, one can expect for all such micro-scale 
physical processes some very interesting interplay between effects 
of turbulence and of thermal noise.  

Here we have considered only non-magnetized molecular fluids, but 
more generally thermal noise could play an important role in the 
turbulence of magnetized plasmas at resistive scales, as already 
suggested in 1961 by Betchov \cite{betchov1961thermal}. In particular,
the kinematic magnetic dynamo regime in a turbulent plasma at high magnetic Prandtl numbers 
is a close analogue of the high-Schmidt turbulent mixing which we have studied in this work. 
Much past theoretical work on the high magnetic Prandtl-number dynamo 
\cite{kulsrud1992spectrum,schekochihin2001structure,schekochihin2002spectra,schekochihin2002small}
is based on the soluble Kazantsev model \cite{kazantsev1968enhancement}, 
which is the exact analogue for a passively advected magnetic field of 
the Kraichnan model for a passively advected scalar \cite{kraichnan1968small,
kraichnan1974convection}. The kinematic dynamo eigenfunction was found 
in these studies to be peaked at the resistive scale, where thermal 
electric-field noise must appear acting on the magnetic field 
according to the general fluctuation-dissipation relation. Furthermore, 
thermal random stresses must act on the advecting velocity at the even larger 
viscous scale. Recently, the theory of thermal fluctuations in a plasma 
has been developed both for linearized dynamics \cite{krommes2018projection} 
and as well for the full nonlinear dynamics \cite{feliachi2021dynamicalA,feliachi2021dynamicalB}.
One can anticipate that there will be significant modifications of the 
predictions of dynamo theories that neglect such noise. It would be 
interesting to investigate the effect of thermal noise on the origin 
and evolution of primordial magnetic fields in cosmology 
\cite{kulsrud2008origin,subramanian2016origin}. 

\acknowledgements 
We thank D. Bandak, J.~B. Bell, F. Bouchet, \red{A. Donev}, A. L. Garcia, N. Goldenfeld, 
\red{T. Gotoh}, A.~A. Mailybaev, and A. Nonaka for useful discussions on the subject of this work. 
We thank also the Simons Foundation for support of this work with Targeted
Grant No. MPS-663054, “Revisiting the Turbulence Problem Using Statistical Mechanics.”

\appendix

\section{Fluctuation-Dissipation Theorem For the Concentration Field}\label{FDT}

We show here as an application of the phenomenological fluctuation-dissipation 
theorem \cite{dezarate2006hydrodynamic,zubarev1983statistical,morozov1984langevin,
espanol2009microscopic}  that the multiplicative noise term in the stochastic equation 
\be\label{Passive111a}
\partial_t c+{\bf u\bdot\grad}c =\grad\bdot  \left(D_0 \grad  c+\sqrt{2mD_0 \rho^{-1} c(1-c)} \; {\boldsymbol\eta}_c({\bf x}, t) \right),
\ee
is the unique expression which is local in $\bx$ so that the equilibrium statistics $P_{eq}[c]$ of the 
equation are given by the Boltzmann-Einstein formula
\begin{eqnarray}\lb{EBdist} 
 P_{eq}[c] &\propto& e^{S/k_B},
 \end{eqnarray}
where $S$ is the thermodynamic entropy. Here the appropriate entropy $S$ is the ideal entropy of mixing 
(\cite{callen1960thermodynamics}, Appendix D.6) 
\be\label{mixingS}
S[c]=-{\rho k_B\over m}  \int d^3x\ \Big(c({\bf x}) \ln c({\bf x})+ (1-c({\bf x}))\ln(1-c({\bf x})\Big) 
\ee
 with particle mass $m$ and fluid density $\rho.$ Note here that 
we have assumed equal masses for the two species of particles, so that 
mass concentration $c$ and molar concentration $n$ in this case coincide. 
 A local equilibrium distribution has been assumed in which
the statistics in each subvolume is determined by the specific entropy $s(c(\bx))$ of the local concentration field 
$c(\bx)$, then integrated against $dM=\rho \, d^3x$ to give the total entropy. 
 
 For the purpose of formal calculations, it is useful to rewrite the stochastic equation by inserting a delta 
 function $\delta^3(\bx-\by)$ and integrating over $\by,$ as: 
\be\label{Passive111}
\partial_t c+{\bf u\bdot\grad}c =D_0 \triangle c+
\int d^3y\  g^a[{\bf x, y}; c]\, \eta_{c,a}({\bf y}, t), 
\ee
with the definition 
\bea 
\hspace{-20pt} 
&& g^a[{\bf x, y}; c]:= \partial_{x_a}\Big[ \sqrt{{2m D_0\over \rho} c({\bf x})(1-c({\bf x}))  } \delta({\bf x-y}) \Big]. 
\lb{ga-def}
\eea
It is important to stress that all ``delta functions''  in this expression and also in the covariance \eqref{etac} of the 
white-noise $\boeta_c$ should be interpreted as {\it cutoff delta-functions}
\be \delta^3_\Lambda(\bx) = \frac{1}{V}\sum_{|\bk|<\Lambda} e^{i\bk\bdot\bx} \lb{cutoff-del} \ee
where $V$ is the domain volume and $\Lambda$ is some high-wavenumber cut-off. 
See \cite{zubarev1983statistical}. Here the cutoff $\Lambda$ should be taken $\lesssim 1/\lambda_{mfp},$ 
the inverse of the mean-free path length. Physically, fluctuating hydrodynamic equations such as 
\eqref{Passive111} should not be interpreted as continuum stochastic partial differential equations 
but instead as low-wavenumber effective field theories. 
 
To obtain the Fokker-Planck equation for the probability distribution $P[c]$ corresponding to the 
Langevin equation eq.(\ref{Passive111}), we convert from Stratonovich to It$\bar{{\rm o}}$
calculus. The noise-induced drift term is
\begin{eqnarray}\nonumber
&&{1\over 2}\iint d^3y \, d^3z  \; g^a [{\bf z}, {\bf y}; c]{\delta g^a[ {\bf x}, {\bf y}; c]\over \delta c({\bf z})},\end{eqnarray}
Because of locality in $\bx$ it is easy to check that 
$$ {\delta g^a[ {\bf x}, {\bf y}; c]\over \delta c({\bf z})}=\delta^3(\bz-\by) G^a[\bx,\by;c] $$
with $G^a$ independent of $\bz.$ But in that case 
\bea  && \int d^3z  \; g^a [{\bf z}, {\bf y}; c]\delta^3(\bz-\by) \cr
 && = \sqrt{{2m D_0\over \rho} c({\bf y})(1-c({\bf y}))}  \, (\partial_{y_a} \delta^3)(\bzed) \ = \ 0. 
\eea 
Therefore, in this particular problem, the noise-induced drift vanishes and It$\bar{{\rm o}}$ and Stratonovich forms 
of the equation are identical. We used above the crucial fact that 
$$ \grad\delta^3_\Lambda(\bzed) = \frac{1}{V}\sum_{|\bk|<\Lambda} i \bk\  =\bzed, $$
which will be exploited also in the following calculations. 

Because of the identity of It$\bar{{\rm o}}$ and Stratonovich here, we obtain easily the Fokker-Planck equation 
\begin{eqnarray}\nonumber
\partial_t P[c]&=&-\int d^3x \ {\delta \over\delta c({\bf x})}\Big[ \Big( -{\bf u}\bdot\grad c +D_0 \triangle c \Big) P[c] \Big]\\\nonumber
&&+ \frac{1}{2} \iint d^3x  \, d^3y \ {\delta^2 \over \delta c({\bf x}) \delta c({\bf y}) } \Big( {\cal D}[{\bf x, y}; c] P[c] \Big),\\\label{FP100}
\end{eqnarray}
where 
\begin{eqnarray}
{\cal D}[{\bf x, y}, c]= \int d^3z  \ g^a[{\bf x, z}; c]g^a[{\bf y, z}; c]\label{diff100}
\end{eqnarray}
 is the probability diffusion coefficient. We must now show that the Einstein-Boltzmann distribution $P_{eq}[c]$
 is the stationary distribution of the Fokker-Planck equation when the noise is chosen as in \eqref{Passive111},
 \eqref{ga-def}.  
  
Note that the  contribution from the first probability drift term vanishes because 
\begin{eqnarray}\nonumber
&&\int d^3x  {\delta\over \delta c({\bf x}) }\Big({\bf u}\bdot\grad_x c({\bf x}) P_{eq}[c]  \Big)\\\nonumber
&=& \int d^3x  \Big({\bf u}\bdot\grad_x \delta^3_\Lambda({\bf 0})
+ \frac{\rho}{k_B}({\bf u}\bdot\grad_x c({\bf x})) s'(c({\bf x})) \Big)P_{eq}[c]\\\nonumber
&=& \frac{\rho}{k_B}\int d^3x  \, {\bf u}\bdot \grad _x s( c({\bf x}))P_{eq}[c] \ = \ 0,\\\nonumber
\end{eqnarray}
where in the second line we used 
\bea  {\delta\over \delta c({\bf x})} P_{eq}[c] &=&  -\frac{\rho}{m} [\ln c(\bx)-\ln(1-c(\bx)]P_{eq}[c] \cr
&=& \frac{\rho}{k_B} s'(c(\bx))P_{eq}[c] \lb{dPeq-dc} \eea
and in the last line incompressibility $\grad\bdot  {\bf u}=0$ was used. 

Next we note using \eqref{ga-def} and \eqref{diff100} that an explicit expression for the 
probability diffusion coefficient follows: 
\begin{eqnarray}\lb{diff200} 
{\cal D}[{\bf x, y}, c]&=& \int d^3z  \ g^a[{\bf x, z}; c] g^a[{\bf y, z}; c]\\\nonumber
&=& {2mD_0\over \rho} \partial_{x_a}\partial_{y_a}\Big[ c({\bf x})(1-c({\bf x}))\delta^3({\bf x-y}) \Big] \\\nonumber 
&=& -{2mD_0\over \rho} \partial_{x_a}\Big[ c({\bf x})(1-c({\bf x}))\partial_{x_a} \delta^3({\bf x-y}) \Big].
\end{eqnarray}
A calculation similar to the proof of It$\bar{{\rm o}}$-Stratonovich identity gives
$$ \int d^3y  {\delta\over \delta c({\bf y})} {\cal D}[{\bf x, y}, c] =0. $$

Finally, using \eqref{dPeq-dc} and \eqref{diff200} 
\begin{eqnarray}\nonumber
&& \frac{1}{2} \int d^3y  {\delta\over \delta c({\bf y})} \Big({\cal D}[{\bf x, y}, c] P_{eq}[c]\Big)  \\\nonumber
&& \hspace{30pt}  = \frac{1}{2} \int d^3y \ {\cal D}[{\bf x, y}, c]  {\delta P_{eq}[c] \over \delta c({\bf y})}  \\\nonumber
&& \hspace{30pt}  = -\frac{\rho}{2m}\int d^3y  \ {\cal D}[\bx-\by,c]  \\\nonumber
&& \hspace{80pt} \times  \big(\ln c(\by)-\ln(1-c(\by)\big)P_{eq}[c]  \\\nonumber
&& \hspace{30pt} =D_0  \partial_{x_a} \int d^3y \  c({\bf x}) (1-c({\bf x})) \delta^3({\bf x-y}) \\\nonumber
&& \hspace{80pt} \times {1\over c(\by)(1-c({\bf y}))}  \partial_{y_a} c({\bf y}) P_{eq}[c]\\\nonumber
&& \hspace{30pt} = D_0 \Delta c \; P_{eq}[c].
\end{eqnarray}
The second drift contribution in the first line of eq.(\ref{FP100}) is thus exactly cancelled by the  
diffusion contribution in the second line when $P=P_{eq}.$ It is clear from this calculation that 
the multiplicative factor $\sqrt{2mD_0 \rho^{-1} c(\bx)(1-c(\bx))}$ is the unique local function 
of the concentration field which can be chosen to multiply the noise term in \eqref{Passive111a} so that 
exact cancellation between drift and diffusion terms is obtained, guaranteeing that $P_{eq}$ is stationary. 

Notice, however, that $P_{eq}$ given by \eqref{EBdist} is not the only stationary 
distribution for the stochastic dynamics described by \eqref{Passive111a}, because
that equation conserves the integral
\be M_1 = \rho \int d^3x\, c(\bx,t) \lb{M1} \ee 
which represents the total mass of species 1 of the mixture. Conservation of the 
integral \eqref{M1} will hold for any boundary conditions on scalar flux 
which conserve mass, such as periodic or zero-flux. In that case, there is 
a 1-parameter family of invariant distributions of the form 
\begin{eqnarray}\lb{EBdist2} 
 P_{eq}^{\lambda} [c] &\propto& e^{S/k_B+\lambda M_1}.
\end{eqnarray}
Comparison with standard equilibrium thermodynamic relations (see 
Appendix \ref{AppMix}) reveals that $\lambda=\mu/k_BT,$ where $\mu$ is 
the chemical potential per mass which is thermodynamically conjugate 
to the concentration $c.$ Its value thus determines the mean concentration
through the relation $\mu=-T s'(\bar{c})$ or $\lambda=-s'(\bar{c})/k_B,$ with $\bar{c}=1/2$ for $\lambda=0.$

The small Gaussian fluctuations $c'(\bx)$ around the mean 
value $\bar{c}$ can be obtained from the formula \eqref{EBdist2} for 
the distribution $P_{eq}^{\lambda}[c]$ by substituting $c(\bx)=\bar{c}+c'(\bx)$
and expanding to quadratic order. Using 
$$ {\delta^2 S[c]\over \delta c({\bf x})\delta c({\bf x'})} =-{k_B \rho\over m}
{\delta^3({\bf x-x'} )\over c(\bx)(1-c({\bf x}))}    ,$$
the result is 
$$P_{eq}^{\lambda} [c]\simeq  \exp{\Big(-{1\over 2}{\bar{\rho}\over m}{\int d^3x 
\ (c'({\bf x}))^2 \over  \bar{c}(1-\bar{c})}\Big)}. $$
It follows that the second order correlation is given by 
$$\langle c'({\bf x}) c'({\bf x}')\rangle
= \frac{m}{\bar{\rho}} \bar{c}(1-\bar{c})\delta_\Lambda^3(\bx-\bx'). $$
Fourier transforming and using the definition \eqref{Scc-def} gives the 
equilibrium structure function 
$$  S_{cc}(k)=\frac{m}{\bar{\rho}} \bar{c}(1-\bar{c}) $$
which is independent of wavenumber $k.$  Using the result \eqref{F1} 
for $(\partial \mu/\partial c)_{T,p}$ from Appendix \ref{AppMix}, we 
see that this special case for an ideal mixture of equal mass 
particles agrees with the general result \eqref{Scc-eq}. 


\section{High $Sc$ Asymptotics}\label{S2c}

We give here the detailed derivation of the equation \eqref{LangevinPassive1a} for the reader who 
is interested in the mathematical details. To simplify the notation, 
in this appendix we shall use $\bf v,\bf u,\bw$ instead of ${\bf v}_{\theta},{\bf u}_{\theta},\bw_\theta$, respectively,
and likewise $c'$ and $c$ will be denoted instead as $c$ and $c_\gamma$ (to remind that the latter 
depends linearly on $\bgamma$). Our analysis follows closely that of DFV in 
\cite{donev2014reversible}, Appendix A, and related works 
\cite{khasmiskii1963principle,kurtz1973limit,papanicolaou1976some,gardiner1984adiabatic,pavliotis2008multiscale}, 
so we shall be terse. 

The forward Kolmogorov operator ${\bf L}$ which corresponds to the Langevin equations (\ref{Momentum200}) 
and (\ref{Passive200}) and which evolves an arbitrary functional $F[\bv,c]$ is the sum of three terms that are 
ordered in inverse powers of $\epsilon$ as ${\bf L}={\bf L}_0+ {\bf L}_1 \epsilon^{-1} +{\bf L}_2 \epsilon^{-2}$: 
\begin{eqnarray}\nonumber
    {\bf L}_0 F&=& {1\over 2} {\cal V}_0 \int d^3x \; ({\cal P}_{ij} \triangle v_j) {\delta F\over \delta v_i({\bf x})}\\\nonumber
        &+&{1\over 2}\iint d^3x\,  d^3x' \ {\cal P}_{im}{\cal P}_{jn}'  ({\cal V}_{kl}({\bf x, x'})\, \partial_k v_m({\bf x}) \partial_l' v_n ({\bf x'}) )\\\nonumber
        && \hspace{80pt}  \times   {\delta^2 F\over \delta v_i({\bf x})\delta v_j ({\bf x'})}\\\nonumber
        &+&  {1\over 2} \iint d^3x \, d^3x'\  {\cal P}_{im}{\cal P}_{jn}' \Big( v_k({\bf x}) v_l({\bf x'})
                  \, \partial_k\partial_l' {\cal V}_{mn}({\bf x, x'}) \Big) \\\nonumber
        && \hspace{80pt} \times {\delta^2 F \over  \delta v_i({\bf x})\delta v_j({\bf x'})}  \\\nonumber 
        &+& (D_0+{\cal U}_0 ) \int d^3x\  \triangle c({\bf x}) \cdot {\delta \over \delta c({\bf x})}\\\nonumber 
       &+&{m D_0\over \rho} \int d^3x \  c_\gamma({\bf x}) (1-c_\gamma ({\bf x})) \cdot \triangle  {\delta^2 F\over \delta c({\bf x})^2}\\\nonumber
      &+&{1\over 2}\iint d^3x \,d^3x'\  {\cal U}_{ij}({\bf x, x'}) \\\nonumber 
      && \times ( \partial_i c({\bf x})+\gamma_i)(  \partial'_j c({\bf x'})+\gamma_j){\delta^2 F\over \delta c({\bf x})\delta c({\bf x'})}\\\nonumber
    &+& \int d^3x  \; s({\bf x}, t) {\delta F\over \delta c({\bf x})}\\\label{L0}
    &+&{1\over 2}\iint d^3x \, d^3x'\ S(|{\bf x-x'}|/L){\delta^2 F \over \delta c({\bf x})\delta c({\bf x'})},
\end{eqnarray}

\begin{eqnarray}\label{L_1}
    {\bf L}_1 F&=&-\int d^3x  \; u_i (\partial_i c+\gamma_i) {\delta F\over \delta c({\bf x})},
\end{eqnarray}
and
\begin{eqnarray}\nonumber
    {\bf L}_2&=& \nu \int d^3x \; ({\cal P}_{ij}\triangle v_j) {\delta F\over \delta v_i({\bf x})}  
    \\\label{L_2}
    &+&\   \frac{\nu k_B T}{\rho}  \int d^3x  \; {\cal P}_{ij}\triangle 
    {\delta^2 F\over\delta v_i({\bf x})\delta v_j({\bf x})}.
    \end{eqnarray}
%
%
    
Denote by $(\tilde{\bf v}({\bf x}, t),\tilde c({\bf x}, t) )$ the solution of (\ref{Momentum200}) and (\ref{Passive200}) 
with initial conditions $(\tilde{\bf v}({\bf x}, 0),\tilde c({\bf x}, 0) )=( {\bf v}({\bf x}),c({\bf x}))$ and consider the functional
\be\label{FunctionalF}
  G[\bv,c, t]\equiv \langle F[\tilde\bv(\cdot,t),\tilde c(., t)]\rangle
\ee
 where $\langle . \rangle$ denotes the expectation value over the realization of noise terms $\boldsymbol\eta$ and ${\boldsymbol\eta}_c$. 
The expectation defines a time-dependent functional $G$ of the initial conditions which satisfies the backward Kolmogorov equation:
\be\label{G1}
\partial_t G={\bf L}_0 G+\epsilon^{-1} {\bf L}_1 G+\epsilon^{-2} {\bf L}_2 G, \quad\mathrm{}\quad   G\Big|_{t=0}=F.\ee
One considers this equation in the limit $\epsilon\rightarrow 0$. Expanding the solution $G$ as
$$G=G_0+\epsilon G_1+\epsilon^2 G_2+\dots,$$
and substituting this relation in (\ref{G1}) and collecting terms of increasing power in $\epsilon$, one finally obtains 
\begin{eqnarray}\nonumber
&&{\bf L}_2 G_0=0,\\\nonumber
&&{\bf L}_2 G_1=-{\bf L}_1 G_0,\\\nonumber
&&{\bf L}_2 G_2=\partial_t G_0-{\bf L}_0 G_0-{\bf L}_1 G_1,\\\label{G2}
&&\dots .
\end{eqnarray}

Because ${\bf L}_2$ is the Markov generator of the equilibrium fluctuating hydrodynamics 
equation \eqref{Momentum100} in the text, which defines an ergodic process, the first equation in eq.(\ref{G2}), i.e., ${\bf L}_2 G_0=0$, 
indicates that $G_0$ is a functional of $c({\bf x})$ only and does not depend on both $\bf v(x)$ and $c({\bf x}):$
$$G_0=G_0[c].$$ 

The second equation in (\ref{G2}) requires a solvability condition as its RHS must be in the range of ${\bf L}_2$. 
Because $({\rm Ran}\,{\bf L}_2)^\perp={\rm Ker}\,{\bf L}_2^*,$ this is equivalent to the statement that the expectation of 
${\bf L}_1 G_0$ must vanish when averaged with respect to the invariant Gibbs measure of $\tilde {\bf v}({\bf x}, t)$ evolving 
under \eqref{Momentum100}, i.e.,
\be
P_{eq}(\tilde{\bf v})={1\over Z} \exp{\Big( -{\rho\over  2k_B T}\int d^3x \, \tilde v^2 }   \Big) \delta^3\Big( \int d^3x \, \rho \tilde {\bf v} \Big)\delta(\grad\bdot  \tilde{\bf v}). 
\ee
Denoting the expectation with respect to this measure by $\langle f\rangle_{\bf v}$, the solvability condition becomes
$$0=\langle \hat{\bf L}_1 G_0\rangle_{\bf v}=-\int d^3x \; \langle {\bf u(x)}\rangle_{\bf v}\bdot(\grad c({\bf x})+{\boldsymbol\gamma}) {\delta G_0\over \delta c({\bf x})}, $$
which is satisfied because $\langle{ \bf v(x)}\rangle_{\bf v}=0.$ 
The second equation in (\ref{G2}) can now be solved for $G_1$:
\be\label{G3}
G_1=-{\bf L}_2^{-1} {\bf L}_1 G_0,
\ee
where ${\bf L}_2^{-1}$ is the pseudo-inverse of the operator ${\bf L}_2$. 

The third equation in (\ref{G2}) also requires a solvability condition, which using 
$\langle {\bf L}_0 G_0\rangle_{\bf v}={\bf L}_0 G_0 $ and (\ref{G3}), can be written as
\begin{eqnarray}\nonumber
\partial_t G_0&=&\langle {\bf L}_0 G_0 \rangle_{\bf v}+\langle {\bf L}_1 G_1\rangle_{\bf v}\\\nonumber
&=&{\bf L}_0 G_0-\langle {\bf L}_1 {\bf L}_2^{-1} {\bf L}_1 G_0  \rangle_{\bf v}.
\end{eqnarray}
Because the operator ${\bf L}_1$ defined in \eqref{L_1} is linear in ${\bf u(x)}=\sigma\star{\bf v(x)},$
one can use 
\begin{eqnarray}\nonumber
{\bf L}_2^{-1} {\bf u(x)}&=&{\bf L}_2^{-1} {\boldsymbol\sigma}\star{\bf v(x)}={\boldsymbol\sigma}\star {\bf L}_2^{-1}{\bf v(x)}\\\nonumber
&=&-{\boldsymbol\sigma}\star \int_0^\infty d\tau e^{\tau {\bf L}_2} {\bf v(x)}\\\label{L2u}
&=&-{\boldsymbol\sigma}\star \int_0^\infty d\tau \langle \tilde{\bf v}({\bf x}, \tau) \rangle,
\end{eqnarray}
where $\tilde{\bf v}({\bf x}, \tau)$ denotes the solution of \eqref{Momentum100} 
with initial condition $\tilde{\bv}(\bx, 0)={\bf v(x)}$ and expectation $\langle . \rangle$ 
is the same as in (\ref{FunctionalF}). This solution, using \eqref{Momentum100}, can be written as 
\begin{eqnarray}\nonumber
&& {\tilde{\bf v}}({\bf x}, \tau)=\exp{\Big(-\tau \nu \bcA \Big)}   {\bf v(x)}\\\nonumber
&&+ \int_0^\tau d\tau' \; \exp{\Big[ -(\tau-\tau')\nu \bcA \Big]}\grad\bdot   
\Big( \sqrt{ 2 \nu k_B T \rho^{-1}} {\boldsymbol\eta}(\tau')\Big),\\\label{tilde-v2}
\end{eqnarray}
in terms of the Stokes operator $\bcA=-\bcP\triangle$. The second term has a zero average 
and does not contribute to the expectation in (\ref{L2u}). Combining relations (\ref{L2u}) 
and (\ref{tilde-v2}), we find
\begin{eqnarray}\nonumber
{\bf L}_2^{-1} {\bf u(x)}&=&-\nu^{-1} \bG_\sigma \star {\bf v(x)},
\end{eqnarray}
where ${\bf G}_\sigma={\boldsymbol\sigma}\star {\bf G}$ is the convolution of the smoothing kernel ${\boldsymbol\sigma}$
with the Oseen tensor $\bf G$  (Green's function for the Stokes flow).
It follows that
\begin{eqnarray}\nonumber
 &&-\langle{\bf L}_1 {\bf L}_2^{-1}  {\bf L}_1 G_0 \rangle_{\bf v}
    =\iint d^3x \,d^3x' (\grad c({\bf x})+{\boldsymbol\gamma}) \\\nonumber
    && \bdot{\delta\over \delta c({\bf x})} \left( {1\over 2} {\bf R(x, x')} \bdot \grad' c({\bf x'})+{\boldsymbol\gamma}){\delta G_0\over \delta c({\bf x'})}\right)\\\nonumber
      &=& {1\over 2} \iint d^3x\, d^3x' \ (\grad c({\bf x})+{\boldsymbol\gamma}) \bdot {\bf R(x, x'})\bdot (\grad' c({\bf x'})+{\boldsymbol\gamma})\\\nonumber
    && \hspace{80pt} \times {\delta^2 G_0\over \delta c({\bf x})\delta c({\bf x'})} \\\nonumber
    &+& \int d^3x  \grad\bdot  \left( {1 \over 2 } {\bf R(x, x)}\bdot(\grad c({\bf x})+{\boldsymbol\gamma})  \right) {\delta G_0\over \delta c({\bf x})}:=(\delta {\bf L}_0)G_0.
\end{eqnarray}
The operator $\delta {\bf L}_0$ which emerges from this last calculation is the generator of the Markov random 
process corresponding to the following It$\bar{{\rm o}}$ stochastic differential equation for the concentration field 
\begin{eqnarray}\nonumber
\partial_t c &=& \grad\bdot  \left( {1\over 2}{\bf R(x, x)}\bdot(\grad c +{\boldsymbol\gamma})\right) 
-\bw\bdot (\grad c+\bgamma) 
\end{eqnarray}
where $\bw$ is the Gaussian random velocity, white-noise in time with spatial covariance $\bR(\bx,\bx')$ which is 
given by equation \eqref{SmoothVelocityCovariance101} in the main text. 

The solvability condition thus yields the limiting equation for $G_0[c]$ as $\epsilon\to 0$
$$\partial_t G_0= ({\bf L}_0 + \delta{\bf L}_0)G_0$$ 
which can be recognized immediately as the backward Kolmogorov equation for the It$\bar{{\rm o}}$ equation
\begin{eqnarray}\nonumber
\partial_t c &=&(D_0+{\cal U}_0)\triangle c+\grad\bdot  \left( {1\over 2}{\bf R(x, x)}\bdot(\grad c +{\boldsymbol\gamma})\right) \\\nonumber
    &&-({\bf w}+{\bf u}_{T})\bdot(\grad c+{\boldsymbol\gamma}) +s({\bf x}, t)+ s_0({\bf x}, t)\\\nonumber
   &&+\grad\bdot  \left(\sqrt{2mD_0 \rho^{-1}c_\gamma(1-c_\gamma)} {\boldsymbol \eta}_c({\bf x}, t) \right),
\end{eqnarray}
which is exactly eq.(\ref{LangevinPassive1}).

\section{Equation for the Correlation Function 
in the Isotropic Kraichnan Model}\label{AppKraichnanCorrelation}


The equation \eqref{iso-V} in the text can be written explicitly as 
\be {\cal V}_{ij}(r)= K(r)\delta_{ij}+\partial_{r_i}\partial_{r_j} H(r) \lb{C1} \ee 
where $-\triangle H(r)=K(r).$ Here $K(r)$ is any positive-definite, 
radially-symmetric, smooth function, which means that it can be written as
a Fourier transform $K(r)\equiv \int d^d{\bf k} \ e^{i\bk\bdot\br}\, E(k)$ 
with $E(k)$ a positive, radially-symmetric, rapidly-decaying spectrum. 
For any radially symmetric function $H(r)$ it is easy to check that 
\be  \partial_{r_i}\partial_{r_j} H(r)=J(r)\delta_{ij}+ r J'(r) \hat r_i\hat r_j \lb{C2} \ee 
with $J(r)= H'(r)/r,$ so that taking a trace gives 
\be  -K(r)=\triangle H(r)= d\cdot J(r)+r J'(r) = \frac{1}{r^{d-1}}\frac{d}{dr}\left(r^d J(r)\right) 
\lb{C3} \ee
in $d$ dimensions. Integration over $r$ yields the formula \eqref{H'/r} in the text. 

Using \eqref{C2},\eqref{C3} in \eqref{C1}, we get
\begin{eqnarray}\label{KraichnanCovariance100}
  {\cal V}_{ij}(r) =[K(r)+J(r)]\delta_{ij}- [K(r)+ d\cdot J(r)]\hat r_i\hat r_j.
 \end{eqnarray}
and substituting this expression into the equation \eqref{correlation1} in the text yields 
\begin{eqnarray}\nonumber\partial_t{C}&=&\Big[\triangle K(r)+\triangle J(r)\Big]\triangle {C}
-\Big[\triangle K(r) + d\cdot\triangle J(r) \Big] {\partial^2{C}\over\partial r^2}\\\nonumber
&&   +2 D_0\triangle{C}+S\left( {r\over L}\right),
\end{eqnarray}
where $\triangle J(r)=J(0)-J(r),$ etc. Using the standard formula for the radial Laplacian
\be \triangle C={1\over r^{d-1}}\frac{\partial}{\partial r}\left(r^{d-1} {\partial {C}\over\partial r}\right)
={d-1\over r}\partial_r{C}+\partial_r^2{C}
\lb{C5} \ee 
then gives further 
\begin{eqnarray}\nonumber
\partial_t{C}&=&{d-1\over r} (\triangle K(r)+\triangle J(r) )\partial_r {C} \\\nonumber
&&+(1-d)\triangle J(r) \partial_r^2{C}+2 D_0\triangle C+S\left( {r\over L}\right) \\\nonumber
&=& -{d-1\over r^{d-1}}{\partial\over\partial r}\left(\triangle J(r)r^{d-1} {\partial{C}\over \partial r}\right)
+2 D_0\triangle C+S\left( {r\over L}\right),
\end{eqnarray}
where we employed again \eqref{C3} to get the second equality. Combining with the 
radial Laplacian \eqref{C5} gives the final result  
\begin{eqnarray}\label{2ndCorrKraichnan101}
\partial_t{C}&=&{1\over r^{d-1}}{\partial\over\partial r}\left(\Big[ 2D_0  - (d-1)\triangle J(r)\Big] r^{d-1} {\partial{C}\over \partial r}\right)\cr
&&  +S\left( {r\over L}\right),
\end{eqnarray}
which coincides with \eqref{2CorrelationKraichnan} in the text.

\section{Renormalized Diffusivity from Thermal Velocity Fluctuations}\label{AppJtheta}

We here derive the scale-dependent diffusivities \eqref{Jtheta},\eqref{expJ} arising from advection by 
thermal velocity fluctuations. The covariance $\bR(\bx,\bx')$ defined in \eqref{SmoothVelocityCovariance102}
can be evaluated for $d=3$ homogeneous, isotropic statistics in the form \eqref{iso-V} or \eqref{C1}, with 
\bea 
K_\theta(r) &=& \frac{1}{(2\pi)^3} \int d^3{\bf k}\, e^{i{\bf k.r}}  \,  
|\widehat{\sigma}(k)|^2 \frac{2 k_B T}{\eta k^2} \cr
&=&\frac{k_BT}{\pi^2\eta} \int_0^\infty dk\, \frac{\sin(kr)}{kr} 
|\widehat{\sigma}(k)|^2. \label{SmoothVelocityCovarianceIso}
\eea
The function $J_\theta(r)$ can then be obtained from the integral 
\eqref{H'/r}.  We now obtain concrete results for the two specific choices 
of filter kernel considered in the main text. 

With the choice of kernel \eqref{DonevKernel} used by DFV, 
\eqref{SmoothVelocityCovarianceIso} becomes after the 
change of variables $x=k\sigma$
\be
K_\theta(r) =  {k_BT\over  \pi^2\eta} {1\over r} \int_0^\infty  
dx\,  {x^4\, \sin(xr/\sigma)
\over x(1+x^2)((\sigma/L)^4+x^4)}. \label{Kfunction1}
 \ee
Although we have worked out the result for finite $L,$ we present here  
only the limit case $L\to\infty$ which gives 
\bea 
K_\theta(r) &=&  {k_BT\over  \pi^2\eta} {1\over r} \int_0^\infty  
dx\,  {\sin(xr/\sigma)
\over x(1+x^2)} \cr 
&=& {k_BT\over 2\pi \eta} 
{1-e^{-r/\sigma}\over r}
\label{Kfunction1}
\eea
using \cite{Erdelyi1954}; formula 2.2(20). 
Substituting this expression into the definition \eqref{H'/r} of 
$J_\theta(r)$ gives by simple integration by parts 
$$
J_\theta(r) =\red{-} {k_BT\over 2\pi \eta\sigma}\left( { {1\over 2}}{1\over {r\over \sigma}} + {e^{-r/\sigma}\over \left({r\over \sigma}\right)^2}-{1-e^{-r/\sigma} \over \left({r\over \sigma}\right)^3}\right) 
$$
and thus the result (\ref{Jtheta}) stated in the text. 
 
We consider next the exponential kernel given by \eqref{exp-sig} in the text, 
or $\widehat\sigma(k)=e^{-k \sigma/\pi}.$ With this choice,  
\bea 
K_\theta(r) &=& {k_BT\over \pi^2\eta} \int_0^\infty dk \frac{\sin(kr)}{kr}
e^{-2\sigma k/\pi} \cr 
&=& {k_BT\over \pi^2 \eta r } \arctan \Big( {\pi r\over 2\sigma} \Big),
\eea
using \cite{Erdelyi1954}, formula 2.4 (1). Substituting into (\ref{H'/r}), 
after integration by parts and some straightforward algebra, yields
$$J_\theta (r)=-{k_BT\over 2\pi^2 \eta r^3}\Big[ 
\Big( {4\sigma^2\over \pi^2}+r^2\Big) \arctan \Big({\pi r\over 2\sigma}\Big)-{2\over \pi} \sigma r \Big]$$
and thus the result (\ref{expJ}) stated in the text. 

\section{Numerical Methods for Plots}\label{AppPlots}

In this appendix, we describe our numerical method in \textsc{Matlab} to plot the concentration spectrum 
for our exact solution (\ref{FFunction102})-(\ref{FFunction104}). From the formula (\ref{exactE}) for 
$E_c(k)$ in terms of $F(k),$ we need to evaluate ${\rm fi}(z)$ and its derivative ${\rm fi}'(z)=-{\rm gi}(z)$. Since the functions {\tt cosint} and {\tt ssinint} 
in \textsc{Matlab} evaluate the cosine and sine integral functions ${\rm Ci}(z)$
and ${\rm si} (z)={\rm Si}(z)-\frac{\pi}{2},$ respectively, the most obvious 
method would be to use \eqref{fi} for ${\rm fi}(z)$ and the analogous
result 
\be
{\rm gi}(z)=-{\rm Ci}(z)\cos z-{\rm si}(z)\sin z; \lb{gi} 
\ee 
see \cite{abramowitz2012handbook}, formula 5.2.7. Unfortunately, this approach
does not work in the asymptotic regime of interest, with $z$ near the imaginary 
axis and of large magnitude. In this region $\cos z,$ $\sin z$ both grow exponentially, but these growing contributions cancel identically in 
${\rm fi}(z),$ ${\rm gi}(z),$ which instead decay. Numerically, evaluating 
these functions using formulas \eqref{fi},\eqref{gi} leads 
to large loss of significance errors in the region of interest. 

We have overcome this problem by alternative expressions for 
${\rm fi}(z),$  ${\rm gi}(z)$ in terms of 
Tricomi's confluent hypergeometric function $U(a, b, z),$ as
\be
 {\rm fi}(z)={i\over 2}\Big(U(1,1,i z)-U(1,1,-i z)\Big), \lb{fiU} \ee
\be {\rm gi}(z)={1\over 2}\Big(U(1,1,i z)+U(1,1,-i z)\Big). \lb{giU} 
\ee
The Tricomi function decays for large $z$ near the real axis,
so that this representation avoids inaccuracy from large cancelling contributions and $U(a, b, z)$ is simply evaluated 
with the function {\tt kummerU} in \textsc{Matlab}. 

The formulas \eqref{fiU},\eqref{giU} can be derived from the standard integral 
representation for Tricomi's function: 
$$U(a, b, z):={1\over \Gamma(a)} \int_0^\infty e^{-zt} t^{a-1} (1+t)^{b-a-1} dt,$$
for ${\rm Re}(z)>0$, ${\rm Re}(a)>0$ and with the Gamma function $\Gamma(a)$. 
See \cite{abramowitz2012handbook}, formula 13.2.5. Taking $a=b=1$, and 
$z\rightarrow \pm iz$, we find 
\be U(1, 1, \pm iz)=\int_0^\infty {e^{\mp it}\over z+t}dt= {\rm gi}(z)\mp i\, {\rm fi}(z), \lb{Ugf} \ee
where we used (\ref{fi-def}) and the corresponding integral formula 
\be {\rm gi}(z)=\int_0^\infty \frac{\cos t}{t+z} dt = \int_0^\infty \frac{t e^{-zt}}{1+t^2} dt, 
\quad {\rm Re}(z)>0. \lb{gi-def} \ee 
See \cite{abramowitz2012handbook}, section 5, formula 5.2.13 and 
\cite{oldham2010atlas}, section 38:13. The formulas \eqref{fiU}, 
\eqref{giU} follow directly from \eqref{Ugf}.  

\section{Thermodynamics of Binary Mixtures}\lb{AppMix} 

We briefly review here the results on thermodynamics of binary mixtures 
required in the main text. We start with the first law of thermodynamics
in the form 
$$ du=T ds-p\, dv +\mu_0 d\nu_0+\mu_1 d\nu_1 $$
where $u=U/M$ is specific energy, $s$ is specific entropy, $v=V/M$ is specific 
volume, and $\nu_i=N_i/M,\ i=0,1$ are the specific particle numbers 
of the two species (solvent, solute). Mass fractions or mass concentrations 
of the two species are defined by 
$$c:=c_1=m_1\nu_1=M_1/M, \quad c_0=m_0\nu_0=M_0/M.  $$
From $c_0+c_1=1$ one then easily obtains 
$$ du=T ds-p\, dv +\mu\, dc $$
where the chemical potential per mass is given by 
$$ \mu = \frac{\mu_1}{m_1} - \frac{\mu_0}{m_0} $$ 
Cf. \cite{landau1959fluid}, Ch.VI, \S 57. 
One can also introduce the molar fractions or molar concentration 
$n_i=N_i/N,$ $i=0,1,$ which are easily related to the mass concentrations by  
$$ n=\frac{m_0 c}{m_0c+m_1(1-c)} $$
with $n:=n_1.$

An {\it ideal mixture} by definition is one in which the chemical 
potential of each component in solution satisfies 
$$ \mu_i= \mu_i^\emptyset(T,p)+ k_BT \ln(n_i) $$
where $\mu_i^\emptyset$ is the chemical potential of the pure substance. 
Note then that 
$$ \mu = \mu^\star(T,p)+ \frac{k_BT}{m_1} \ln(n) 
-\frac{k_BT}{m_0} \ln(1-n). $$
A straightforward calculation gives
\be \left( \frac{\partial\mu}{\partial c}\right)_{T,p}=
\frac{k_BT}{c(1-c)[m_0 c+m_1(1-c)]}. \lb{F1} \ee

Chemical potentials of non-ideal mixtures are generally written 
in the form 
$$ \mu_i= \mu_i^\emptyset(T,p)+ k_BT \ln(n_i f_i) $$
where $f_i$ is the {\it activity coefficient} which takes into account the 
non-ideality of the solution. E.g. see \cite{tyrrell2013diffusion}. 
Note from the Gibbs-Duhem relation 
$\nu_0 d\mu_0 + \nu_1 d\mu_1 = -s dT + v dp$ and from the condition 
$n_0 + n_1=1$ that, at constant $T$,$p$,
$$ n_0 d(\ln f_0)+ n_1 d(\ln f_1)=0.   $$
Defining 
$$ B_i= 1+ \frac{d\ln f_i}{d\ln n_i}\Big|_{T,p}, \quad i=0,1 $$
it then follows that $B_0=B_1:=B.$
Furthermore, from this definition, at constant $T,$ $p,$
$$ \frac{d}{dc}\ln(n_if_i)= \frac{d}{dc}\ln(n_i) \cdot B $$
and thus for a non-ideal mixture
\be \left( \frac{\partial\mu}{\partial c}\right)_{T,p}=
\frac{B\cdot k_BT}{c(1-c)[m_0 c+m_1(1-c)]}. \ee

\section{Survey of Experiments on Turbulent High-Schmidt Mixing}\label{survey}

As a convenience for readers, we here briefly survey experiments known to us on 
high Schmidt-number turbulent advection. These experiments all differ considerably 
from each other, both in the turbulent flows considered and also in the fluid mixtures 
employed, which include water-fluorescein \cite{miller1996measurements, williams1997mixing, jullien2000experimental}, salt-water \cite{gibson1963universal, grant1968spectrum}, and 
ink in water\cite{nye1967scalar}. We additionally consider here experiments 
which studied turbulent mixing of temperature fluctuations at high Prandtl numbers,
in order to expand our view of the range of parameters which can be practically achieved. 
We shall briefly describe each experiment and the physical parameters stated in the paper. 
In addition, some further parameters could be calculated with the reported quantities 
and with data extracted from the published figures, and we describe our methods for this. 
We shall discuss the main experiments of which we are aware, in chronological order.  


We start with the experiment performed by Gibson \& Schwartz \cite{gibson1963universal} who used a single-electrode conductivity probe in a bridge circuit to measure the spectra and decay of  homogeneous fields of both concentration and temperature behind a grid in dilute salt water at $Re\simeq 10^4$ and who reported Batchelor spectrum in the viscous-convective range. Here $\nu\simeq 10^{-2} {\rm cm}^2/{\rm s}$ and $D\simeq 1.5\times 10^{-5} {\rm cm}^2/{\rm s}$, thus $Sc\simeq 666.7$. The Batchelor scale can be obtained using $\kappa_B=\kappa_K\sqrt{Sc}$ in terms of the Kolmogorov wavenumber $\kappa_K$ (which is denoted by $k_s$ and given in Table.1 in \cite{gibson1963universal}). For $6$ different runs in this series of experiments, we find $\kappa_B\simeq 761$; $2631$; $4257$; $2505$; $1437$ and $3444\,{\rm cm}^{-1}$, respectively, for CM1 through CM17 in Table.1. Using $\gamma=D/\ell_B^2\equiv \kappa_B^2 D$, for $6$ runs CM1 through CM17, we find $\gamma\simeq 8.7$; $103.8$; $271.8$; $94.1$; $31$ and $177.9\ {\rm s}^{-1}$, respectively. The injection rate of concentration fluctuations is given by $\chi={3\over 2}{U\over x} \overline{\theta^2}$ with the variance of concentration or temperature fluctuations denoted as $\overline{\theta^2}$, where velocity $U$ and distance from the grid $x$ are given in Table 1 for different runs. With  $0.57\leq {U\over x}\leq 4.7$, and $\overline {c^2}\sim 10^{-12}$ from Fig.(1), we estimate $\chi\sim 10^{-12}-10^{-11}\;s^{-1}$.

In another set of experiments, Nye \& Brodkey \cite{nye1967scalar} studied commercial blue ink in water flowing through a pipe with a fibre optic light probe and reported a full $1.5$ decades of $1/k$ spectrum, starting near the velocity spectrum cutoff, which was observed to be at about $0.1 \kappa_K$ with Kolmogorov wavenumber $\kappa_K$. 
The diffusivity of the dye was given as $D=2.6\times 10^{-6}\;{\rm cm}^2/{\rm s}$ and viscosity (of water) is $\nu\simeq 10^{-2}\;{\rm cm}^2/{\rm s}$, hence the Schmidt number should be around $Sc\simeq 3800$. The Kolmogorov wavenumber is given as $\kappa_K\simeq 62\;{\rm cm}^{-1}$, thus the Batchelor wavenumber is given by $\kappa_B\red{\equiv \ell_B^{-1}}=\kappa_K\sqrt{Sc}\simeq 3822 \;{\rm cm}^{-1}$ using which we also find $\gamma\simeq  D/\ell_B^2 \simeq 38\;{\rm s}^{-1}$. 
The energy dissipation rate $\varepsilon$ can be estimated using $\varepsilon\simeq \gamma^2 \nu \simeq 14.4\;{\rm cm}^2/{\rm s}^3$ which also agrees with $\varepsilon\simeq \nu^3 \kappa_K^4$ as expected. Finally, using 
\cite{nye1967scalar}, Fig.4 for the scalar spectrum $E_c(k)$, we estimated $\int k^2 E_c(k) dk$ for the three cases presented, by extracting data for $E_c(k)$ and then numerically integrating $k^2 E_c(k)$, and found that the scalar injection rate should be of order $\chi\simeq 10^{-4}-10^{-2}\;{\rm s}^{-1}$ in this set of experiments.

An experiment of Grant et al. \cite{grant1968spectrum} measured temperature and velocity fluctuations in the open sea and a tidal channel and they reported observing Batchelor's spectrum over at least one decade in the viscous-convective range. In these experiments, the injection rate for temperature fluctuations varied from $\chi\simeq 7.2\times 10^{-8}\;{\rm degC}^2/{\rm s}$ to $\chi\simeq 5.2\times 10^{-4}\;{\rm degC}^2/{\rm s}$; see Table.1 in \cite{grant1968spectrum}. Since these experiments dealt with temperature field, in order to compare the corresponding 
injection rate (in units of ${\rm degC}^2/{\rm s}$) to those corresponding to a concentration field (in units of $(\% {\rm concent.})^2/{\rm s}$), we converted the reported rates. Extracting data from \cite{grant1968spectrum}, 
Figs. 6-11, we calculated the temperature fluctuations, which is given in terms of the spectrum $\psi(k_1)$ as $\langle \Delta T^2\rangle\simeq \int \psi(k_1) dk_1$. For Figs. 6-11, respectively, we get $\langle \Delta T^2\rangle \sim 10^{-3}$; $10^{-2}$; $10^{-3}$; $10^{-4}$; $10^{-4}$; $10^{-4}\,{\rm degC}^2$. Therefore, the quantity $\chi/\langle \Delta T^2\rangle$ will be in the range $\chi_{min}/\langle \Delta T^2\rangle_{max}$ through $\chi_{max}/\langle \Delta T^2\rangle_{min}$ which turns out to be of order $\sim10^{-6}-1\;{\rm s}^{-1}$. Also shown in Table 1 of \cite{grant1968spectrum}
is the energy dissipation rate $\varepsilon$, which varies in the range $4.4\times 10^{-4}-5.2\times 10^{-1}$ (${\rm cm}^2/{\rm s}^3$). Taking viscosity of order $\nu\sim 10^{-2}\;{\rm cm}^2/{\rm s}$ for water, we can use $\gamma\sim\sqrt{\varepsilon/\nu}$ to estimate $\gamma$. Above values for $\varepsilon$ translate into values in the range 
$\gamma\sim 0.2\;{\rm s}^{-1}$ through $\gamma\sim 7.2\;{\rm s}^{-1}$.

Miller \& Dimotakis \cite{miller1996measurements} used a mixture of water and fluorescein, with diffusivity $D\simeq 5.2\times 10^{-6}\;{\rm cm}^2/{\rm s}$, to investigate the temporal, scalar power spectra of high Schmidt number turbulent jets with Reynolds number of order $Re\simeq 10^4$ and Schmidt number $Sc\simeq 1.9\times 10^3$. At the smallest scales, the measured spectra were reported not to exhibit Batchelor's $1/k$ power-law behaviour, but, rather, seemed to be approximated by a log-normal function, over a range of scales exceeding a factor of $40$, in some cases. At $x/d=305$ ($x$ distance from the injection nozzle and $d$ nozzle diameter), the Kolmogorov scale was reported as $\ell_K\simeq 2.57\times 10^{-2}\;{\rm cm}$. Using $\ell_K\simeq \Big({\nu^3\over\varepsilon} \Big)^{1/4}$, and taking $\nu\simeq 10^{-2}\;{\rm cm}^2/{\rm s}$) for the viscosity of water, we get kinetic energy dissipation rate $\varepsilon\simeq 2.29\;{\rm cm}^2/{\rm s}^3$. Using $\ell_B\simeq\ell_K/\sqrt{Sc}$, we find the Batchelor scale $\ell_B\simeq 5.8\;\mu$m. Finally, using  $\varepsilon\simeq \gamma^2\nu$, \red{or equivalently using the relation $\gamma=D/\ell_B^2$, we find $\gamma\sim 15\;{\rm s}^{-1}$.} In this set of experiments, unfortunately, we were unable to obtain a certain estimate for the scalar injection rate.

Williams et al. \cite{williams1997mixing} employed fluorescein dye in a quasi-two-dimensional turbulent flow to investigate the Batchelor regime at Schmidt number around $Sc\simeq 2000$. They reported that the spectrum falls below $k^{-1}$ at wavenumbers lower than expected from theory. In order to estimate $\kappa_B$, the authors first considered the competing effects on a dye structure due to the stretching produced by the large-scale flow and also due to the dissipation by diffusion and found $\kappa_B/2\pi\simeq 35\,{\rm cm}^{-1}$. They also estimated $\kappa_B$ by measuring $\gamma$: to determine the rate of strain tensor on a regular grid, the authors obtained the velocity derivatives from their velocity field measurements, then diagonalized this tensor at each location and finally ensemble-averaged over space and time. The strain rate determined in this way turns out to be $\gamma\simeq 1.0\,{\rm s}^{-1}$. Using $E_c(k)={\chi\over \gamma k}
\exp\left(-D k^2/\gamma\right)$, with $D=5\times 10^{-6}\;{\rm cm}^2/{\rm s};$ and $\gamma=0.5(\varepsilon/\nu)^{1/2}$ (and defining $\kappa_B$ as the wavenumber for which the exponent is $-1$) the authors found $\kappa_B\simeq 70\,{\rm cm}^{-1}$. The authors reported a spectral slope steeper than Batchelor's value $-1$ and more so with increasing wavenumber, while including a Gaussian tail did not improve the fit. The inclusion of an exponential tail is reported to yield a satisfactory fit but only when unphysical parameters are used. The authors fit the data shown in their Fig. 17(b) to a spectrum of the form $E_c(k)=C k^{-1} \exp{(-k\sqrt{\kappa/\gamma_{e\! f\! f}})}$ and found $\gamma_{e\! f\! f}\simeq 1.1\times 10^{-3}\,{\rm s}^{-1}$, which is drastically different from their directly measured value of $\gamma\simeq 1.0\,{\rm s}^{-1}$. Putting all this together, the value $\gamma\simeq 10^{-3}\, {\rm s}^{-1}$ for the least strain rate seems possible, but quite small and difficult to reconcile with 
the value $\gamma\sim 1.0\, {\rm s}^{-1}$ from direct velocity derivative measurements. 
As for the scalar injection rate, by extracting data from \cite{williams1997mixing}, Fig.22 for $E_c(k)$, 
we estimated $\chi\simeq 2D \int k^2 E_c(k) dk\sim 10^{-5}\;{\rm s}^{-1}$.

Jullien et al. \cite{jullien2000experimental} employed an electromagnetically driven two-dimensional flow of a water-fluorescein mixture, in which Batchelor's $1/k$ spectrum was observed,  with exponential tails for the probability distributions of the concentration and concentration increments and logarithm-like structure functions. In this set of experiments, the flow was statistically stationary, whereas the concentration field was in a freely-decaying 
quasi-equilibrium regime starting from an initial circular 5$\,$cm blob of fluorescein. The results were also 
confirmed by a simulation where the scalar advection equation was solved using observed values of the advecting velocity. The trajectories of $6\times 10^5$ particles, located initially in a disk, $5\,$cm in diameter, were calculated by integrating the experimentally measured velocity field and used to estimate the evolved concentration spectrum at negligible diffusivity. For this experiment, we used Fig.2 (b) and Fig.(3) in \cite{jullien2000experimental} to extract data for $E_c(k),$ from which we estimated $\chi=2D\int E_c(k)k^2 dk$ by numerical integration. In this way, $\chi$ turns out to be of order $2\times 10^2\;{\rm s}^{-1}$, roughly in agreement with the inset of \cite{jullien2000experimental}, Fig.2. Also, with given $\kappa_B=2800\,{\rm cm}^{-1}$, \red{and the relation $\kappa_B=\sqrt{\gamma/\kappa}$ used in \cite{jullien2000experimental} with diffusivity $ \kappa\simeq 10^{-6} \, {\rm cm^2/s} $}, we find $\gamma\simeq 7.8\,{\rm s}^{-1}$.

\red{Finally, in the most recent experiment of which we are aware, Iwano et al. \cite{iwano2021power}
studied concentration fluctuations of the fluorescent dye Rhodamine 6G in an axisymmetric turbulent 
water jet at the Schmidt number $Sc\simeq 2.9\times 10^3$ and Reynolds number $Re\simeq 2.0 \times 10^4$.
The authors observed the $1/k$ Batchelor spectrum in the viscous-convective range although they also report that the spectrum does not perfectly follow this scaling due to ``a small bump in the range''. The Batchelor and Kolmogorov length scales in this experiment are given as $\eta_B=2.6\;{\rm \mu m}$ and $\eta_K=142\;{\rm \mu m}$, respectively. 
For the injection rate, we have $\gamma\simeq \eta_B^{-2} D\simeq 59 \;{\rm s}^{-1}$, using $D\simeq 4.0\times 10^{-6} {\rm cm^2} {\rm s}^{-1}$  \cite{Gendron2008}. The latter result can be used to obtain the energy dissipation rate; $\epsilon\simeq \nu \gamma^2 =Sc\times D\times \gamma^2\simeq 40.4 \;{\rm cm^2} {\rm s}^{-3}$. As for the scalar injection rate, $\chi$, by extracting data from \cite{iwano2021power}, Fig.3 for $E_c(k)/C_0^2$, we numerically estimated the quantity $\int k^2 {E_c(k)\over C_0^2} dk$ cutting off the the integral at the Batchelor wavenumber. For different values of the initial concentration $C_0$, given as $1$, $10$, $100$, $1000$, the value of this integral lies roughly between $ 10^4$ and $ 10^5$. Using $\chi=2D C_0^2\int k^2 {E_c(k) \over C_0^2}dk$, we can estimate the range of $\chi$ values for different choices of $C_0$ in this experiment as $10^{-4}\;{\rm s}^{-1} \lesssim \chi \lesssim 10 \;{\rm s}^{-1}.$}

\section{GCF's in a 2D Quiescent Fluid}
\label{2DGCF}
We obtain here the steady-state concentration correlation function $C(r)$  and the corresponding spectrum $E_c(k)$
for a 2D fluid, with velocity field in thermal equilibrium  and with concentration fluctuations 
generated by a stochastic source \eqref{SourceCorrelation} that is white-noise in time. 
Similar to the $3D$ case discussed in \S\ref{ourmodel}, we will solve the general expression 
\eqref{2CorrelationKraichnan} for the $2$-point correlation function and then obtain the spectrum 
by Fourier transformation. 
However, the analysis is somewhat more difficult in 2D than in 3D. First, the effective 
diffusivity $D_{e\!f\!f}$ is well-known to be logarithmically divergent in 2D 
(e.g. see \cite{donev2014reversible} and \eqref{2D-Deff} below). Thus, the calculation
requires an IR cutoff $L$ in addition to a UV cutoff $\sigma,$ and we shall be concerned 
with evaluating $C(r)$ only for $\sigma\ll r\ll L.$ Second, Fourier transforms 
of isotropic correlation functions do not reduce to a Fourier cosine transform as in 3D,
but instead require in 2D a more difficult Hankel transform. 

The covariance $\bR(\bx,\bx')$ defined in \eqref{SmoothVelocityCovariance102} can be evaluated for homogeneous, isotropic statistics in $d=2$ dimensions using the $2D$ version of the expression given by the first line of \eqref{SmoothVelocityCovarianceIso} in terms of $K_\theta(r)$, defined by (\ref{iso-V}), and $J_\theta(r)$, given by \eqref{H'/r}. We have
\bea 
K_\theta(r) &=& \frac{1}{(2\pi)^2} \int d^2{\bf k}\ e^{i{\bf k.r}}  \,  
|\widehat{\sigma}(k)|^2 \frac{2 k_B T}{\eta k^2} \cr
&=&\frac{k_BT}{\pi \eta} \int_0^\infty dk\, \frac{J_0(kr)}{k} 
|\widehat{\sigma}(k)|^2,
\eea which is a Hankel transform, with the Bessel function of the first kind $J_0(kr)$; see e.g., \cite{bateman2006higher}, 7.3.1 (2). 
We will use an isotropic filter $\bsigma=\sigma {\bf I}$ whose Fourier transform is given by
\be\label{2DKernel} 
\widehat\sigma(k):={k\over \sqrt{k^2+L^{-2} }}e^{-{\sigma\over 2}\sqrt{k^2+L^{-2}}},
\ee with IR cut-off $L$ and UV cutoff $\sigma\ll L$. Aside from required properties for such a smoothing filter, we have made this particular choice due to the fact that it leads to tractable Hankel transforms. It follows that
\be\label{2DK}
K_\theta(r)={k_BT\over \pi \eta}\int_0^\infty dk\, J_0(kr) {k\over k^2+L^{-2} }e^{-\sigma\sqrt{k^2+L^{-2}}}.
\ee This particular example has not been presented in standard tables of Hankel transforms, as far as we are aware. However, 
it can  be obtained from formula 8.2 (24) in \cite{Erdelyi1954II}, which in our notation reads
$$\int_0^\infty {k e^{-\alpha \sqrt{k^2+L^{-2}}} \over \sqrt{k^2+L^{-2}}}J_0(kr) dk={e^{-\sqrt{\alpha^2+r^2}/L } \over  \sqrt{r^2+\alpha^2}},$$
by integrating over the parameter $\alpha$ from $\sigma$ to infinity. Integrating the LHS of the above expression over $\alpha$, from $\sigma$ to $+\infty$, is straightforward and gives the integral (\ref{2DK}). On the other hand, the integral of the RHS of the above expression over $\alpha$ can be written as
\bea\nonumber
\int_\sigma^\infty {e^{-\sqrt{\alpha^2+r^2}/L } \over  \sqrt{r^2+\alpha^2}}d\alpha &=&\int_0^\infty {e^{-\sqrt{\alpha^2+r^2}/L } \over  \sqrt{r^2+\alpha^2}}d\alpha\\\nonumber
&&-\int_0^\sigma {e^{-\sqrt{\alpha^2+r^2}/L } \over  \sqrt{r^2+\alpha^2}}d\alpha,
\eea where the integral, from $\sigma$ to $+\infty$, can be obtained using \cite{Erdelyi1954} formula 1.4 (27) while the remaining integral in the second line can be estimated by making use of the approximation $\int_0^\sigma { \over } d\alpha { e^{-{\sqrt{r^2+\alpha^2}/ L} }\over  \sqrt{r^2+\alpha^2}}\simeq {\sigma\over r}e^{-r/L}$ for $\sigma\ll r$. Putting all this together, we find
\be\label{2DK2}
K_\theta(r)\simeq {k_BT\over \pi \eta}\Big[ K_0\Big({r\over L} \Big)-{\sigma\over r}e^{-r/L}\Big],
\ee where $K_0(x)$ (not to be confused with $K_\theta (r)$) is the modified Bessel function of the second kind; 
see e.g., \cite{bateman2006higher}, 7.2.2. The term proportional to $\sigma$ is much smaller than the first term involving 
$K_0(r/L)$ for $\sigma\ll r\ll L,$ but we retain it here because the analogous small term gave rise in 3D 
to a non-analytic contribution to $C(r)$ which was responsible for the GCF's. However, we remark in advance that the 
contribution from this term will be found in 2D to be negligible compared with that from the Bessel function. The major difference from 3D 
is that the larger contribution to $C(r)$ from the Bessel function 
is also non-analytic in $r$ in 2D. 

With $K_\theta(r)$ at hand, we can now proceed to obtain $J_\theta(r)$ (not to be confused with Bessel function $J_0(kr)$) using its definition given by (\ref{H'/r}) with $d=2$:
\be
J_\theta(r)={k_BT\over \pi\eta}{L\over r^2}\Big[r K_1\Big({r\over L} \Big)-L+\sigma(1-e^{-r/L})  \Big].
\ee To get this result, we have used the relation $\int z K_0(z)dz=-z K_1(z)$ in terms of the modified Bessel function of the second kind $K_1(z)$; see \cite{bateman2006higher}, 7.14.1 (3). Note that \eqref{2DK2} is valid only for $r\gg \sigma$ while to calculate $J_\theta(r)$, we need to integrate from $0$ to $r.$ Nevertheless, the error in this calculation is of order $(\sigma/r)^2$, which can be easily verified by writing $J_\theta(r)={-1\over r^2}\Big[\int_0^{N\sigma} \rho K_\theta(\rho) d\rho+\int_{N\sigma}^r \rho K_\theta(\rho) d\rho \Big]$ in terms of a fixed $N\gg 1$ such that our asymptotic relations used to evaluate the second integral remain valid. Even for a large $N$, we can take $\sigma$ small enough such that the first integral gives a negligible contribution of order $K_\theta(0)(N\sigma/r)^2$ assuming that for $r\ll N\sigma$, $K_\theta(r)\rightarrow K_\theta(0)$. The function $K_\theta(0)$ has only a logarithmic dependence on $\sigma/L$, as shown immediately below.

In order to calculate $J_\theta(0)$, we note that in $2D$ the expression \eqref{H'/r} implies $J_\theta(0)=-{1\over 2}K_\theta (0)$. We can write
\bea
K_\theta(0)&=& {k_BT\over \pi \eta}\int_0^\infty {k dk\over k^2+L^{-2}}e^{-\sigma\sqrt{k^2+L^{-2}}}\\\nonumber
&=&{k_BT\over \pi\eta}E_1\Big({\sigma\over L}\Big).
\eea The first line in the above expression in fact defines the exponential integral function $E_1(x)=-{\rm Ei}(-x)=\int_x^\infty {e^{-\kappa}d\kappa\over\kappa}$ as can be seen by a simple change of variable as $\kappa=\sigma \sqrt{k^2+L^{-2}}$; see \cite{bateman2006higher}, 9.7 (1). Thus we find $J_\theta(0)={-K_\theta(0)\over 2}={-k_BT\over 2\pi\eta}E_1(\sigma/L)$ and consequently
\bea\nonumber
\triangle J_\theta(r)\simeq{k_BT\over \pi\eta}\Big[&& -{1\over 2} E_1\Big({\sigma\over L}\Big)+\Big({L\over r}\Big)^2-\Big({L\over r}\Big)K_1\Big({r\over L}\Big)\\\nonumber
&&-{\sigma L\over r^2}(1-e^{-r/L}) \Big].
\eea
Next, for $r/L\ll 1$, we use the asymptotic relations $K_1(r/L)\simeq L/r+I_1(r/L)\ln (r/2L)$ where $I_1(x)$ is the modified Bessel function of the first kind; see \cite{bateman2006higher}, 7.2.5, (37), and also $I_1(r/L)\simeq r/2L$; \cite{bateman2006higher}, formula 7.2.2 (12). In addition, we expand the exponential function to the first order in $\sigma/r\ll 1$; $e^{-r/L}\simeq 1-r/L$, to write the final expression for $\triangle J_\theta(r)$ as
\bea
\triangle J_\theta(r)\simeq {k_BT\over \pi\eta}\Big[&& -{1\over 2}E_1\Big({\sigma\over L}\Big)-{1\over 2}\ln \Big({r\over 2L}\Big)
-{\sigma\over r}\Big], \cr 
&& \hspace{40pt} \sigma\ll r\ll L. \label{2DdeltaJ}
\eea

 With $\triangle J_\theta(r)$ at hand, we proceed to calculate the steady-state correlation function $C(r)$ given by the integral 
\eqref{2CorrelationKraichnan}. Analogous to the $3D$ calculation presented in \S\ref{ourmodel}, however, it is easier to work with the derivative $\partial_r {C}(r)$ and also, since $r\ll L$, to take  $S(r/L)\simeq S(0)=2\chi$. Hence, we proceed by inserting (\ref{2DdeltaJ}) in $\partial_r C(r)$, obtained by taking the derivative of (\ref{2CorrelationKraichnan}): 
\be \partial_r C\simeq {-\chi r\over 2}\Big[D+{k_BT\over 4\pi\eta}\ln\Big({r\over 2L}\Big)+{k_BT\over 2\pi\eta}{\sigma\over r} \Big]^{-1}.
\lb{dCdr-S} 
\ee
Here we have defined the ``renormalized diffusivity'' $D$ as
\be
D:=D_0+{k_BT\over 4\pi\eta}E_1\Big({\sigma\over L}\Big)
\doteq D_0-\gamma+{k_BT\over 4\pi\eta}\ln\left(\frac{L}{\sigma}\right)
\lb{2D-Deff} 
\ee
with bare diffusivity $D_0$ and $\gamma=-\Gamma'(1)$. Factoring out a $D$ from the 
square bracket ``$[\cdot]$'' in \eqref{dCdr-S}, note that both the logarithmic term $\propto \ln(r/L)/\ln(\sigma/L)$
and last term containing $\sigma/r$ in this expression are small and we can expand using $1/(1+x)\simeq 1-x$ for small $x$: 
\be\nonumber
\partial_r C(r)\simeq-{\chi r\over 2 D}\Big[1-{ k_BT\over 4\pi\eta D}\ln\Big({r\over 2L}\Big)-{k_BT  \over 2\pi\eta D} {\sigma\over r}\Big].
\ee Integrating this expression over $r$, from $N\sigma$ to $r$, we find
\bea\nonumber
C(r)=-{\chi\over 2D}\Big[&&{1\over 2}r^2-{k_BT\over 16\pi\eta D}r^2 \left(2\ln\Big({r\over 2L}\Big)-1\right) \\\label{2DC(r)}
&&-{k_BT\over 2\pi\eta D}\sigma r \Big]+const.
\eea
The result is very much like \eqref{GCF100} obtained in 3D, except that the middle term containing 
$\ln(r/2L)$ is non-analytic, whereas the corresponding term in 3D was proportional to $r^4$ and analytic. 
We note also that the term proportional to $\sigma$ has a negative sign, opposite to that in 3D.

We can obtain the spectrum using the general expression $E_c(k)={k\over 2}\int_0^\infty r C(r) J_0(kr) dr$ in $2D$. Performing this Hankel transform, the last term in (\ref{2DC(r)}), which is linear in $r$ and $\sigma$, yields a negative contribution:
\bea\label{1/k^2}
{k\over 2} {\chi\sigma\over 2D}{k_BT\over 2\pi \eta D}\int_0^\infty r^2J_0(kr) dr=-{\chi\sigma\over 2D}{k_BT\over 4\pi\eta  D}k^{-2}.
\eea Here, we have evaluated the integral by introducing an IR cut-off $\mu$; $\lim_{\mu\rightarrow 0}\int_0^\infty r^2 e^{-\mu r} J_0(kr) dr$, which can be calculated using \cite{Erdelyi1954II}, formula 8.2 (20).
The quadratic term in (\ref{2DC(r)}), that is ${-\chi\over 4D}\Big(1+{k_BT\over 8\pi\eta D}\Big)r^2$ will contribute only a rapidly decaying spectrum. This can be easily confirmed by directly taking the integral $\int_0^\infty r^3 J_0(kr) dr$ using e.g., \cite{Erdelyi1954II}, formula 8.2 (20), once again introducing an exponential IR cut-off $\mu$ and taking the limit $\mu\rightarrow 0$, which gives a vanishing contribution. 

Finally, to obtain the contribution to the spectrum of the logarithmic term in (\ref{2DC(r)}), we need to evaluate 
\bea\nonumber
\int_0^\infty r^3 \ln(r/2L) J_0(kr) dr&=&\int_0^\infty r^3 \ln(r) J_0(kr) dr\\\nonumber
&&-\ln(2L) \int_0^\infty r^3  J_0(kr) dr.
\eea The integrals on the RHS both can be evaluated using a general property of Hankel transforms, which gives the Hankel transform (of arbitrary order $\nu$) of function $x^m f(x)$, $m=0, 1, 2,\dots$, in terms of the $m$th order derivative of the Hankel transform (of order $\nu+m$) of function $f(x)$; see \cite{Erdelyi1954II}, 8.1 (3). This general property allows us to write the two integrals as
$$\int_0^\infty r^3 J_0(kr) dr = 
\Big({d\over kdk}\Big)^3\Big[ k^3 \int_0^\infty J_3(kr) dr \Big]$$
and 
$$\int_0^\infty r^3\ln r  J_0(kr) dr = 
\Big({d\over kdk}\Big)^3\Big[ k^3 \int_0^\infty \ln r J_3(kr) dr \Big]$$
in terms of the Bessel function $J_3(z)$ of the first kind.
These integrals can then be evaluated using 
$$\int_0^\infty J_3(kr) dr=k^{-1} $$
from \cite{Erdelyi1954II}, 8.5 (3) and 
$$\int_0^\infty \ln(r) J_3(kr) dr=(2\psi(2)-\ln(k^2/4))/2k,$$
from \cite{Erdelyi1954II}, 8.6 (25), for $\mu=-1/2,$ $\nu=3,$
where $\psi(z)=\Gamma'(z)/\Gamma(z)$ is the logarithmic derivative of the Gamma 
function $\Gamma(z).$ It is then readily seen that only the logarithmic term 
yields a non-vanishing contribution to the required integral:  
\bea \int_0^\infty r^3 \ln(r/2L)J_0(kr) dr&=&
-{1\over 2}\Big({d\over k dk} \Big)^3\Big[ k^2\ln(k^2/4)\Big] \cr 
&=&4 k^{-4}\eea
and thus the contribution of the logarithmic term in \eqref{2DC(r)} to the 
spectrum is given by 
$$ {\chi\over 2D}{k_BT\over 16\pi\eta D} k \int_0^\infty r^3 \ln(r/2L)J_0(kr) dr
= {\chi\over 2D}{k_BT\over 4\pi\eta D} k^{-3}.$$  
 
 Combining this result with \eqref{1/k^2}, therefore, we find the $2D$ spectrum 
of the concentration field to be 
\bea
E_c(k)&=&{\chi\over 2D}{k_BT\over 4\pi\eta D}(k^{-3}-\sigma k^{-2}), \cr
&& \hspace{40pt} 1/L\ll k\ll 1/\sigma. 
\eea
It is interesting that the term proportional to $\sigma$ has the power-law 
scaling $k^{-2}$, just as in 3D. However, the sign of this term is 
{\it negative} in 2D and it does not become sizable compared with the first 
term until $k\simeq 1/\sigma,$ when our asymptotic approximations break down.
In contrast to 3D where the leading term in $C(r)$ was analytic in $r$
and gave a contribution to the spectrum very rapidly decaying in $k$, this term 
is now non-analytic because of the $\ln r$ and produces a power-law. 
Note that the resulting spectrum is again in close correspondence with the 
prediction of linearized fluctuating hydrodynamics. Using the relation 
in 2D between the spectrum and the structure function,
$$E_c(k)=\frac{1}{2} \frac{1}{(2\pi)^2}2\pi k S_{cc}(k)=\frac{1}{4\pi} k S_{cc}(k) $$
and using the relation $\chi=D(\grad c)_{e\! f\! f}^2$ just as in 3D yields
$$ S_{cc}(k)\simeq \frac{1}{2} \frac{k_BT}{D\eta} |\grad c|_{e\! f\! f}^2 k^{-4}. $$
This agrees with the prediction of linearized fluctuating hydrodynamics up 
to a factor of $1/2.$ Based on the results in 2D and in 3D we may conjecture
that the result of the DFV theory for the case of a random concentration gradient 
$\grad c$ with homogeneous, isotropic statistics and the prediction of linearized 
fluctuating hydrodynamics for a fixed mean concentration gradient $\grad\bar{c}$ will agree 
in any space dimension $d$, up to a factor of $(d-1)/d.$

\newpage

\bibliography{HighSchmidt}

\end{document}